\scrollmode
\documentclass[useAMS,usenatbib, proof, a4paper]{mn2e}
\usepackage{bbding}
\usepackage{natbib}
\usepackage{graphicx}
\usepackage{times}
\usepackage{amssymb}
\usepackage[fleqn]{amsmath}
\usepackage[draft]{hyperref}
\usepackage{wasysym}
\usepackage{textcomp}
\usepackage{enumitem}
\usepackage[latin1]{inputenc}
\newcommand{\beq}{\begin{equation}}
\newcommand{\eeq}{\end{equation}}

\newcommand{\lsim}{\ \raise
-2.truept\hbox{\rlap{\hbox{$\sim$}}\raise5.truept\hbox{$<$}\ }}
\newcommand{\gsim}{\ \raise
-2.truept\hbox{\rlap{\hbox{$\sim$}}\raise5.truept\hbox{$>$}\ }}
\newcommand{\simsim}{\ \raise
-2.truept\hbox{\rlap{\hbox{$\sim$}}\raise5.truept\hbox{$\sim$}\ }}

\def\gtorder{\mathrel{\raise.3ex\hbox{$>$}\mkern-14mu
                \lower0.6ex\hbox{$\sim$}}}
\def\ltorder{\mathrel{\raise.3ex\hbox{$<$}\mkern-14mu
                \lower0.6ex\hbox{$\sim$}}}

\def\aj{AJ}                   
             
\def\apj{ApJ}                 
\def\apjl{ApJL}                
\def\apjs{ApJS}

\def\aap{A\&A}

\def\mnras{MNRAS}

\title[Pre-Main-Sequence Stars in the Tarantula Nebula]{Hubble Tarantula Treasury Project -- VI. Identification of Pre--Main-Sequence Stars using Machine Learning techniques}

\author[V. F. Ksoll et al.]
{Victor F. Ksoll,$^{1,2,}$\thanks{E-mail: v.ksoll@stud.uni-heidelberg.de}
	Dimitrios A. Gouliermis,$^{1,3,}$\thanks{E-mail:
		gouliermis@uni-heidelberg.de}
	Ralf S. Klessen,$^{1}$
	Eva K. Grebel,$^{4}$
	\newauthor
	Elena Sabbi,$^{5}$
	Jay Anderson,$^{5}$
	Daniel J. Lennon,$^{6}$
	Michele Cignoni,$^{7}$
	Guido de Marchi,$^{8}$
	\newauthor
	Linda J. Smith,$^{9}$
	Monica Tosi,$^{10}$
	and Roeland P. van der Marel$^{5}$	
\\
\\
	$~~^1$Institut f\"ur Theoretische Astrophysik, Zentrum f\"ur Astronomie der Universit\"at Heidelberg, Albert-Ueberle-Str.\,2, 69120 Heidelberg, Germany\\
	$~~^2$Interdisciplinary Center for Scientific Computing,
	University of Heidelberg, Mathematikon, Im Neuenheimer Feld 205, 69120
	Heidelberg, Germany \\
	$~~^3$Max Planck Institute for Astronomy, K\"{o}nigstuhl\,17,
	69117 Heidelberg, Germany \\
	$~~^4$Astronomisches Rechen-Institut, Zentrum f\"ur Astronomie der Universit\"at Heidelberg, M\"{o}nchhofstr.\,12-14, 69120 Heidelberg, Germany\\
	$~~^5$Space Telescope Science Institute, 3700 San Martin Drive,
	Baltimore, MD 21218, USA \\
	$~~^6$ESA -- European Space Astronomy Center, Apdo. de Correo 78,
	E-28691 Associate Villanueva de la Cañada, Madrid, Spain \\
	$~~^7$Department of Physics, University of Pisa, Largo Pontecorvo 3, I-56127 Pisa, Italy\\
	$~~^8$European Space Research and Technology Centre, Keplerlaan 1, 2200 AG Noordwijk, Netherlands\\
	$~~^9$European Space Agency and Space Telescope Science Institute, 3700 San Martin Drive, Baltimore, MD 21218, USA\\
	$~~^{10}$INAF-Osservatorio Astronomico di Bologna, Via Ranzani 1, I-40127 Bologna, Italy
}

\begin{document}

\date{Draft version \today}

\pagerange{\pageref{firstpage}--\pageref{lastpage}} \pubyear{2018}

\maketitle


\begin{abstract}
The Hubble Tarantula Treasury Project (HTTP) has provided an unprecedented photometric coverage of the entire star-burst region of 30\,Doradus down to the half Solar
mass limit. We use the deep stellar catalogue of HTTP to identify all the pre--main-sequence (PMS) stars of the region, i.e., stars that have not started their lives on the main-sequence yet. The photometric distinction of these stars from the more evolved populations is not a trivial task due to several factors that alter their colour-magnitude diagram positions. The identification of PMS stars requires, thus, sophisticated statistical methods. We employ Machine Learning Classification techniques on the HTTP survey of more than 800,000 sources to identify the PMS stellar content of the observed
field. Our methodology consists of 1) carefully selecting the most
probable low-mass PMS stellar population of the star-forming cluster
NGC\,2070, 2) using this sample to train classification algorithms to build a predictive
model for PMS stars, and 3) applying this model
in order to identify the most probable PMS content across the
entire Tarantula Nebula. We employ Decision Tree, Random Forest and
Support Vector Machine classifiers to categorise the stars as PMS and Non-PMS. The Random Forest and Support Vector Machine provided the most accurate models,
predicting about 20,000 sources with a candidateship probability higher than 50
percent, and almost 10,000 PMS candidates with a probability higher than
95 percent. This is the richest and most accurate photometric catalogue of extragalactic PMS candidates across the extent of a whole star-forming complex.
\end{abstract}


\begin{keywords}
	Hertzsprung--Russell and colour--magnitude diagrams --
	stars: pre-main-sequence --
	star clusters: individual: NGC\,2070 --
	methods: data analysis, statistical --
	Magellanic Clouds
\end{keywords}


\begin{table*}
	\centering
	\caption{Filter wavelength coverages, used HST instruments, and available data in individual filters out of the 822,204 total stars of the HTTP catalogue. Sources with photometric flags \citep[as described in][]{2016ApJS..222...11S} higher than 2 in certain filters are considered as non-detections. The last column refers to the fraction of stars out of the total amount detected in the respective filter.}
	\label{tab:HST_filters}
	\begin{tabular}{lrrrcr}
		\hline
		\hline
		Filter & $\lambda_{\mathrm{mean}}$ [\AA] & $\lambda_{\mathrm{min}}$[\AA] & $\lambda_{\mathrm{max}}$[\AA] & HST Instrument & Available Data \\
		\hline
		F275W (UV) & 2377.6 & 1990 & 2980 & WFC3\_UVIS1 & 46215 (5.6\%)\\
		& 2363.9 & 1990 & 2968 & WFC3\_UVIS2 & \\
		F336W (U)  & 3358.5 & 3014 & 3707 & WFC3\_UVIS1 & 151679 (18.5\%)\\
		& 3358.6 & 3014 & 3707 & WFC3\_UVIS2 & \\
		F555W (V)  & 5396.7 & 4584 & 6209 & ACS\_WFC  & 409042 (49.8\%)\\
		& 5397.5 & 4584 & 6209 & ACS\_WFC  & \\
		F658N ($\mathrm{H}_\alpha$) & 6584.1 & 6510 & 6659 & ACS\_WFC & 132496 (16.1\%) \\
		F775W (R)  & 7729.7 & 6804 & 8632 & ACS\_WFC & 657279 (79.9\%) \\
		& 7730.6 & 6804 & 8632 & ACS\_WFC & \\
		& 7660.1 & 6869 & 8571 & WFC3\_UVIS1 & \\
		& 7658.4 & 6869 & 8571 & WFC3\_UVIS2 & \\
		F110W (J)  & 11623.8 & 8832 & 14121 & WFC3\_IR & 617129 (75.1\%)\\
		F160W (H)  & 15392.3 & 13854 & 16999& WFC3\_IR & 618508 (75.2\%)\\
		\hline
		\hline
	\end{tabular}
\end{table*}

\section{Introduction}\label{s:intro}
Giant star-forming regions, the signposts of star formation across whole giant molecular clouds (GMCs), are one of the major birthplaces of stars in a galaxy. The youthfulness of these regions is shown by their blue massive stars, located at the bright part of the main sequence in the colour-magnitude diagram (CMD). However, also the stars of low and intermediate masses in these regions hold important information about the star formation process. These stars are not yet fully-formed, and since they do not fuse hydrogen in their cores, they do not appear on the main sequence like their massive counterparts \cite[][]{2012fees.book.....S}. These so-called {\sl pre--main-sequence} (PMS) stars are still under formation through contraction and accretion and they occupy the faint red part of the CMD \citep[][]{2005fost.book.....S}. Assuming a typical stellar initial mass function (IMF), PMS stars with masses up to a few M$_{\odot}$ account for almost half of the total stellar mass budget of a young star cluster. Consequently, identifying and studying these stars improves our understanding of clustered star formation by parametrizing its properties, such as the star formation efficiency, rate, duration, and the low-mass end of the resulting stellar IMF. The characterisation of PMS stars, with T\,Tauri stars being prototypical examples, can be achieved with spectroscopic measurements of usually only a small number of objects \citep[see, e.g.,][and references therein for the spectral features of these stars]{2011psf..book.....B}. However, characterizing whole ensembles of such stars with spectroscopy is not practically feasible for star-forming regions in the Galaxy, nor even possible for those outside the Galaxy. The study of rich samples of faint PMS stars has, thus, to rely on deep photometric measurements. 

\begin{figure}
	\centering
	\includegraphics[width=1.\columnwidth]{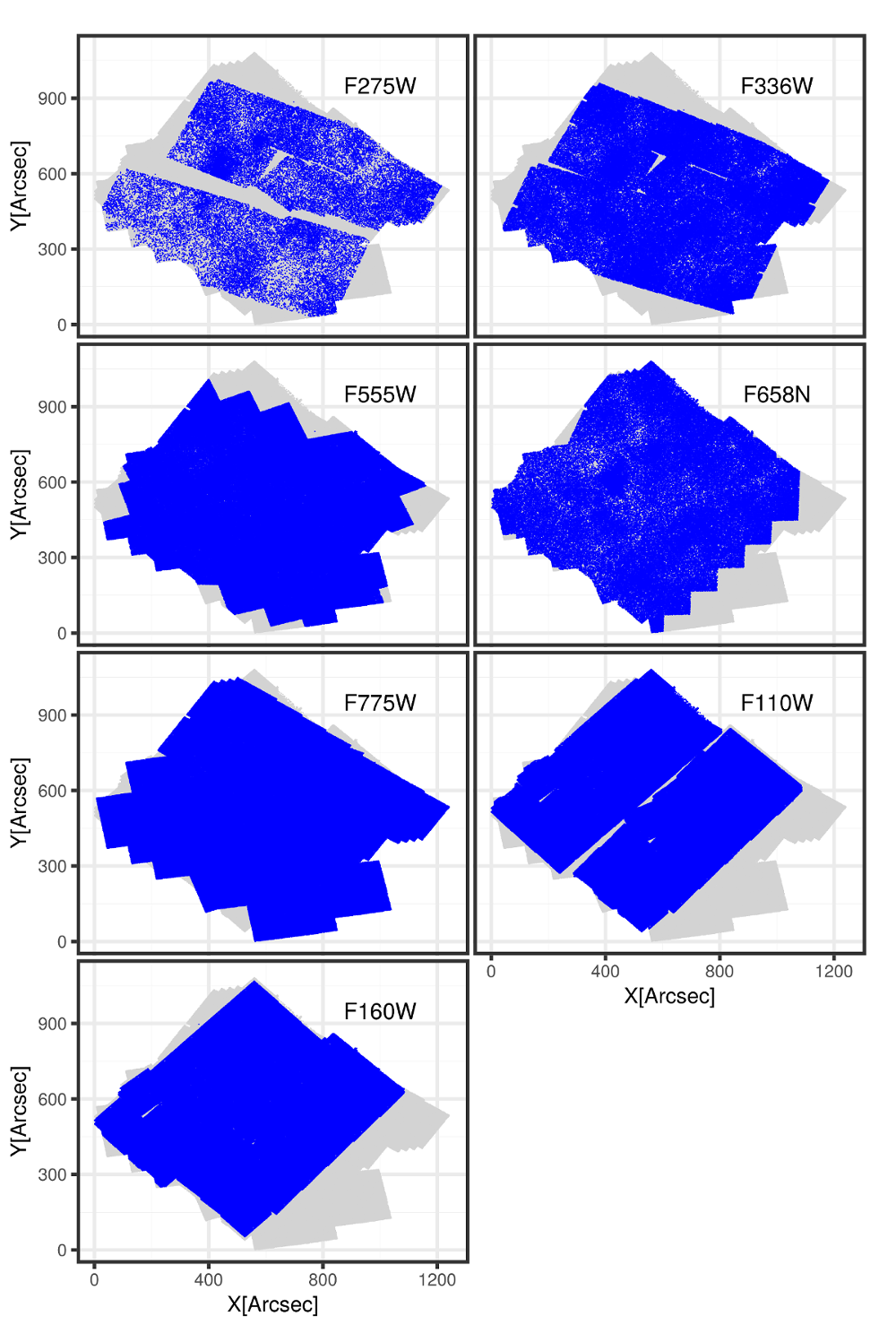} 
	\caption{Overview of the spatial coverage in the individual HTTP filters. The grey points indicate the objects not covered in the respective filter, while the blue ones mark objects with a detection.
		\label{fig:http_spatial_coverage1}}
\end{figure}

More than a decade ago several studies showed the exceptional ability of the {\sl Hubble Space Telescope} to detect faint PMS stars in the Magellanic Clouds, the satellite galaxies of our Milky Way \citep{Brandner2001, Gouliermis2006LH52, Nota2006, Sabbi2007}. These are the only extragalactic PMS stars we can resolve and they are extremely useful in understanding star formation at GMC length-scales because of their cospatial, rich samples, unobscured by the dusty Galactic disk. The Magellanic Clouds, due to their exceptional star formation activity at much lower extinction and stellar field contamination than our Milky Way are, thus, very attractive targets for the study of 
ensembles of PMS stars. Nevertheless, there are various observational and physical constraints that introduce difficulties to the identification and characterisation of PMS stars from photometry alone. The main issue is the dislocation of the stars from their theoretical CMD positions, introducing a spread of the PMS stars in the CMD along both the brightness and colour axes. Among the effects that produce the PMS spread on the CMD, differential reddening, variability, and excess emission due to circumstellar disks are considered the most important \citep{2012SSRv..169....1G}. Moreover, observational limitations, such as unresolved binarity, stellar crowding and photometric uncertainties introduce a `mixing' of the PMS stars' CMD positions. 

These issues can be mitigated in (almost) single-age individual young clusters and associations \citep[][]{Gouliermis2007LH95, Gouliermis2010ESA}, allowing the investigation of their star formation history \citep[e.g.,][]{MasseyHunter1998, DaRio2009LH95, Cignoni2010}, stellar IMF \citep{Sabbi2008, Schmalzl2008, DaRio2010LH95}, structure \citep{Schmeja2009, Gouliermis2014}, and star formation rate \citep{Hony2015} from their apparently coeval PMS stellar populations. However, the PMS spread across the CMD introduces a significant difficulty in disentangling these stars from the evolved stellar populations of the galactic field or older clusters across giant star-forming complexes. These structures of gas and stars host multiple star-forming centers, still embedded in their gaseous natal environments, with a significant amount of evolved field populations being projected onto their field-of-view (FoV). 

A typical example of such regions is the {\sl Tarantula Nebula} in the Large Magellanic Cloud (LMC), the natal environment of 30\,Doradus, the most impressive starburst in the Local Group. The complexity of the region, indicated in various previous investigations, is revealed at its full extent with the deep panchromatic {\sl Hubble Space Telescope} imaging of the whole nebula from the {\sl Hubble Tarantula Treasury Project} \citep[HTTP;][]{2013AJ....146...53S, 2016ApJS..222...11S}. The HTTP field contains different populations of various ages, so that an overlap between the turn-on, i.e., the locus in the CMD where the PMS joins the ZAMS \citep{Cignoni2010}, and 
faint giant/subgiant regions of the CMD occurs. These factors become evident in the particularly rich (in stellar numbers) CMD of the survey, constructed from measurements in the F555W and F775W (V- and R-equivalent) filters. The wide broadening of the upper main-sequence (UMS) stars and the elongation of the red clump {(RC)} (which in theory should appear nearly circular) provide clear evidence for significant differential extinction \citep[see also][]{2016MNRAS.455.4373D}. In this CMD it is also practically difficult to distinguish the faint lower main-sequence (LMS) field stars from the faint PMS stars, as these two populations strongly overlap in the low brightness regime. Because of these effects that prevent the clear identification of the various populations based on their CMD positions, the distinction of the faint PMS stars requires the use of sophisticated statistical approaches. In this study we develop a classification methodology by employing {\sl Machine Learning} techniques in order to perform a robust identification of the most probable PMS stars in the Tarantula Nebula. The clear sample of faint PMS stars in combination with the young bright stellar population of the UMS will provide an unprecedented stellar dataset to investigate the complexity of the star formation process along the most interesting star-forming region in the Local Group.

Machine learning, i.e., the study of algorithms that can learn from and make predictions on data, has introduced a variety of statistical tools to the astronomical research. These tools are designed to solve problems of regression, classification, and clustering \citep[see, e.g.,][for recent applications]{Beaumont2014, Dieleman2015, Elorrieta2016}. While the first two types of problems are addressed with {\sl supervised} learning processes, the third requires {\sl unsupervised} methods. We employ classical Machine Learning classification techniques {on the HTTP photometry} in order to determine the most successful in identifying the most probable PMS stars across the nebula, based on prior information retrieved from our data of the most prominent faint PMS stellar sample in our FoV. This sample comprises the stellar members of the giant HII region, where the young stellar cluster NGC\,2070, host of the starburst cluster R136, resides.

This paper is structured as follows. In Sect.\,\ref{s:datadescription} we give a short description of the photometric dataset of HTTP. In order to account for the differential reddening of the region in our identification, we apply a correction for extinction to the photometric measurements of all stars in Sect.\,\ref{s:extcorr}. This correction is estimated in terms of the proximity of PMS candidates to UMS stars, for which reddening is being determined from their dislocation unreddened from the main sequence. In preparation of our experiments on various classification methods we build a so-called training dataset, which is made with a careful selection of the LMS and PMS stars included in the area of NGC\,2070, as well as the ``contaminating'' evolved field populations (Sect.\,\ref{s:trainset}). This is necessary because of the partial overlap between the LMS and PMS stars in the CMD and the extended star formation history of R136 of several Myr \citep{Hunter1995}. The classification of the observed stellar populations in terms of supervised machine learning based on the training dataset takes place in Sect.\,\ref{chap:Classification}, where the most robust algorithms for the identification of faint PMS stars are established, and the final dataset of the best PMS candidates across the whole Tarantula nebula is constructed. A summary and 
future prospects concerning this study are given in Sect.\,\ref{s:summary}.

\begin{figure}
	\centering
	\includegraphics[width=0.95\columnwidth]{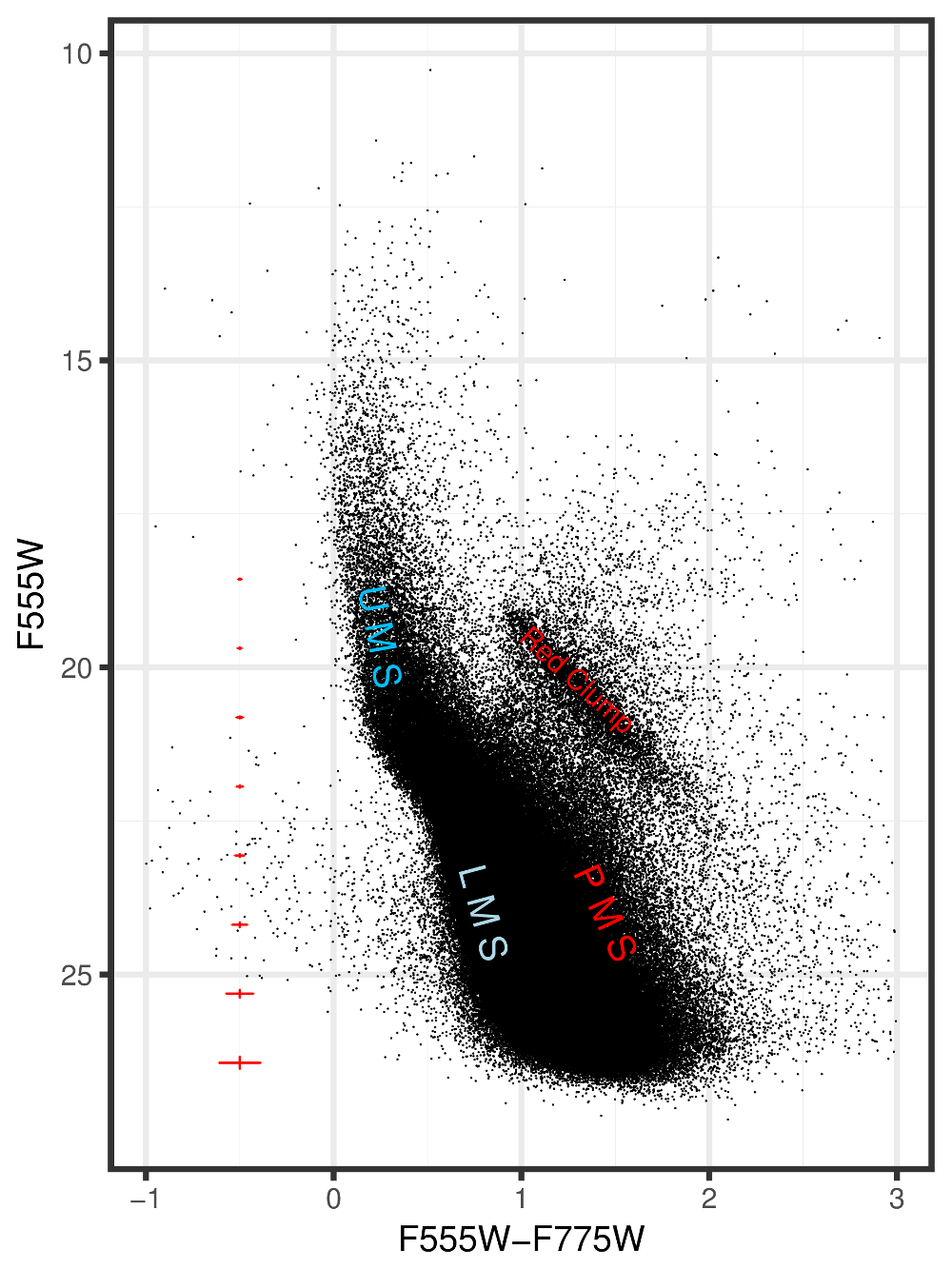}
	\caption{The CMD of the data obtained in the optical F555W and F775W (V-, R-equivalent) filters. It is the richest CMD of the HTTP survey in terms of spatial coverage. For reference representative errorbars of the photometric errors are overplot in red {and the coloured labels indicate the approximate locations of populations of interest}.} 
	\label{fig:HTTP_CMD}
\end{figure}

		\begin{figure*}
			\centering
			\includegraphics[width=0.49\textwidth]{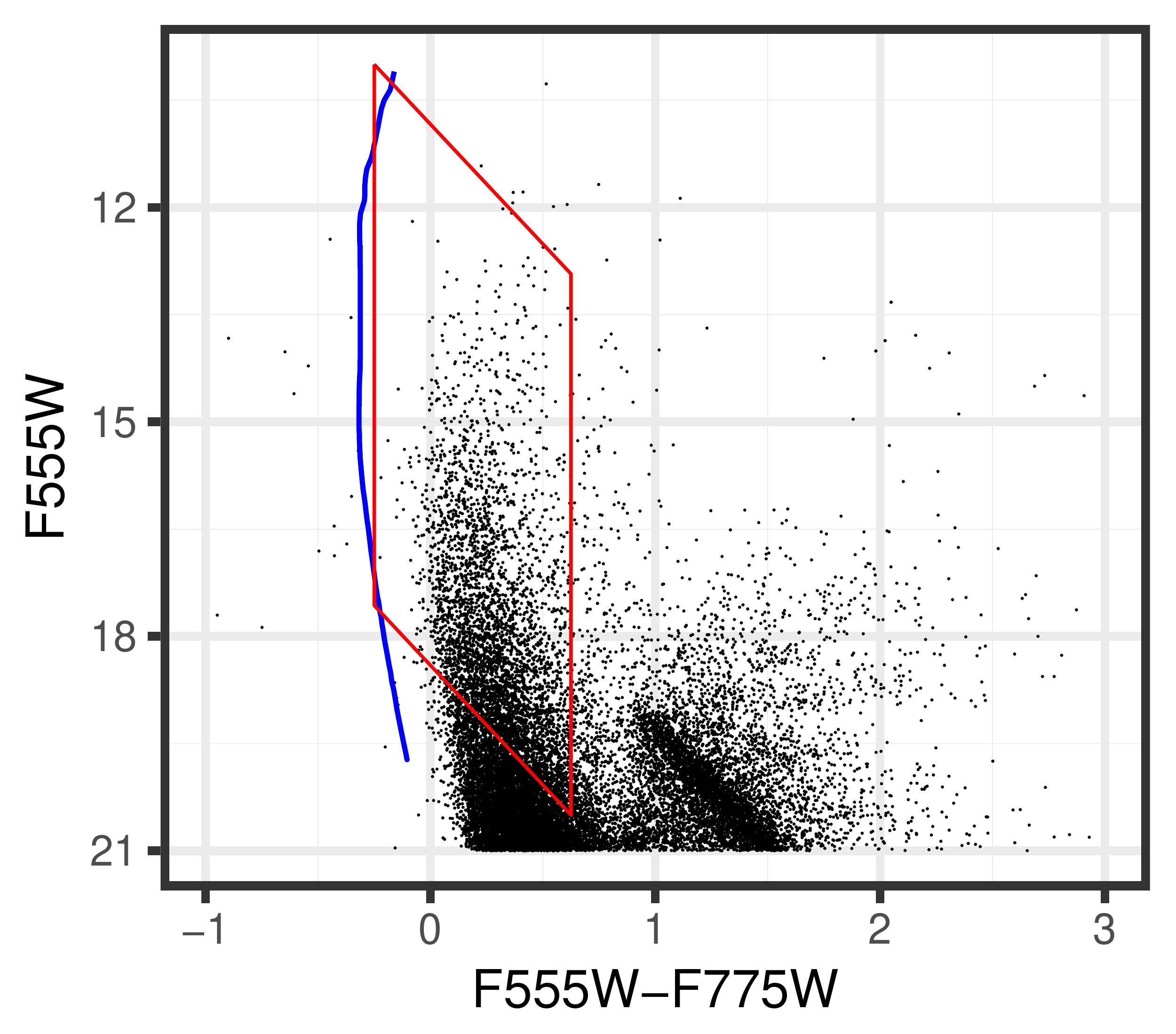}
			\includegraphics[width=0.49\textwidth]{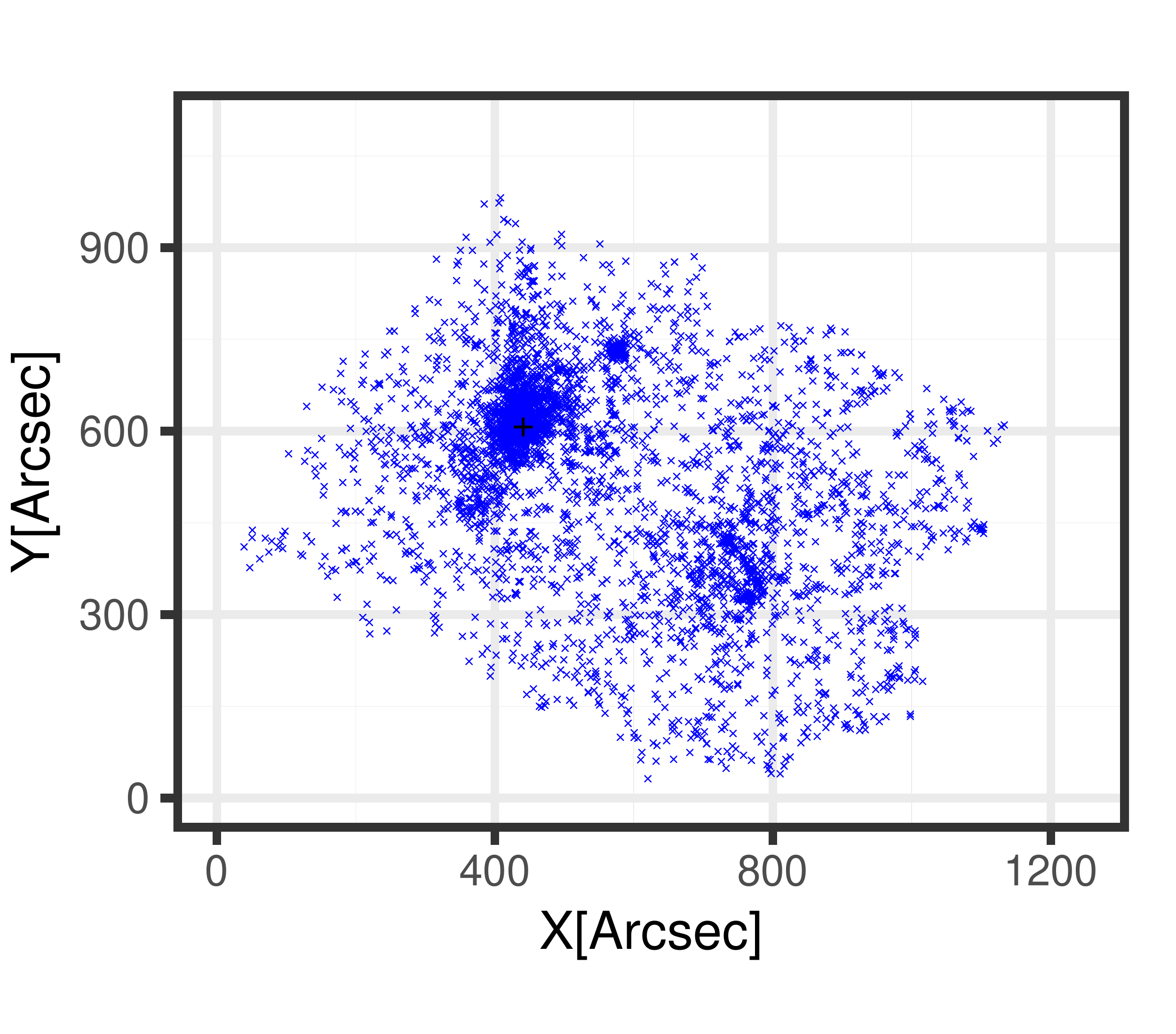}
			\caption{{\em Left:} The bright part of the optical CMD with the region, where the UMS stars are selected for the measurement of their extinction enclosed by the polygon in red. The blue line corresponds to the upper part of the {\em un-reddened} Zero Age Main-Sequence (ZAMS), corrected for the distance of the LMC. The top and bottom sides of the polygon follow the direction of the reddening vector. {\em Right:} The spatial distribution of the UMS stars selected within the polygon for the determination of the extinction across the HTTP field-of-view. The black cross marks the centre of R\,136 ($\mathrm{{R.A.}_{J2000}} = 05^\mathrm{h}\,38^\mathrm{m}\,42^\mathrm{s}.3$, $\mathrm{{DEC}_{J2000}} = -69^\circ\,06'\,03''.3$) to be used later in the selection of our training sample.}
			\label{fig:http_extinction}
		\end{figure*}

\section{Data description}\label{s:datadescription}

HTTP is a high spatial resolution stellar photometric survey of the Tarantula Nebula (the nebula of the starburst of 30\,Doradus). It has a high dynamic range in spectral coverage, extending from the near-ultraviolet (NUV) to the near-infrared (NIR) part of the spectrum \citep{2013AJ....146...53S, 2016ApJS..222...11S}. Its spatial coverage extends across the whole region of the nebula of $\sim\,16 \times 13$ arcmin$^2$, corresponding to $240 \times 190$ pc$^2$ at the distance of the Large Magellanic Cloud (LMC, $(m - M)_0 = 18.55 \pm 0.05$\,mag). This region includes the clusters 
NGC\,2070 (with the starburst cluster R\,136 at its core), 
NGC\,2060 and Hodge\,301 (see, e.g., Fig.\,1 in \citealt{2016ApJS..222...11S}). The observations were obtained with the {\sl Advanced Camera for Surveys} (ACS) and {\sl Wide Field Camera 3} (WFC3) on board the {\sl Hubble Space Telescope} (HST) in filters corresponding to a wide range of broad and narrow bands of $0.27 - 1.6\,\mu$m (Table \ref{tab:HST_filters}). An overview over the available data in the respective filters is given in Table \ref{tab:HST_filters}, where it can be seen that there are significantly less data available in the UV, U and $\mathrm{H}_\alpha$ filters than in the remaining four filters, with the most data being available in the R, J and H bands. Figure \ref{fig:http_spatial_coverage1} shows the spatial coverage of the observations in each filter, indicating that filters V and R cover the largest area of the observed field, while especially the UV filter covers only a very sparse area and both infrared filters do not cover the South-Eastern region of the observed area. Figure \ref{fig:HTTP_CMD} shows the optical CMD of the HTTP survey, exhibiting the previously mentioned widened UMS and elongated {RC} caused by differential extinction \citep[see][for a discussion of the reddening in this area and maps]{Haschke2011,2016MNRAS.455.4373D}.

		\begin{figure}
			\centering
			\includegraphics[width=\linewidth]{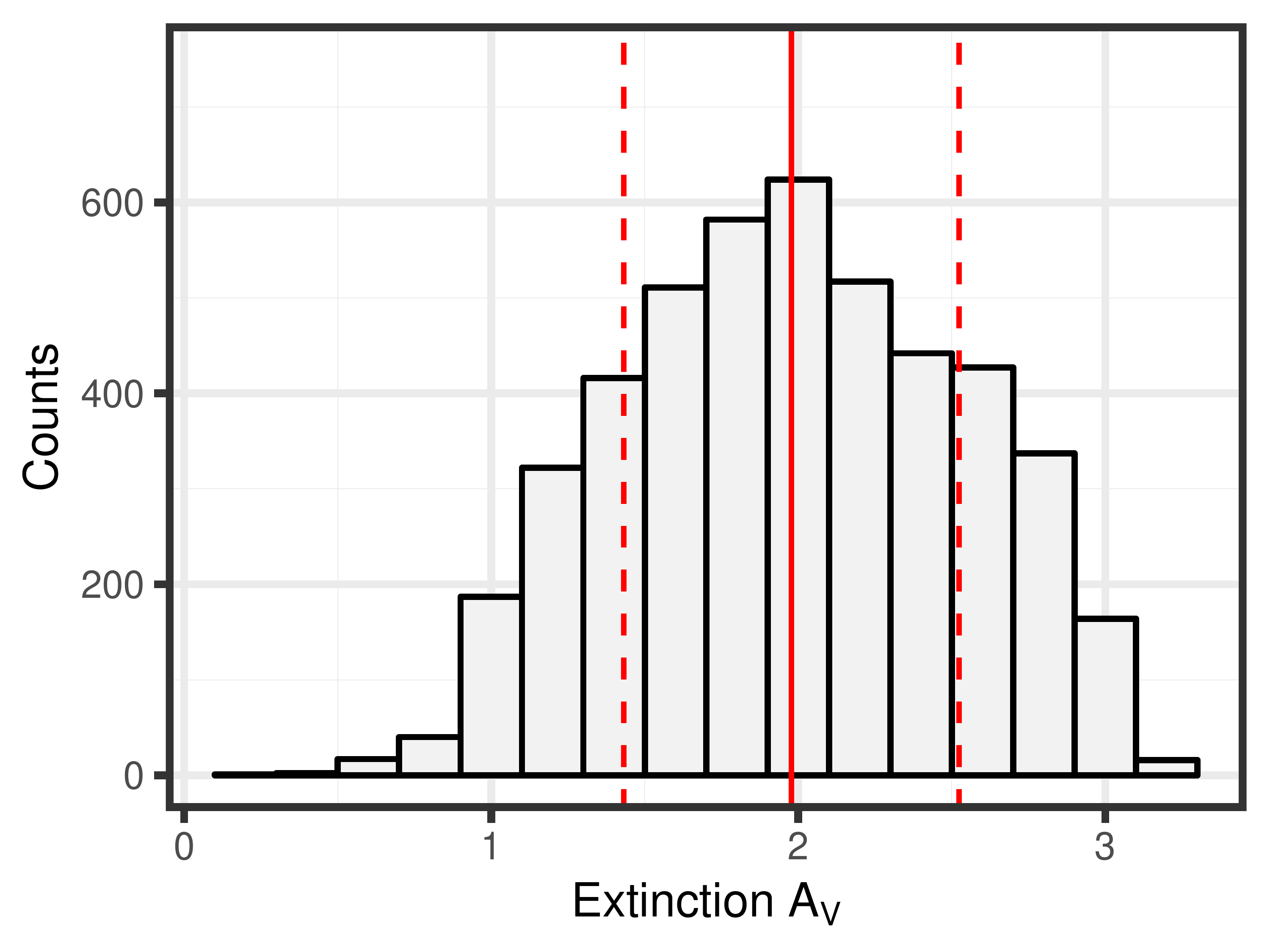}
			\caption{Histogram of the measured extinction values of the selected UMS extinction probes in the HTTP data, where the thick red line indicates the mean $A_\mathrm{v, mean} = 1.98$\,mag and the dotted lines the standard deviation $\sigma_{A_V} = 0.55$\,mag.}
			\label{fig:Ex_Hist}
		\end{figure}

\section{Extinction Correction}\label{s:extcorr}
		
Star-forming regions such as the Tarantula Nebula still include a significant amount of gas from the parental molecular cloud. It is, thus, expected that a considerable amount of interstellar extinction influences the photometric measurements of the observed stars and dislocates them accordingly in the CMD. This phenomenon contributes to the mixing of old and young stars in the CMD and it should be considered in the application of our classification procedure for the identification of PMS stars. We increase, thus, the number of variables by adding to the magnitudes and colour indexes of the stars their reddening, as determined by the extinction measurements of their close-by UMS stars \citep{Panagia2000, Romaniello2002, 2016MNRAS.455.4373D, DeMarchi2017}. 
The use of UMS stars for the extinction correction is based on the fact that the young PMS candidates are more likely to be spatially correlated with the UMS population, than with the LMC field. In addition, earlier extinction studies showed that ``dust is highly localised near the hotter, younger stars" and the average extinction correction for older populations is lower than that for younger \citep{Zaritsky2002}. The use of {RC} stars for the extinction correction of PMS stars would, thus, compromise the accuracy of this correction and the reddening of the related star-forming regions.

		\begin{figure*}
		  	\centering
		  	\includegraphics[width = 0.33\textwidth]{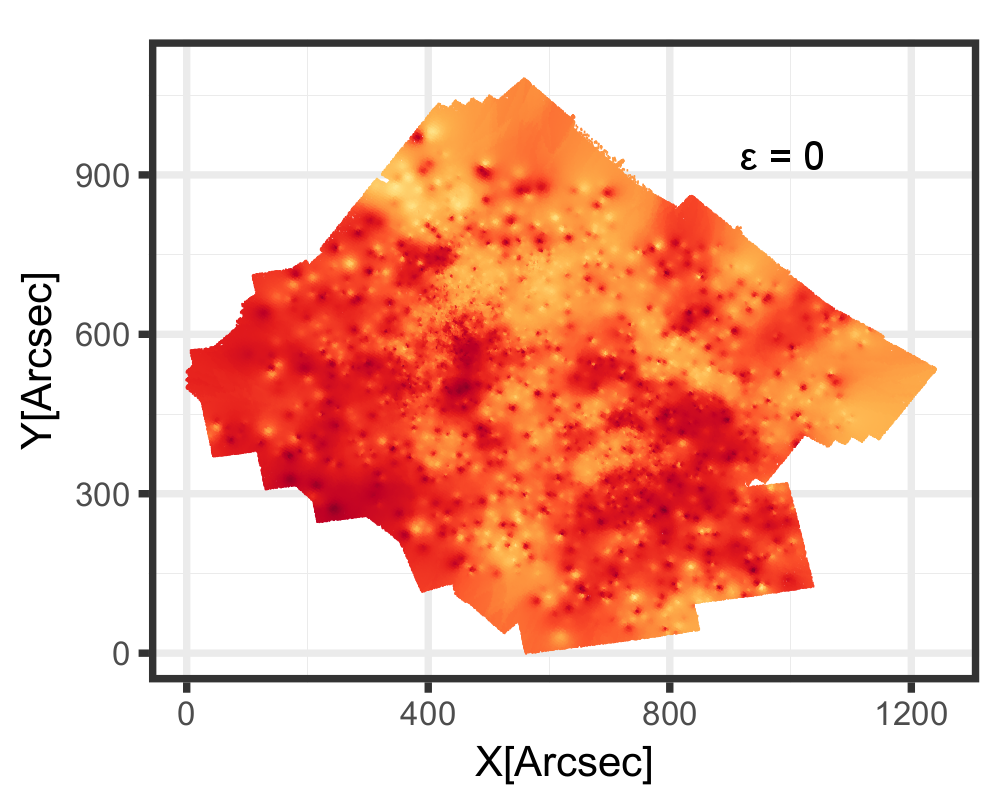}
	  		\includegraphics[width = 0.33\textwidth]{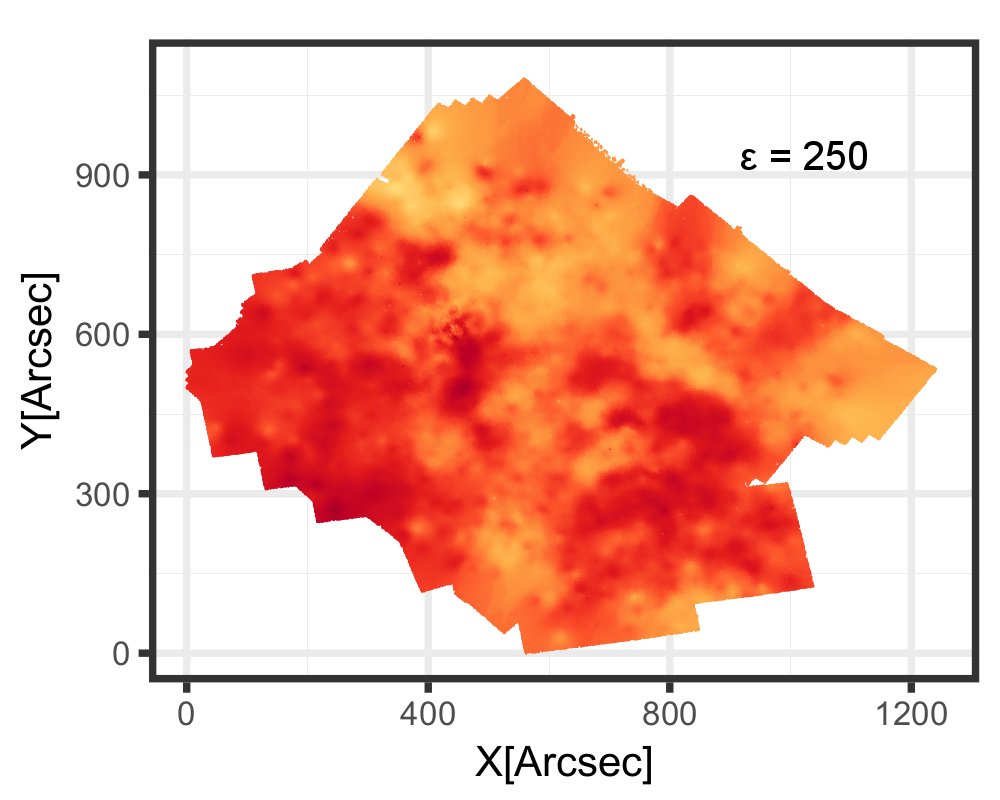}
  			\includegraphics[width = 0.33\textwidth]{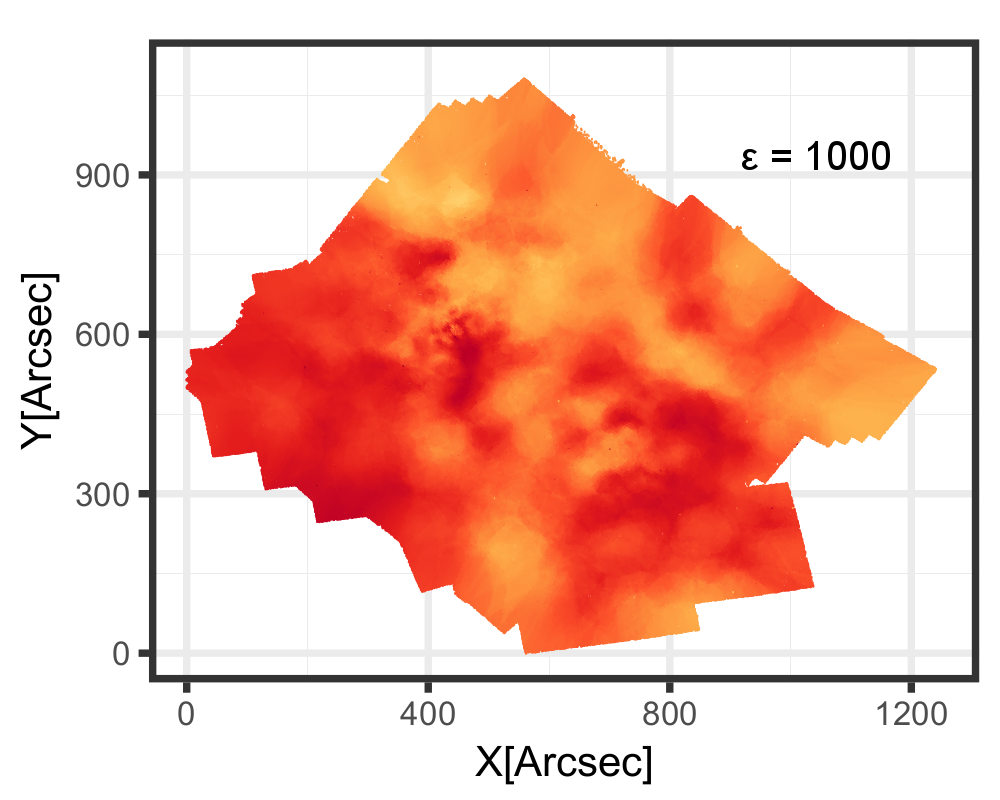}
		  	\caption{A series of extinction maps of the Tarantula Nebula, where each data point is colour-coded according to its assigned extinction value. From left to right, the series shows the influence of the smoothing parameter $\epsilon$, introduced to the distance weight in equation \ref{eq:AV_smoothing} on the extinction map. The respective $\epsilon$ value is noted in the top right corner of each plot. Note that all plots share the colour gradient from Figure \ref{fig:HTTP_ex_map_final}. The adapted extinction law is described in \protect\cite{2016MNRAS.455.4373D}.}
		  	\label{fig:HTTP_ex_map}
		\end{figure*}

We retrieve extinction measurements for the UMS stars in our catalog by ``relocating'' their CMD positions on the main sequence along the known reddening vector of the Tarantula Nebula with slope $R_{555} = A_{555}/E(m_{555}-m_{775}) = 3.35 \pm 0.15$, as determined by \cite{2016MNRAS.455.4373D}, up to its intersection with the zero-age-main sequence (ZAMS). We make use of PARSEC family of evolutionary models \citep{Bressan2012} with a metallicity of $Z=0.08$ for the LMC, corrected for a distance modulus of 18.55\,mag \citep{Panagia1991, WalbornBlades1997, 2016MNRAS.455.4373D}.


Figure \ref{fig:http_extinction} (left) shows the selected UMS extinction probes. This selection is based on the CMD region occupied by these stars, as indicated by the red polygon, entailing 4,605 stars distributed across the entire observed FoV, and identified in the V-band. The selected sample of UMS stars exhibits higher concentrations in the regions corresponding to the clusters 
R136, Hodge 301, and NGC2060, as demonstrated by their map also shown in Figure \ref{fig:http_extinction} (right). We assign extinction values to each non-UMS star as the distance-weighted average of the extinctions $A_{V_{n}}^{\mathrm{(UMS)}}$ of its $N$ nearest UMS neighbours according to 
		\begin{equation}
		A_V = \sum_{n=1}^{N} w_n A_{V_{n}}^{\mathrm{(UMS)}}
		\label{eq:av1}
		\end{equation}
		with weights
		\begin{equation}
		w_i = \frac{1}{d_i^2 + \epsilon^2}\frac{1}{\sum_{n=1}^{N}\frac{1}{d_n^2 + \epsilon^2}},
		\label{eq:AV_smoothing}
		\end{equation}
		{where $d_i$ denotes the Euclidean distance in pixels to the $i$-th} nearest UMS neighbour and $\epsilon$ is a smoothing parameter (also given in pixels), which we introduce in order to reduce the dominance of close proximity to a single UMS star in the averaging process. The corresponding weighted standard deviation of the assigned average extinction value is given by
		\begin{equation}
		\delta A_V = \sqrt{\sum_{i = 1}^{N} w_i (A_{V_{i}}^{\mathrm{(UMS)}}-A_V)^2}.
		\label{eq:av_error}
		\end{equation}
This distance weighted $A_\mathrm{V}$ calculation eliminates the possibility of underestimation due to low extinction foreground stars that may be projected by chance in the region of the PMS candidates. Nevertheless such stars represent an insignificant fraction of our stellar sample. The extinction measurements of the UMS probes are summarised in Figure\,\ref{fig:Ex_Hist}, where it is shown that these stars have a mean extinction of $A_\mathrm{v, mean} = 1.98$\,mag with a standard deviation of $\sigma_{A_V} = 0.55$\,mag.\\

\begin{figure*}
			\centering
			\includegraphics[width = 0.75\textwidth]{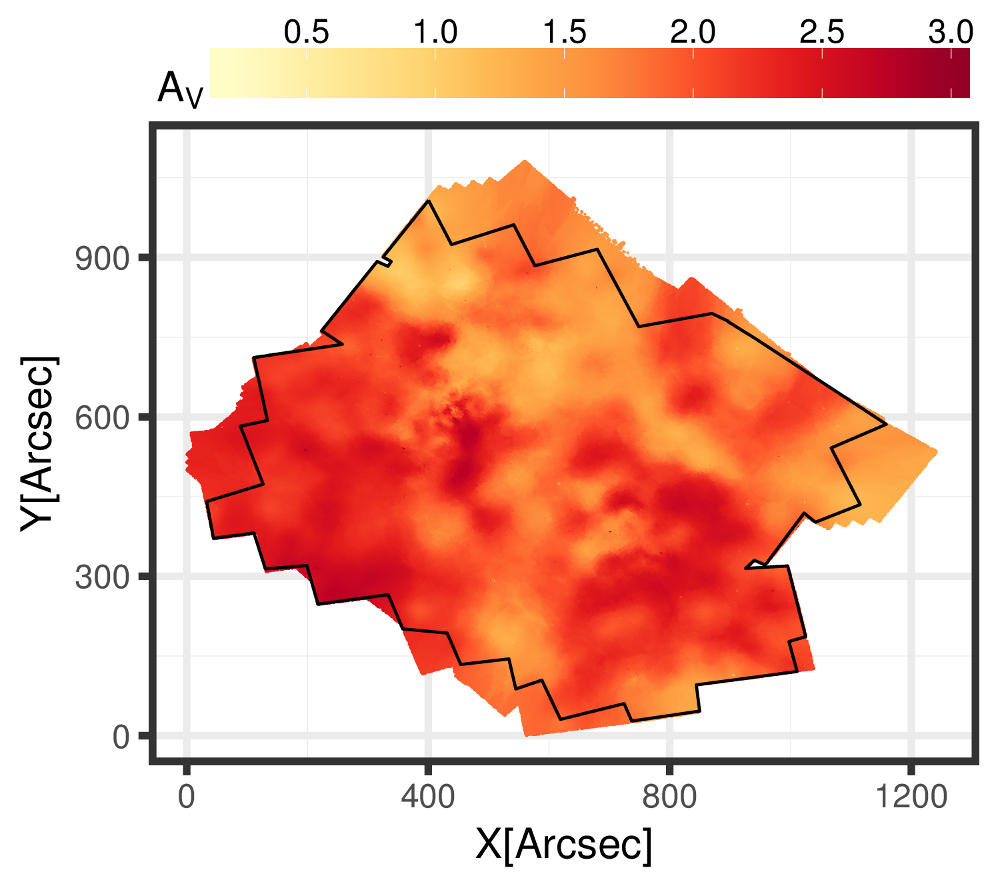}
			 \caption{Extinction map of the Tarantula Nebula, where each data point is colour-coded according to its assigned extinction value, using a smoothing parameter of $\epsilon = 500$. The black outline indicates the area that is covered both in V and R, i.e., the area we considered for our further analysis.}
			\label{fig:HTTP_ex_map_final}
		\end{figure*}

For the assignment of extinction values we use the $N=20$ nearest UMS neighbours. Figure\,\ref{fig:HTTP_ex_map} shows a series of $A_V$ maps, as constructed for various values of $\epsilon$. These maps are generated by colour-coding each data point in the spatial distribution plot according to its assigned (or measured for UMS stars) extinction value. It should be noted that the assigned extinction values in the regions without V-band coverage are biased towards the UMS stars at their borders and, thus, might not necessarily represent the true extinction within these regions. They are shown here only for visualisation, and they are not included in our further analysis. For the final estimation of the extinction corrections, and the construction of the final $A_V$ of the region, we choose a smoothing parameter of $\epsilon = 500$, based on the natural appearance of the constructed extinction map, i.e., a map which is not over-smoothed and still provides spatially detailed $A_V$ measurements (Figure\,\ref{fig:HTTP_ex_map_final}). 
{It should be noted that the assigned $A_V$ measurements of the majority of the stars are found to be insensitive to the chosen value for the smoothing factor, with the relative differences not exceeding $5-10\,\%$ for $\epsilon$ between $0$ and $2000$\,px (in comparison with $\epsilon = 500$\,px).}

The extinction-corrected optical CMD of the Tarantula Nebula is shown in Figure\,\ref{fig:HTTP_ex_corr}. In order to provide a more realistic appearance of this CMD, we applied artificial noise to the corrected positions of our UMS probes, based on small random dislocations along the reddening vector by amounts sampled from a Gaussian distribution with zero mean and the standard deviation of the colour indexes of all stars within the same magnitude range as the selected UMS probes. Our extinction correction demonstrates some over- or under-estimation for the RC stars, as the remaining elongation of the RC shown in Figure\,\ref{fig:HTTP_ex_corr} indicates. However, this does not affect our classification, because the CMD positions of these stars do not overlap with those typically occupied by our target PMS stars.

\section{Building the Training Set}\label{s:trainset}

The machine-learning algorithms applied in this study are based on \textit{supervised learning} techniques, i.e., they infer a function from labeled training data, which consist of a set of training examples (see Appendix\,\ref{app:MLCA}). These techniques require the construction of a labeled {\sl training data set} in order to ``teach" the algorithms, in our case, how to identify PMS stars based on their positions in the CMD. Also, our study aims at the simple distinction between two classes of objects, namely PMS and non-PMS stars, i.e., we address a {\sl binary classification problem}. Considering these, we build our training set so that each star has a label, which indicates whether it is a PMS example or not. We train our algorithms directly on the observational data of the HTTP data set, rather than on synthetic populations. The advantage of this approach is that modeled populations of stars would assume a specific behaviour of observable characteristics, such as binarity, circumstellar extinction, and variability, which affect the theoretical CMD positions of observed PMS stars. On the other hand, using real data allows us to account for these characteristics {\sl intrinsically}, without modeling, and thus without possibly biasing their expected behaviour. The use of real data also allows for the unbiased assessment of various observational limitations, such as crowding and photometric uncertainties, that affect the identification of PMS stars.

\subsection{Selection of the training stellar sample}
	\label{sec:kde_contours}

Due to the aforementioned physical and observational constraints (see also Sect.\,\ref{s:intro}), the identification of PMS stars in large datasets of multiple populations is not trivial, even if we take extinction into account. It is, therefore, important to train our identification algorithms on the most clear stellar training set possible, i.e., on a selected subset of the HTTP catalogue where PMS stars are clearly defined in the CMD. In this study we focus on low-mass PMS stellar populations (with masses up to few M$_\odot$), which are easily confused with low-mass MS (LMS) stars. Our training subset should, thus, comprise large numbers of PMS stars, as well as of LMS stars, and other evolved populations, which are easily distinguishable. Within large star-forming complexes, regions that comprise such stellar samples are those where high concentrations of easily identifiable young stars exist, i.e., young star clusters. We select, thus, our training sub-sample 
from the most densely populated areas of the nebula, i.e., the starburst-cluster R136 and its surroundings. Specifically, we define a squared area centered on R136 ($\mathrm{{R.A.}_{J2000}} = 05^\mathrm{h}\,38^\mathrm{m}\,42^\mathrm{s}.3$, $\mathrm{{DEC}_{J2000}} = $-$69^\circ\,06'\,03''.3$) with a side length of $8^\prime$ ($\sim$\,120\,pc). We construct the surface density map of the region by applying a kernel density estimation with a two-dimensional normal kernel. 

\begin{figure}
		\centering
		\includegraphics[width=\columnwidth]{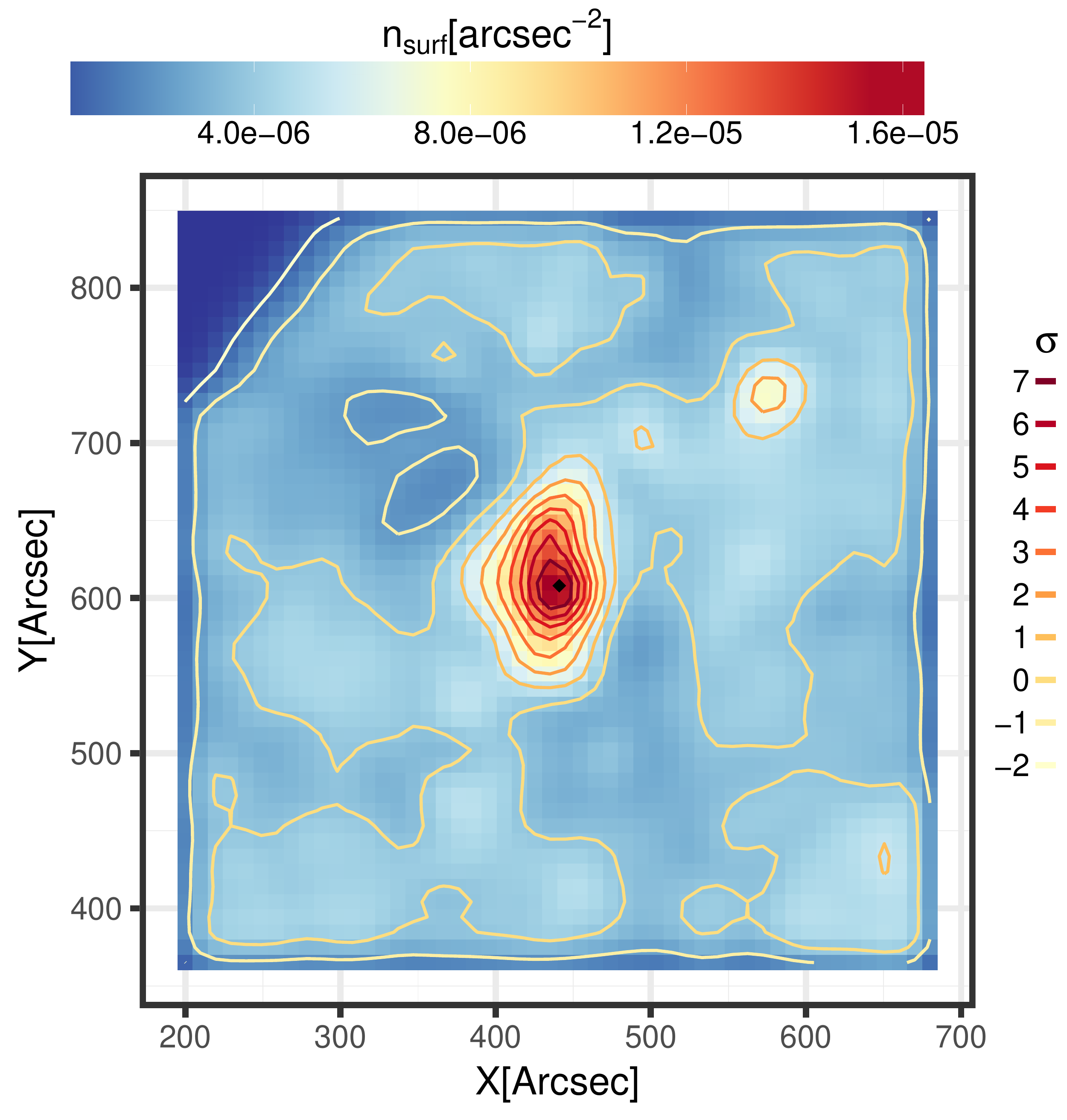}
		\caption{Surface density plot of NGC2070 with overlaid density contours covering the range from $-2\sigma_\mathrm{n}$ to $7\sigma_\mathrm{n}$ in steps of $\sigma_\mathrm{n}$. The black square marks the centre of R136.}
		\label{fig:ngc2070_contours}
	\end{figure}
\begin{figure*}
	\centering
	\includegraphics[width = \textwidth]{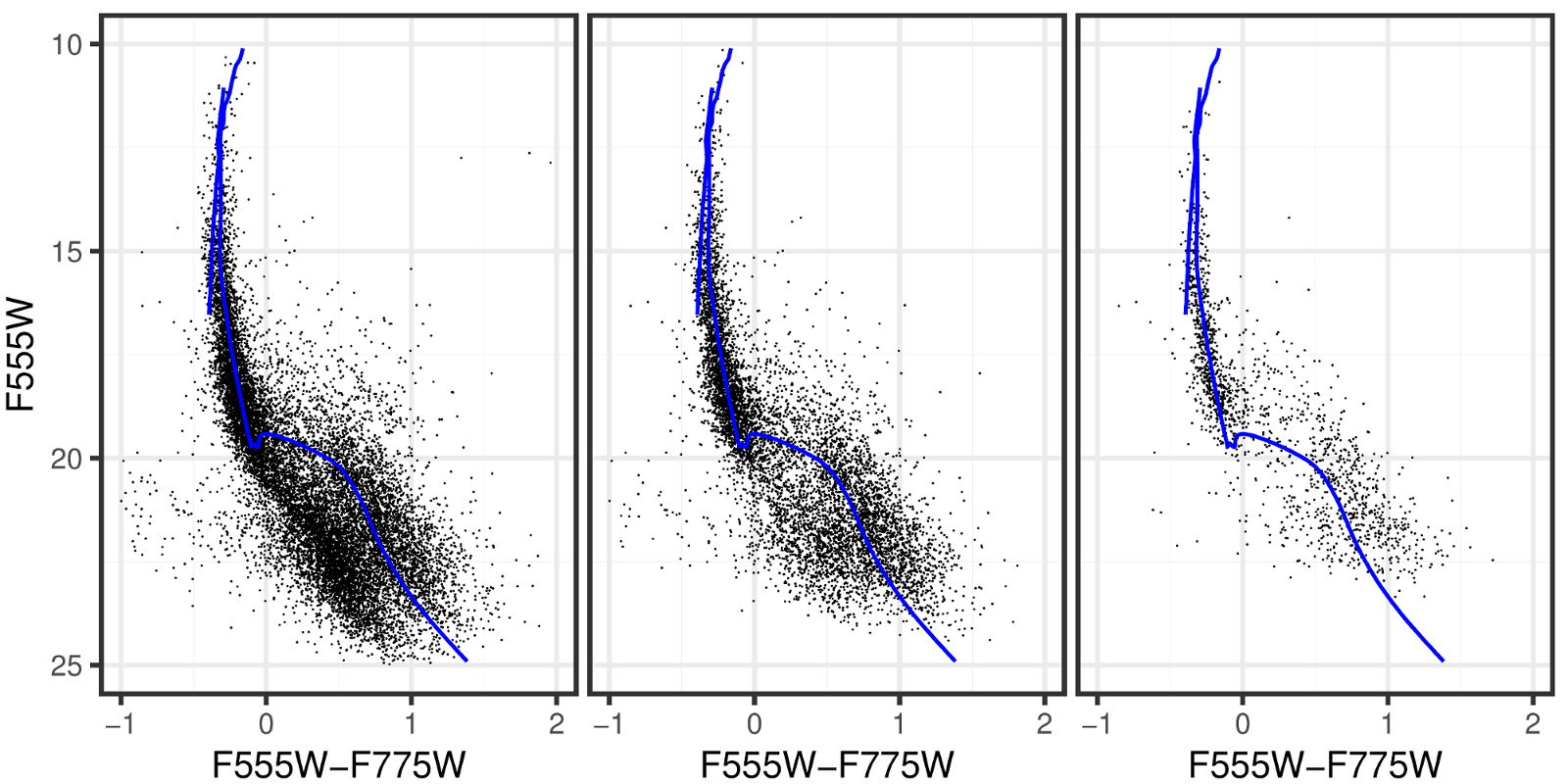}
	\caption{Optical extinction corrected CMDs for the regions within NGC2070 with surface density higher than $1\sigma_\mathrm{n}$ {\em (left)}, $3\sigma_\mathrm{n}$ {\em (middle)} and $7\sigma_\mathrm{n}$ {\em (right)} above the local background density (see Figure\,\ref{fig:ngc2070_contours} for the corresponding areas). The blue line marks the 4\,Myr PARSEC isochrone.}
	\label{fig:sigma_selection_hess}
\end{figure*}

Figure \ref{fig:ngc2070_contours} shows the surface density map of the region with the statistically significant isopleths (contour lines) in steps of $\sigma_\mathrm{n}$, the standard deviation of the map, overlaid. In our analysis we exclude grid points that fall outside the HTTP coverage (top left corner of the map), in order to avoid biases due to artificial zero measurements. As expected, the highest density peak corresponds to the starburst-cluster R136, but we also identify the more evolved cluster Hodge\,301 \citep{Grebel2000, 2016ApJ...833..154C}, north-west of R136, at density levels higher than $1\sigma_\mathrm{n}$ above the average map density. A series of CMDs of stars in the region of NGC\,2070 for three selected significance levels, namely $1\sigma_\mathrm{n}$, $3\sigma_\mathrm{n}$ and $7\sigma_\mathrm{n}$, is shown in Figure\,\ref{fig:sigma_selection_hess}. This figure demonstrates that apart from the young UMS stars occupying the bright part of the CMDs, the high-density cluster region hosts indeed a prominent low-brightness PMS population, which is located at the red part of the CMD and well distinguishable from the LMS blue part of the CMD. Moving to higher densities within the cluster we find continuously less LMS stars, while the highest peak, corresponding to R136, contains almost exclusively PMS stars. The 4\,Myr isochrone from the PARSEC evolutionary models {\citep{Bressan2012}} is also shown in the figure for guidance of where young stellar populations are expected in the CMD. 

		\begin{figure}
			\centering
		  	\includegraphics[width = \linewidth]{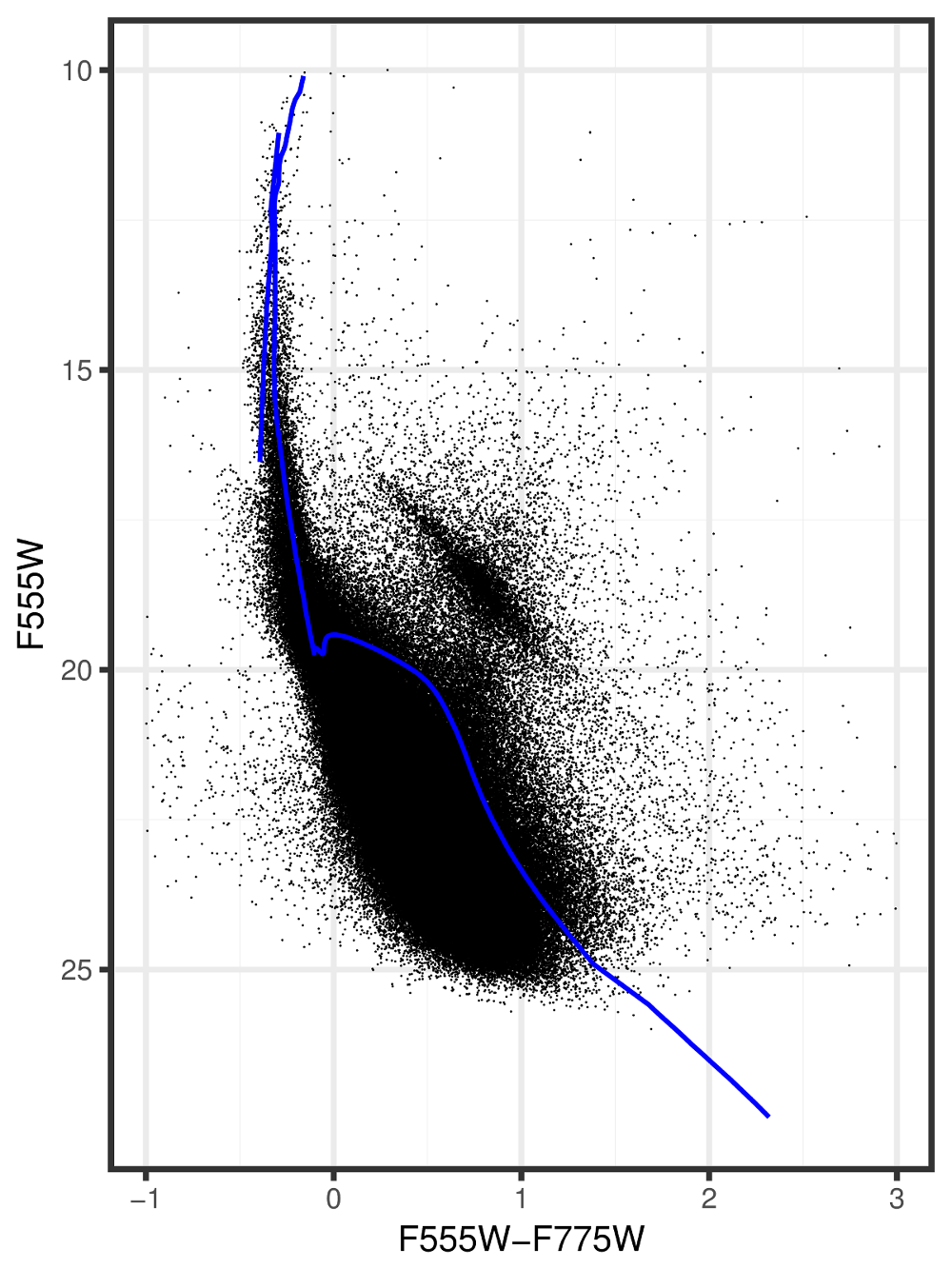}
		  	\caption{The extinction-corrected CMD of the HTTP data with additional artificial noise imposed on the corrected positions of the UMS stars in order to generate a more organic CMD. The blue line corresponds to the 4\,Myr PARSEC isochrone, whose upper part is used as an approximation for the ZAMS to measure the extinction of the UMS stars.}
		  	\label{fig:HTTP_ex_corr}
		\end{figure}

A comparison of the CMD in Figure\,\ref{fig:sigma_selection_hess} with that shown in Figure\,\ref{fig:HTTP_ex_corr} indicates that while it is relatively straightforward to identify the low-mass PMS stars inside 
the giant HII region NGC\,2070, the identification of the same type of stars across the whole HTTP CMD requires a statistical modeling of their positions. Our classification focuses on the performance of this modeling across the whole extent of the HTTP survey. In order to include a fair number of non-PMS examples, i.e., LMS stars, in our training set we select the HTTP subset included within the isopleth of $1\sigma_\mathrm{n}$ above background centered on R136, as shown in Figure\,\ref{fig:sigma_selection_hess}. The two distinct populations in the low-brightness CMD regime, i.e. LMS and PMS stars, are clearly demonstrated in the corresponding Hess diagram, shown in Figure\,\ref{fig:sigma_selection_hess2}. For building the training dataset we restrict the stars to be considered to the low-brightness regime, excluding most of the UMS stars, and we remove some of the very blue and red objects, as we suspect them to have poor photometry. The limits of the CMD region covered by this stellar sample are shown in Figure\,\ref{fig:Em_fit_dat_rc_ellipse}. Considering 
that R136 is a young cluster, the LMS stars in the region are most probably field contaminants, not belonging to the cluster. Under these circumstances our selected subset is optimal in including good training examples of low-mass PMS and non-PMS stars. In the following section, in order to characterise each star as a positive or negative PMS example, we distinguish these two observed populations in a quantitative way.

\subsection{Distinguishing PMS from LMS stars} \label{sec:EM_fit}

\begin{figure}
		\centering
		\includegraphics[width = \linewidth]{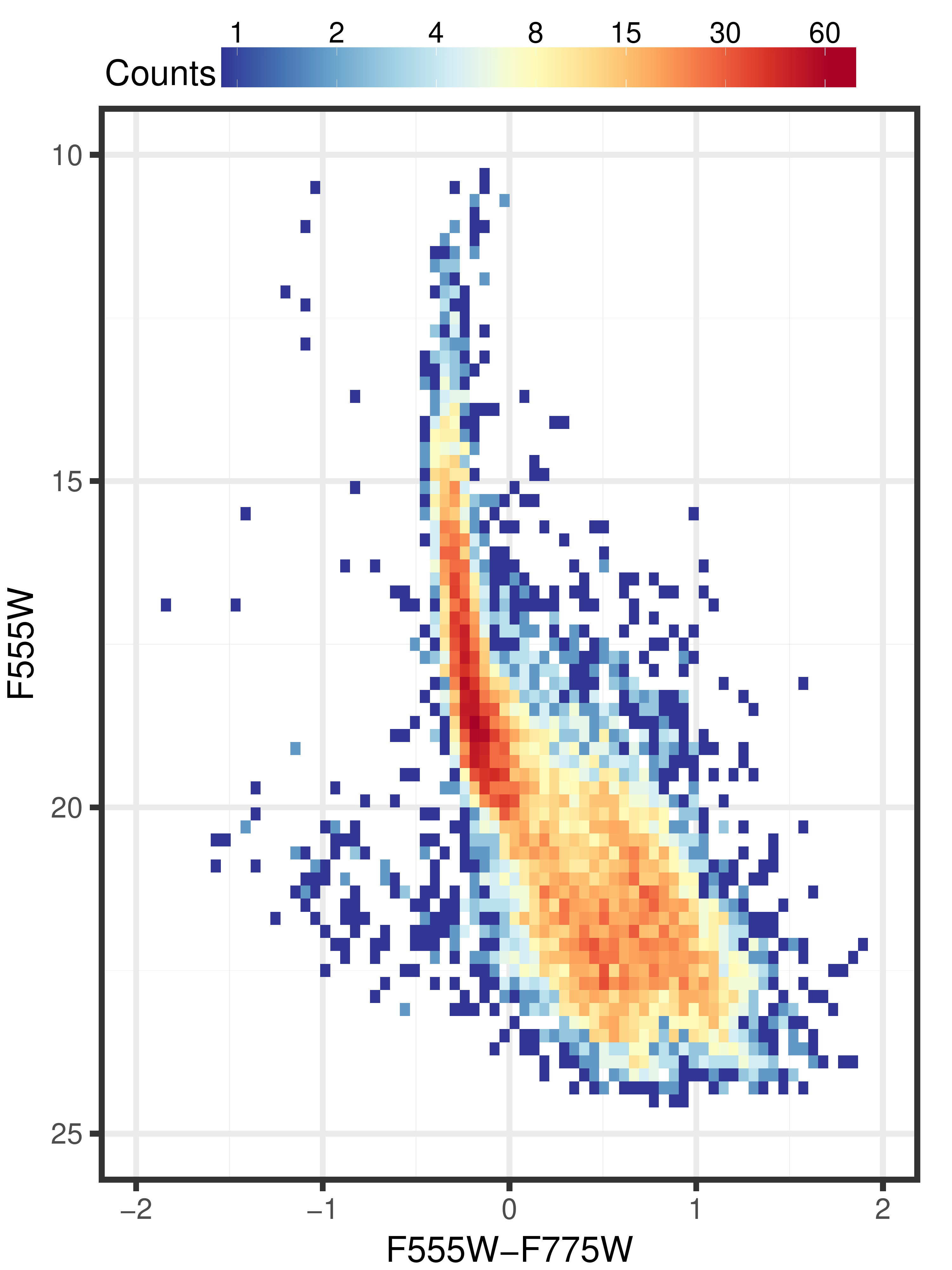}
		\caption{Optical extinction-corrected Hess diagram of the stars in the region of NGC\,2070 included within the $1\sigma_\mathrm{n}$ density significance level (see map in Figure\,\ref{fig:ngc2070_contours}). The diagram uses 75 bins in both coordinate directions with limits $[-2,2]$ in F555W$-$F775W and $[25,10]$ in F555W.}
		\label{fig:sigma_selection_hess2}
	\end{figure}

After selecting the low-brightness stellar sample to be used for the training of our algorithms, we characterise its members as PMS or non-PMS (i.e., LMS) stars, according to their observed extinction-corrected CMD positions. 
A method based on the use of stellar number distributions along cross sections of the faint CMD was proposed by \cite{2012ApJ...748...64G} for separating the LMS from the PMS populations on the CMD. The distributions of well-separated populations show two distinct peaks, the width and the separation of which are found to depend on stellar brightness. We implement this method with {one} modification: We further introduce a reference line in the CMD, which has a slope roughly equal to the gap between the LMS and PMS, as observed in the Hess diagram of Figure\,\ref{fig:sigma_selection_hess2}, and we calculate the distance of each star in the selected sample from this reference line. 

With these modifications we analyse the stellar distance distributions from the reference line, while avoiding to bin the stars, in contrast to \cite{2012ApJ...748...64G}, who analysed the binned stellar number distributions. We use a bimodal Gaussian mixture model to fit the calculated distances of the stars from the reference line and we quantify the fit via maximum-likelihood with the application of the {\sl expectation maximisation} (EM) {\em algorithm} (see Appendix\,\ref{sec:em}), an iterative method to derive maximum a posteriori estimates of parameters in statistical models, where the model depends on 
latent variables\footnote{{Unobserved, hidden variables, usually inferred from observables, like e.g. categories.}} \citep[][]{Dempster77maximumlikelihood}. We choose this method over a simple non-linear least-squares regression, because of its high reliability in converging to a successful fit even in cases where the regression could not.

\begin{figure}
		\centering
		\includegraphics[width = 0.49\columnwidth]{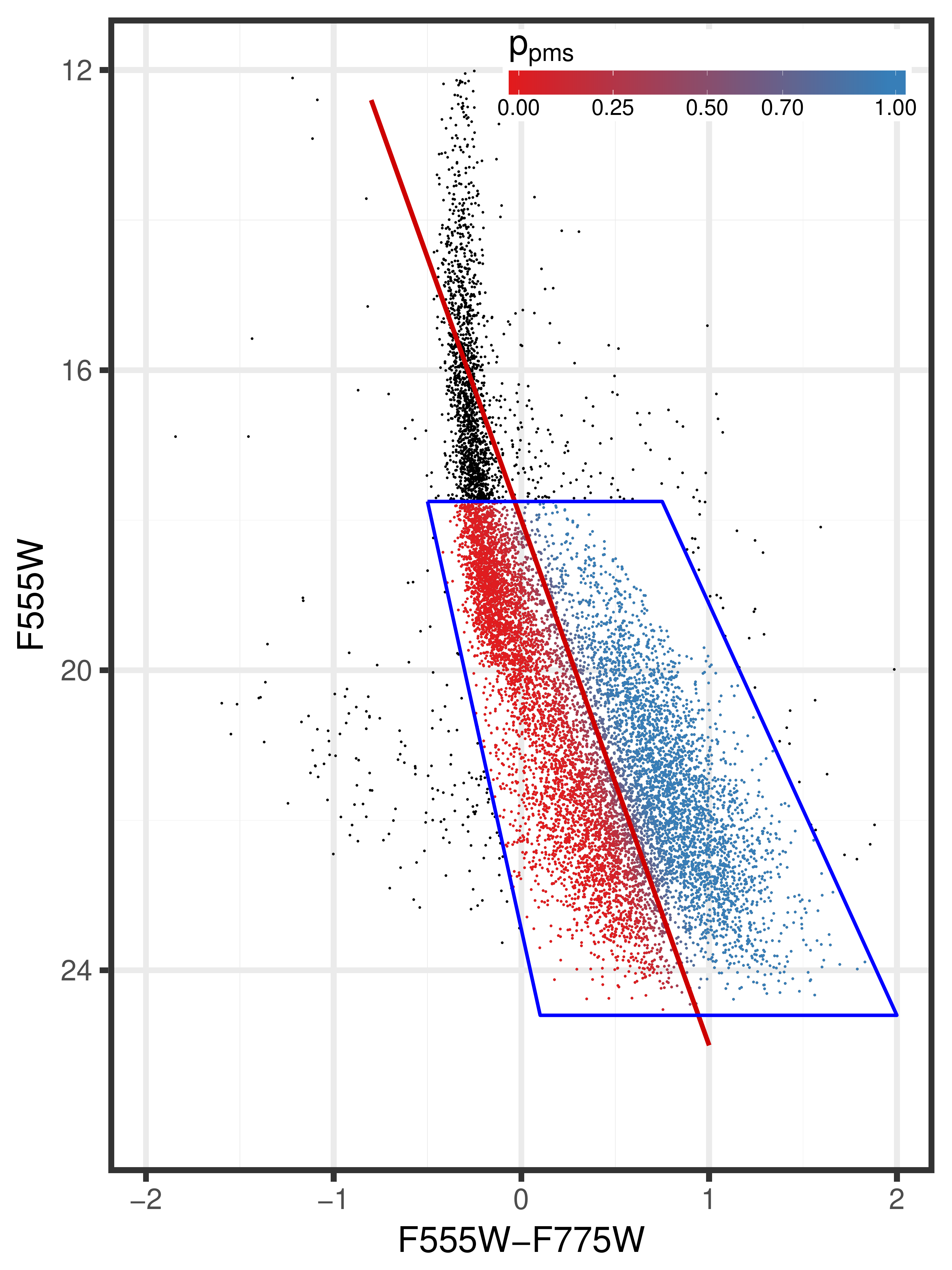}
		\includegraphics[width = 0.49\columnwidth]{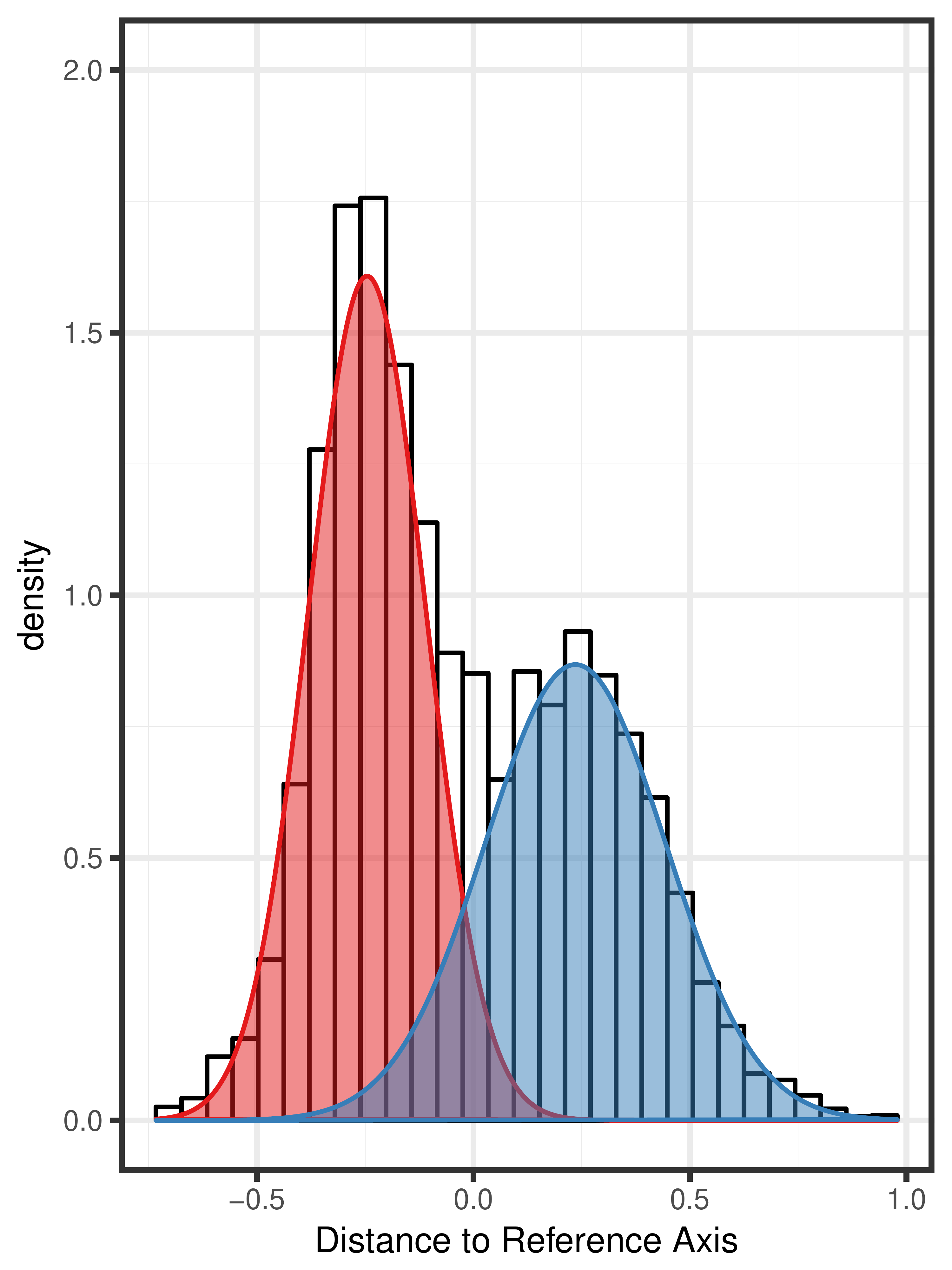}
		\includegraphics[width = 0.49\columnwidth]{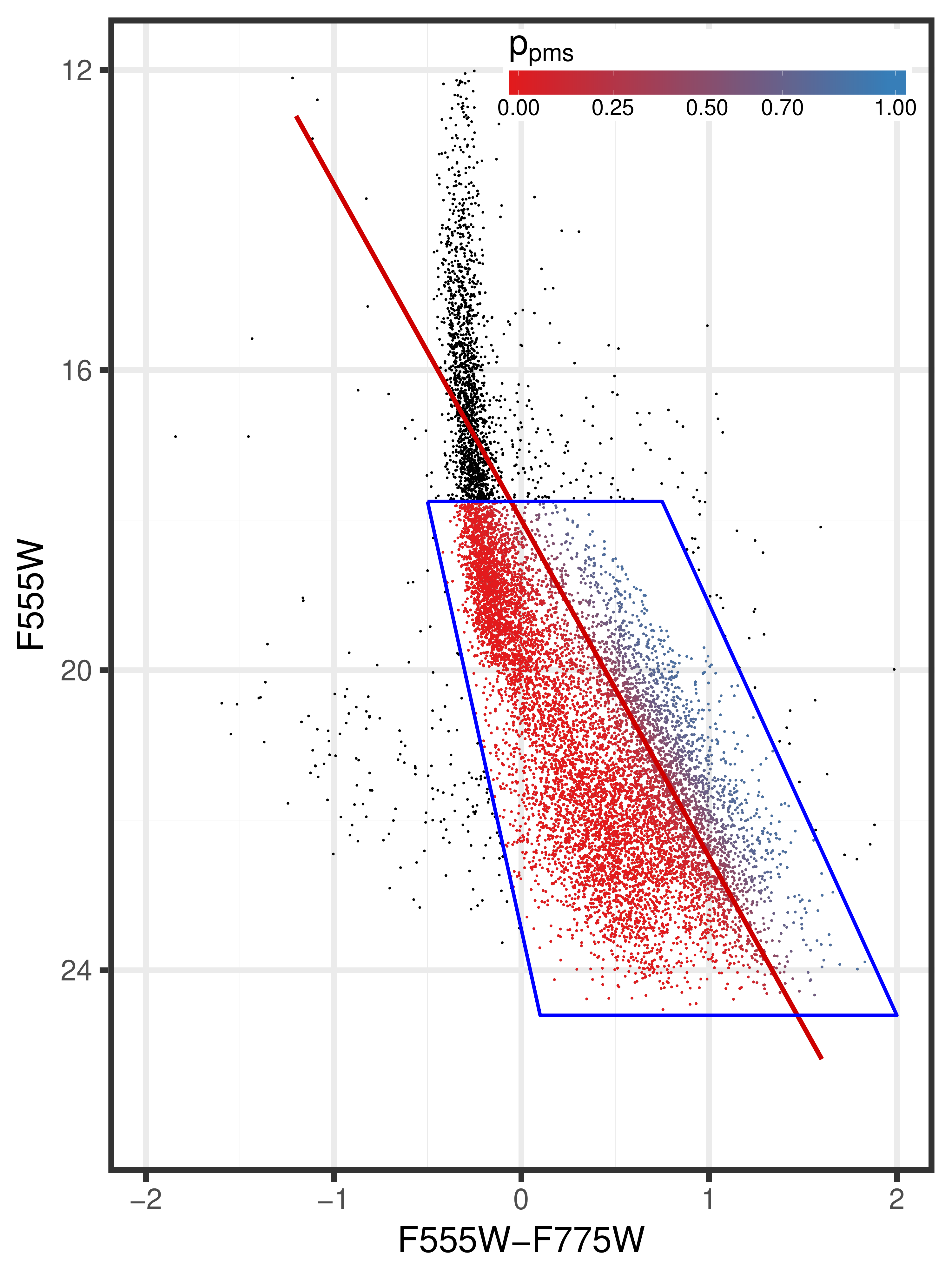}
		\includegraphics[width = 0.49\columnwidth]{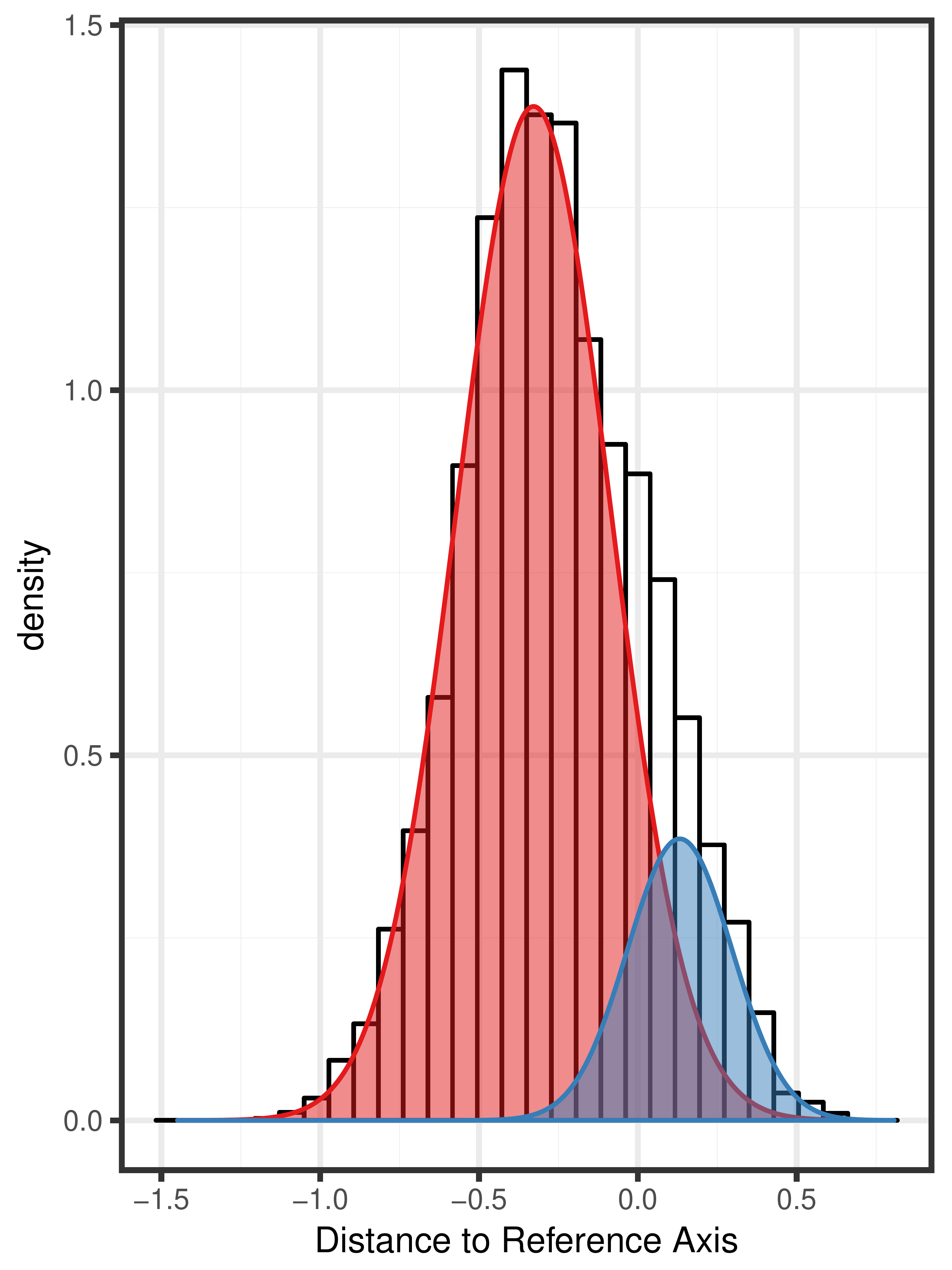}
		\caption{{\em Left:} The optical CMD of our selected training stellar sample with the reference line (separating the observed LMS and PMS parts of the CMD), drawn with a {dark} red line. The blue box indicates the CMD region occupied by the selected sample. {The points within the blue box are colour-coded according to their posterior probability of belonging to the right component of the mixture model displayed in the right panel.} {\em Right:} Histogram of the stellar {(perpendicular)} distances from the reference line with the bimodal Gaussian components of the fit solution overlaid. The panels show the resulting distributions of distances from the reference line for lines with the same intercepts but different slopes. While the outcome of the EM method is found to be independent of the actual position of the reference line, as demonstrated in these plots, it is very sensitive to its slope. It is thus important to identify a reference line the slope of which fits at the best possible degree that of the observed gap between LMS and PMS stars (see Sect.\,\ref{sec:EM_fit}).}
		\label{fig:EM_fits_combined_AxisVariation}
	\end{figure}

The bimodal Gaussian distribution used to model the distances of stars from the LMS-PMS separating line has the form
	\begin{equation}
		G(x) = \sum_{m=1}^{2} \alpha_m \Phi(x;\mu_m, \sigma_m),
	\end{equation}
where $\alpha_m$ denotes the mixing proportions, with the sum of all proportions (in this case two) equal to unity, and $\mu_m$ and $\sigma_m$ are the mean and the standard deviation, respectively, of each of the individual components. Using the model fit by the EM algorithm, one can estimate the posterior probability $\mathrm{p}_{im}$ that a star $i$ belongs to one of the components $m$ of the Gaussian mixture model as 
	\begin{equation}
		\mathrm{p}_{im} = \frac{\alpha_m \Phi(x_i; \mu_m, \sigma_m)}{\sum_{k=1}^{2}\alpha_k \Phi(x_i, \mu_k, \sigma_k)}.
	\end{equation}
With this measure we can distinguish the PMS from the LMS stars in our selected sample, on an individual-star basis by assigning a probability of PMS membership to each star. We can thus set a probability threshold above which all stars are considered as the best PMS examples. In Figure\,\ref{fig:EM_fits_combined_AxisVariation} we show two examples of our test runs of the EM method for different reference lines. The histograms on the right panel are only shown to provide a visualisation of the result, since the fitting process itself requires no binning of the distance measure. The posterior probability $\mathrm{p}_{im}$ of each star being a PMS star is calculated from the Gaussian mixture model component with the larger mean $\mu_m$, i.e., the PMS component of the model.

\begin{figure}
		\centering
		\includegraphics[width = \linewidth]{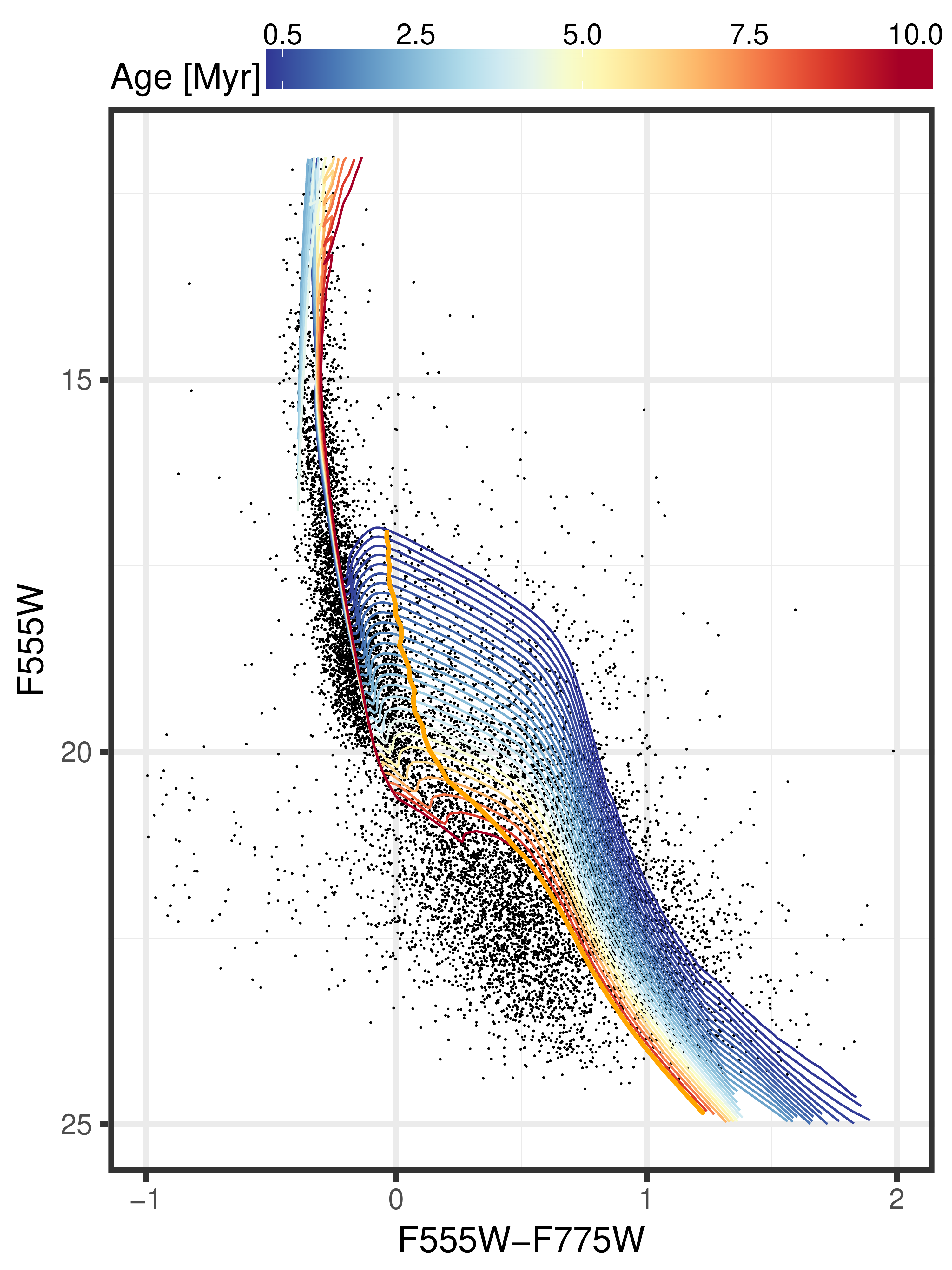}
		\caption{{\em Left:} Optical extinction-corrected CMD of our selected training stellar sample with an overlay of the PARSEC isochrones with ages between 0.5 and 10\,Myr (coloured according to their age) used to approximate the LMS-PMS gap. The thick orange line marks our extrapolated threshold curve between the PMS and LMS.}
		\label{fig:Isochrones_extrapol}
	\end{figure}

The examples of Figure\,\ref{fig:EM_fits_combined_AxisVariation} show that the EM method is quite successful in distinguishing the two separate populations within the low-brightness regime of the CMD {for a given reference line}. Our tests also demonstrated that the outcome of the EM method is independent of the axis intercept of the reference line.  However, as shown in the plots of  Figure\,\ref{fig:EM_fits_combined_AxisVariation} the result of the EM method does depend on the choice of the slope. 
It is thus important to accurately define the reference gap between the LMS and PMS populations in the observed CMD, in order to avoid any potential biases in the application of the EM method. 
{Since representing this gap with a single straight line would provide an unrealistic boundary between LMS and PMS stars, for the application of the EM algorithm we do not consider a single LMS-PMS reference line (as in the examples of Figure\,\ref{fig:EM_fits_combined_AxisVariation}). Instead, we define a threshold curve using a series of PARSEC isochrones, ranging from 0.5 to 10\,Myr, which approximates realistically the observed LMS-PMS gap.}
While there may be somewhat older PMS stars, 
{we select 10\,Myr as the oldest considered age} based on the fact that this limit corresponds to the majority of the star formation history of the region, as specified by previous studies \citep{Hunter1995,2015ApJ...811...76C}. The corresponding isochrone model also nicely traces the observed gap in the low brightness regime {(see orange line in Figure\,\ref{fig:Isochrones_extrapol})}. 

We consider the faint part of the 10\,Myr isochrone up to one stage before its MS turn-on as the best representative line of the LMS-PMS gap, and we extend this line to brighter magnitudes by connecting the points corresponding to the same stage, i.e., to the red of the local minimum before the turn-on, for all the other isochrones. With this process we construct a threshold curve that does not overlap with the UMS, while adequately tracing the gap between the LMS and PMS populations, which we want to quantify. This is demonstrated in Figure\,\ref{fig:Isochrones_extrapol}, where each isochrone model is plotted with a different colour, and the thick orange line indicates the constructed threshold curve between PMS and LMS.

\begin{figure}
		\centering
	\includegraphics[width = \linewidth]{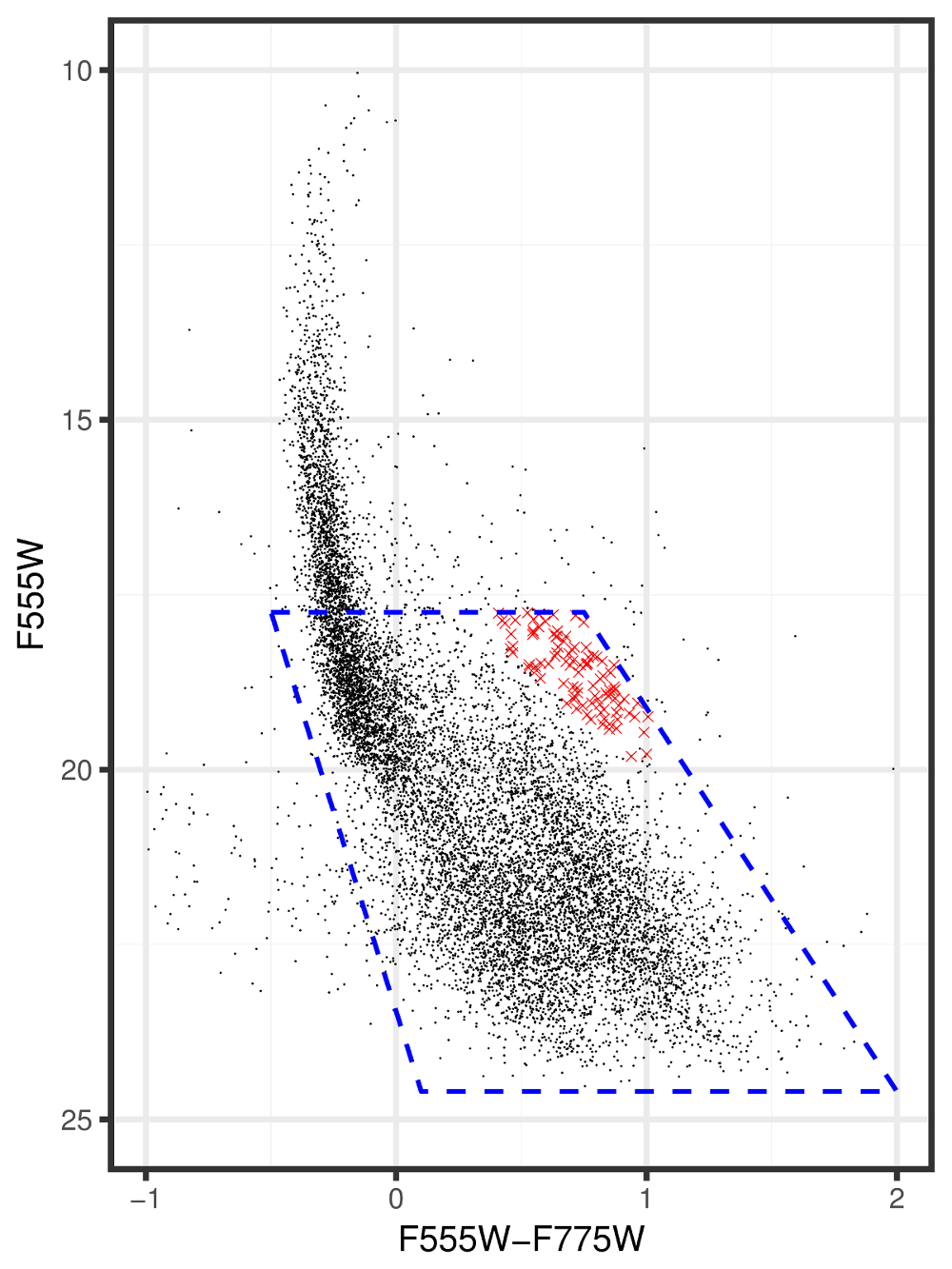}
		\caption{Optical extinction-corrected CMD of the data set used to create the training set. The blue dotted line indicates the data that are used for the EM fitting. The red crosses mark the data points that are excluded from the fitting, because they fall into the {RC} area.}
		\label{fig:Em_fit_dat_rc_ellipse}
	\end{figure}

For the application of the EM algorithm, instead of using a single LMS-PMS reference line, we use the constructed LMS-PMS threshold curve to generate {\sl a series of reference lines} by fitting a line to sequential sets of four points of the curve. We allow for some overlap between the point sets, with three of the brightest points in each set coinciding with the three faintest in the next. With this process we produce 46 different lines with different parameter sets of slopes and intercepts. The determination of the PMS membership probability for the stars in the selected sample is then made with the application of the following steps: (1) We calculate the distances of the stars from each 
{of the 46 reference lines}, (2) we fit the corresponding bimodal Gaussian distribution, (3) we estimate from each model the PMS component membership posterior probability for each star, and (4) we average the results from all Gaussian mixture models for each star. Our selection excludes all stars brighter than $m_{\rm F555} = 17.75$\,mag and the noisy observations to the right of the PMS and left of the LMS population, outside the CMD region limits shown in Figure\,\ref{fig:Em_fit_dat_rc_ellipse}. In our treatment we also do not consider the stars marked with red crosses in this figure as these fall into the {RC} part of the CMD, and they are likely evolved stars. With the methodology described above we avoid contamination by UMS or other more evolved stars and objects with poor photometry during the fitting process. The fit of the Gaussian distribution is repeated 100 times for each parameter set and the resulting PMS component membership probabilities are averaged to reduce the influence of the random initial model parameter guess in the EM algorithm. 

\begin{figure}
		\centering
		\includegraphics[width = \linewidth]{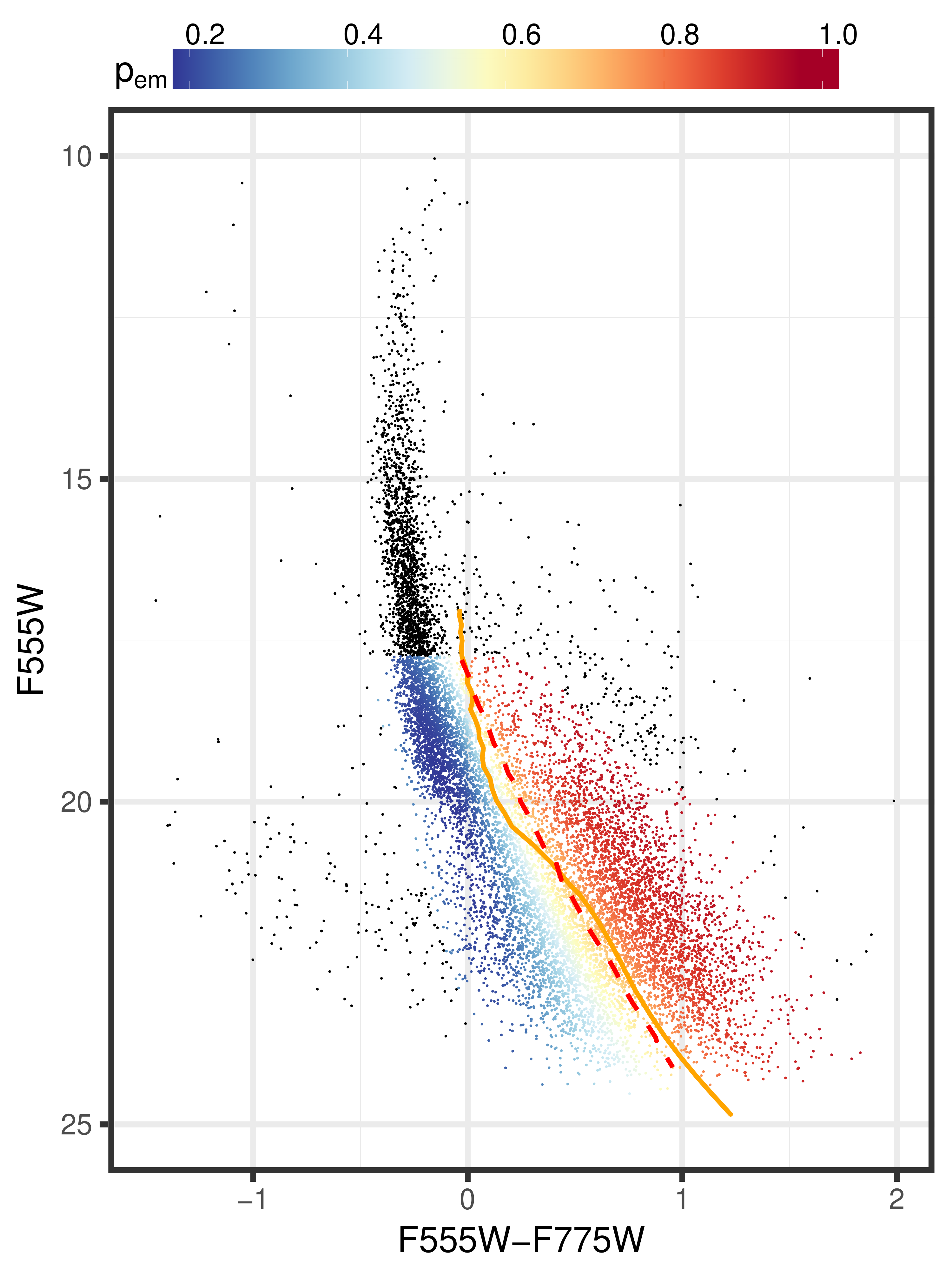}
		\caption{Optical extinction-corrected CMD of the stars in the
		region of NGC 2070 included within the $1\,\sigma_\mathrm{n}$ density significance level (see map in figure \ref{fig:ngc2070_contours}). Stars that were considered during the application of the EM method are coloured according to their estimated PMS membership probability $\mathrm{p_{em}}$. The orange line marks the previously established approximation curve of the LMS-PMS gap (see figure \ref{fig:Isochrones_extrapol}) for reference. The red dotted curve indicates a 70\% probability threshold of $\mathrm{p_{em}}$.} 
		\label{fig:TrainingAllSlopes}
	\end{figure}

The outcome of the application of the EM method is visualised in the CMD of Figure \ref{fig:TrainingAllSlopes}, where each considered star is coloured according to its estimated PMS membership probability ($\mathrm{p_{em}}$). The threshold curve used for the application of the EM algorithm and the curve corresponding to the limit of $\mathrm{p_{em}} = 0.7$ are also shown in the figure (with orange continuous and red dashed lines respectively). The comparison of these lines indicates that a minimum of PMS membership probability of about 70\% provides a reasonable separation between PMS and LMS stars in the training sample. We, thus, tested the construction of our training dataset by using various probability thresholds, starting at $0.7$ up to $0.9$. We assigned as PMS stars those with $\mathrm{p_{em}}$ larger than the chosen threshold, while the remaining stars were characterised as non-PMS. Our subsequent investigation of the performance of the classification algorithms in dependence of the considered $\mathrm{p_{em}}$ thresholds showed that the best PMS candidates in the training set are all stars with $\mathrm{p_{em}} \geq 0.85$ (see Section 5).


\subsection{The Final Training Dataset}\label{sec:final_train_set}

With the implementation of the EM method as described above we established a reasonable dataset of low-brightness true PMS and LMS stars to be used for training the classification algorithms. However, while the CMD area covered by these stars in the region of NGC\,2070 shows a clear distinction between these two stellar types, the bright part of this area in the complete HTTP CMD includes other types of evolved stars, such as {RC} and 
faint giant/subgiant stars. While the contamination of the PMS dataset by these stellar types is not significant, they must be considered in our final training dataset. We, thus, complete the compilation of the training set by ``artificially'' adding examples of these evolved stars as {\sl negative} (non-PMS) examples, so that the classification algorithms can treat them as such. For the {RC} stars we include the previously excluded examples, marked with red crosses in Figure\,\ref{fig:Em_fit_dat_rc_ellipse}, and we assign a PMS membership probability of $\mathrm{p_{em}} = 0$ to them. As we discuss in Sect.\,\ref{sec:kde_contours} we constrained our EM analysis for distinguishing PMS from LMS stars in a well-defined region in the CMD, where prominent members of both populations are located (blue dashed polygon in Figure\,\ref{fig:Em_fit_dat_rc_ellipse}), excluding some sources at the extreme blue and red faint parts of the diagram. We now include these faint uncertain sources in the final training set by assigning to them also zero PMS probability. The reason for this inclusion is to eliminate the danger of misclassifying objects with uncertain measurements as PMS stars.

\subsubsection{Adding Evolved Field Stars in the Training Set}

Apart from the stellar sources discussed in the previous paragraph, an important contaminant of a PMS dataset 
are the old stellar populations of the general LMC field, occupying the 
giant/subgiant branches of the CMD. The 
fainter giant and subgiant stars of the LMC field can roughly coincide, depending on age and reddening, with the bright part of the PMS population. In the case of variable extinction by gas and dust, 
giant and subgiant field stars are distributed along the reddening vector and can overlap with the CMD positions of PMS members of young clusters on the same line of sight. In order to use a training dataset that accounts also for these contaminants we identify typical examples of faint, field 
giant and subgiant stars in regions of the observed HTTP field-of-view, which mostly cover the general LMC field. We select two such regions to account for both high- and low-extinction of the field stars by the nebula. 

In selecting these regions we were aided by a preliminary unrefined classification of PMS stars in the HTTP dataset by employing a {\sl Support Vector Machine} (SVM) algorithm trained on the V- and R-equivalent magnitudes of the stars in the preliminary training set constructed in the previous section. Details on the employment of this method are given in the final application of our classification (Sect.\,\ref{chap:Classification}) and in Appendix\,\ref{app:SVM}. Here, it suffices to note that we applied the SVM method via a 10-fold cross-validation, repeated 5 times and labeling as PMS stars {(label of 1)} those with PMS membership probability, derived with the EM method, higher than the lowest reasonable limit of $\mathrm{p_{em}} = 0.7$. We performed the classification of the HTTP stars with measurements in these two filters and we retrieved a {\sl tentative set of PMS candidates}, i.e. stars with a classification probability $\geq 0.5$, across the whole observed FoV. We constructed the surface density map of this stellar sample in order to identify the regions across the Tarantula Nebula that are mostly devoid of candidate PMS populations, i.e., the regions where the cleanest samples of field stars can be detected. 

\begin{figure*}
		\centering
		\includegraphics[width = 0.39\linewidth]{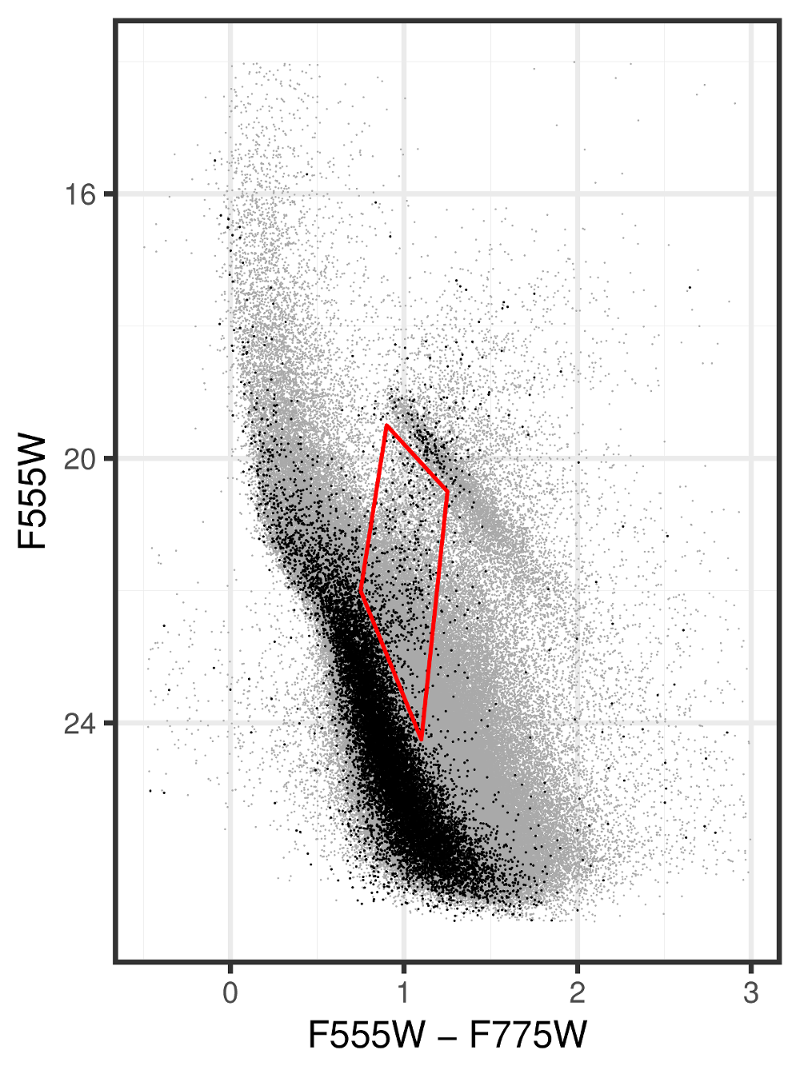}
		\includegraphics[width = 0.39\linewidth]{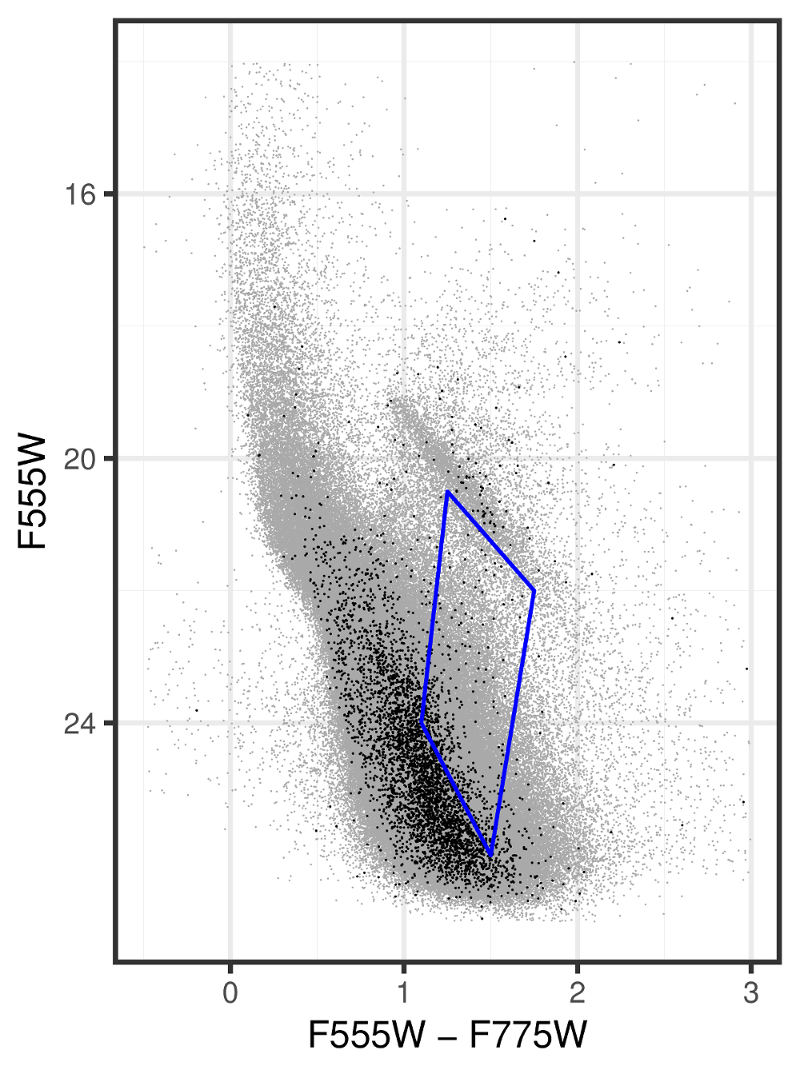}
		\caption{Optical CMD of the HTTP data in gray with an overlay of the CMD of our field star selections, where the low extinction region is shown on the {\em left} and the high extinction region on the {\em right}. The solid red and blue polygons indicate our selections for the low and high extinction, faint giant/subgiant candidates, respectively.} 
		\label{fig:turn_off_selection}
	\end{figure*}

We combined the surface density map of the PMS candidates with the extinction map of Figure\,\ref{fig:HTTP_ex_map_final} to identify the {\sl field regions} within the observed area with both the lowest and highest extinction. We, thus, considered the field contaminants in the whole range of reddening conditions across the Tarantula Nebula. It is interesting to note that our selected low-extinction field region roughly coincides with that defined by \cite{2015ApJ...811...76C} as reference field in recovering the star formation history of NGC\,2070. The identified field stellar populations are depicted in the CMD of Figure\,\ref{fig:turn_off_selection}. We select from this population the 
{most prominent}, faint 
giant/subgiant stellar candidates as enclosed by the red and blue polygons (one from the high-reddening and one from the low-reddening region), and add a bit more than 900 objects to our training set as non-PMS examples, i.e., stars with zero PMS membership probability. 

\begin{figure}
		\includegraphics[width=\linewidth]{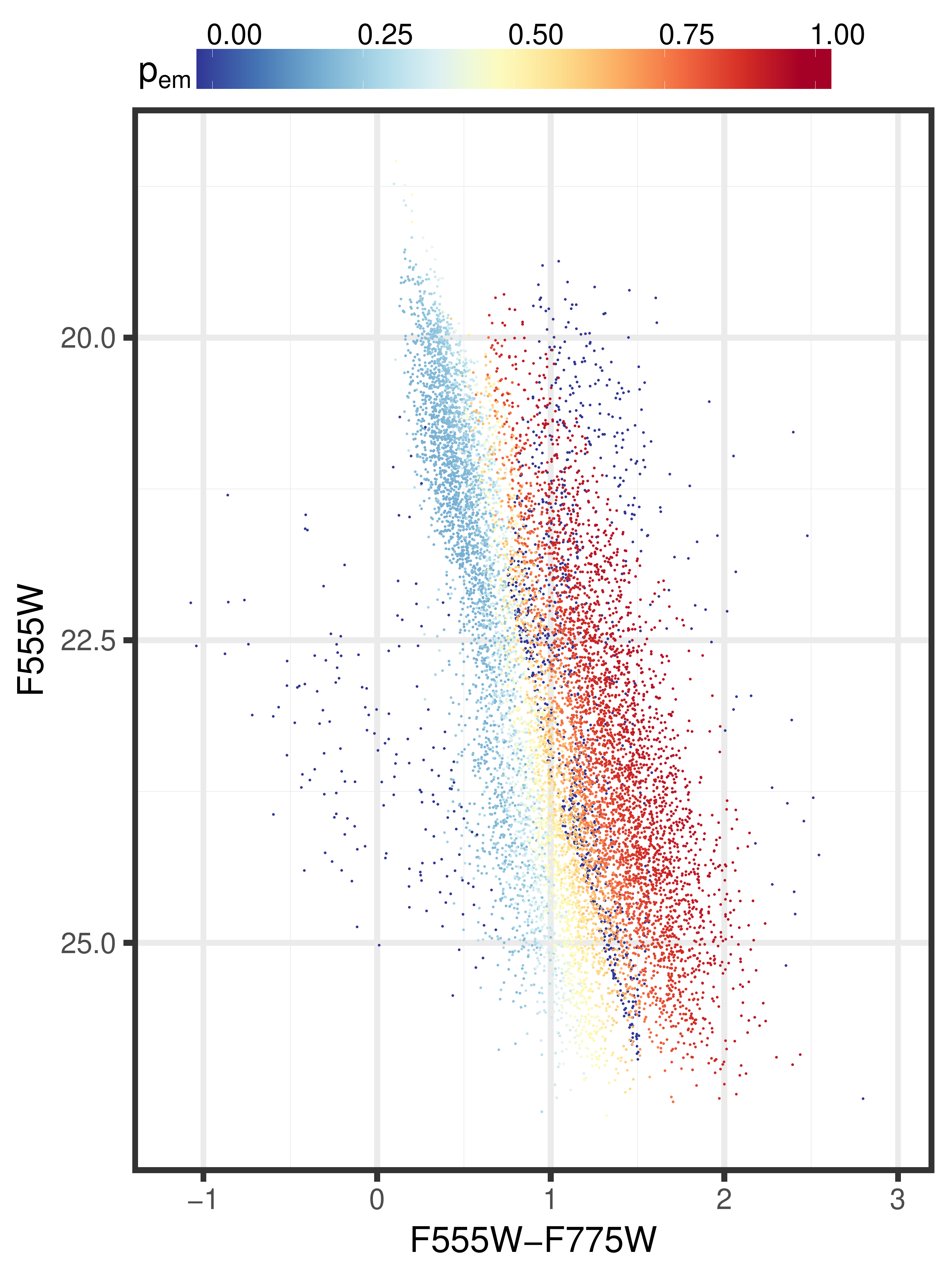}
		\caption{Optical CMD of the final training dataset selection. The data points are colour-coded according to the PMS membership probability defined as described in the text (see Sect.\,\ref{s:trainset}).}
		\label{fig:training_set}
	\end{figure}

With the process described in this and previous sections, we have constructed a training dataset of 10443 stars, containing the best possible examples of (1) evolved field stars, both 
LMS and 
potential giant/subgiant stars, (2) {RC} stars, (3) non-specified stars with poor photometry, and (4) young low-mass PMS stars, which we aim at identifying across the whole FoV. Figure \ref{fig:training_set} shows the part of the CMD on which our training will take place, with an overview of the positions of these populations, coloured according to their $\mathrm{p_{em}}$ probability. As shown in this CMD, our classification is limited to the faint part of the CMD where the PMS stars reside, and therefore, we do not include examples of the UMS stars, as their positions should not overlap with those of the PMS stars. 


{The available data in the training data set in each of the HST filters is summarised in the top part of Table\,\ref{tab:train_parameters}.}
Apart from V- and R-equivalent wavebands, our training set includes substantial data in J- and H-equivalent wavebands (although with a smaller FoV), making the training of our algorithms on these variables also feasible. The remaining three (UV-, U-, and H$\alpha$-equivalent) filters are less helpful for the application of a machine-learning classification due to their significantly smaller coverage, drastically reducing the amount of data to predict on (see Table\,\ref{tab:HST_filters}). As a consequence, we limit our tests to these four filters. Also, before feeding the learning process with the training examples, we need to define this limiting PMS membership probability threshold, $\mathrm{p_{em}}$, to be considered for separating the positive (PMS) instances from the negative (non-PMS) instances in the training dataset. 
{The primary criterion for determining this threshold is the inclusion of the purest possible sample of PMS candidates, reducing, thus, the number of possible false positive examples. 
As we discuss later, we determined the best threshold choice to be $0.85$ based on two additional factors. (1) The need for balance between the numbers of positive and negative examples in the training data set, and (2) the classification performance of the algorithms.}

Concerning the balance between positive-negative examples, 
the lowest reasonable threshold of $\mathrm{p_{em}}= 0.7$ provides about 38\,percent of the training set as positive instances. At the even higher threshold of $\mathrm{p_{em}}= 0.9$ we retain roughly 19\,percent of the dataset in positive examples, which still provides a useful amount of observations. Using an even higher probability threshold would not be practical, because we would limit the training sample to a number of positive instances that would be unrealistically low, and that would introduce a strong imbalance between positive and negative examples, which is not ideal for machine learning. 
Concerning the performance of the classification algorithms, our experiments (Section 5) showed that we achieve the best trade-off between algorithm performance and training set balance by constraining the sample of PMS members in the training set to those with $\mathrm{p_{em}} \geq 0.85$, corresponding to roughly 27\,percent of the total training dataset.

\section{Classification of Pre--Main Sequence Stars}\label{chap:Classification}

After constructing the training dataset to be used for the learning process, we performed various experiments in order to identify the most efficient machine-learning algorithm for the classification of PMS stars in the Tarantula Nebula. Since our training set was constructed from the region of NGC\,2070, our classification will identify the stellar {\sl siblings} of the PMS members of this region, i.e., stars with similar characteristics and star formation history, spread across the whole nebula. In our experiments we tested three popular classification algorithms: Decision Trees, Random Forests (RF), and Support Vector Machines (SVM). Descriptions of the concepts behind these algorithms and references to the related literature are provided in Appendix\,\ref{app:MLCA}. During our early experimentation we also considered the application of the Naive Bayes Classifier, a simple probabilistic classifier based on Bayes' theorem, which assumes (naively) a strong independence between the features \citep[see, e.g.,][]{RussellNorvigAI}.

The success of supervised {machine-learning} modeling is based on the availability of complete sets of observations with as many variables to model on as possible. As a consequence, in the present study a limiting factor in our classification is the amount of available data, the algorithms can be trained on. This translates to both the number of available stars per filter and the number of stars observed in as many filters as possible, i.e., the size of the feature space (see Appendix\,\ref{app:MLCA}). The second aspect is particularly important, because most of the classification algorithms cannot perform a prediction on incomplete feature vectors, i.e., on missing data. With this in mind we optimised our classification for datasets with the best waveband and spatial coverage, i.e., for stars found in the HST wavebands equivalent to standard V, R, J, and H filters. Among the tested methods only {\sl decision trees} can compensate for incomplete feature vectors using the so-called {\sl surrogate splits} (see Appendix\,\ref{app:dtree}). It is, thus, the only algorithm that can predict on all available HTTP filters. For our tests with other algorithms, as mentioned above, we do not take into account measurements in the UV-, U- and H$\alpha$-equivalent filters, due to the significantly large amount of non-detections in these wavebands (as shown in Table\,\ref{tab:HST_filters}). Our experiments with the use of photometric flags as categorical classification variables (to compensate for non-detections) also performed very poorly.

The performance of the algorithms was measured using {three} 
metrics, the {\sl accuracy}, {the {\em balanced accuracy}, both} estimated from the {\sl confusion matrix}, and the area under the {\sl Receiver Operating Characteristic} (ROC) curve, or in short {\sl Area Under the Curve} (AUC), 
{all} described in Appendix\,\ref{app:Performance}. These metrics were calculated on the basis of a {\sl train/test split}, i.e., by splitting the original training dataset in two subsets, one to train the algorithm (``Train'' subset), and one to test its performance (``Test'' subset). This method is very efficient when there is a {sufficiently} large number of records in the training dataset, as in our case. Typically, a 70/30 split, i.e., $\sim$\,70 percent of the training dataset reserved for the Train subset and the rest for the Test, is the most efficient split for training the algorithm in order to avoid {\sl overfitting}, i.e., constructing a general model that can fit a variety of data, and not exclusively those in the training dataset. It is worth noting that the measurements of the predictive power of the classification model must be made on a {\sl held-out} Test set, i.e., the records of the Test set must not be influenced in any way by the instances in the Train set. Therefore splitting the Train/Test sets is an important aspect of the process. 

{For the training of the algorithms on the Train subset, we employed a 10-fold cross-validation (see appendix \ref{app:CV}) in all our experiments.}
Due to differences in the available data that depend on the availability of measurements in various wavebands, the Train/Test subsets may vary from one experiment to the other. {For the sake of direct comparability of the presented results the SVM and RF algorithms are trained and tested on exactly the same subsets for any given experiment. This is of course not a necessity, as the algorithms' performance varies insignificantly for different partitions of the data.} 


The implementation of the algorithms was in the programming language {\sl R}, an environment for statistical computing and graphics \citep{Rlanguage}. 
From early on the {\sl Naive Bayes classifier} shows significantly low performance in our experiments, achieving accuracies of at most $\sim 60\%$, providing thus classification models comparable to random guessing. As a consequence, we do not further discuss in our analysis this algorithm, and we focus on the three remaining methods that proved to provide more accurate results. Random forests, operating by constructing a multitude of decision trees, correct for decision trees' occasional overfitting to the Train dataset \citep{ElOfStatLearn}. The RF algorithm is thus a more efficient choice for our classification. Nevertheless, decision trees can handle missing values in the photometric variables of stars without imputation (by using surrogate splits). We apply, thus, a preliminary classification with decision trees only on the complete set of photometric variables (set no. 1), which includes a large number of missing values, in order to understand how these variables may influence our classification. Table\,\ref{tab:train_parameters} gives a detailed summary of our experiments, listing the sets of different variables (features) combinations used for training and classifying the HTTP stars, the algorithms applied, the available instances in both the training and the whole HTTP datasets in dependence of the variables choice, the performance of the algorithms on the held-out Test dataset in terms of the accuracy, {balanced accuracy} and the AUC, and the amount of the identified PMS candidates\footnote{{In the following sections we only state the accuracy and AUC for simplicity. We refer the reader to Table\,\ref{tab:train_parameters} for the corresponding balanced accuracy.}}

As we discuss later in this section, for the final census of the PMS stars across the Tarantula nebula we operate on the features set no. 3 (as described in Table\,\ref{tab:train_parameters}), because it provides the largest stellar coverage across the whole extent of the observed field, 
thus the richest stellar sample, {and the highest performance scores across the tested methods.}
This sample comprises observations in three variables (V-, R-equivalent magnitudes and extinction), providing the largest amount of HTTP data to classify. 

{Features set no.1 includes all photometric measurements, while set no.2 is equivalent to set no. 3, but without the inclusion of extinction. It is used in order to validate the influence of extinction to the accuracy of our models.} Including the infrared measurements unfortunately limits not only the area that we can investigate (Figure\,\ref{fig:http_spatial_coverage1}), but also drastically reduces the amount of data that can be classified, due to the lower number of stars observed in all four wavebands\footnote{We remind that while decisions trees can deal with missing measurements, the RF and SVM algorithms cannot.}. We nevertheless investigate in features set no. 5 their influence on the performance of the algorithms {(Note that extinction is not included in this features set, as it is intended to highlight the influence of the inclusion of the infrared bands alone)}. Features set no. 4 is intended to give insights on whether the photometric errors in the respective filters could prove to be helpful in the classification approach, or not. {Therefore extinction was not included in this features set.}

\subsection{Classification with Decision Trees}
\label{sec:ClassDtree}
{\sl Decision or classification trees} use flowchart-like structures that break the process of a complex decision into a series of simpler decisions, made upon the input features of the available data (see Appendix\,\ref{app:dtree}). Beginning at the {\sl root node} data flows through if-else {\sl decision nodes} that split the data according to its features. The {\sl branches} indicate the potential choices and the {\sl leaf nodes} the final decisions. Given that the decision tree algorithm can compensate for incomplete feature vectors, our first experiments were made with the application of this algorithm in order to use the complete photometric variables space of the HTTP dataset, which includes a large number of missing values for stars not identified in specific wavebands. We trained decision trees using various maximum tree depths up to 30, at most 5 surrogate splits per node and pre-pruning with complexity degrees of the order of 0.01, {using the {\em Gini Index} as measure of node impurity} (see Appendix\,\ref{app:dtree} for explanations on these parameters). {In order to compare with the other algorithms the results presented for this method are also based on PMS stars with $p_\mathrm{em} \geq 0.85$}. The final decision tree, with an accuracy of 86.11\% and an AUC of 0.845 on the held-out test set, did not achieve the expected performance, 
but it appeared quite promising in providing valuable insight on the importance of certain filters. 

In this tree the two most important variables for primary splits appear to be the measurements in the F555W and F160W filters with minor contributions from those in F775W. Measurements in all remaining filters are only considered for surrogate splits. Comparing with 
Table\,\ref{tab:train_parameters} this is not surprising, since our training dataset contains significantly less records in the F275W, F336W and F658N filters, and therefore these variables are not considered for primary splits. The fact that the filters pair (F555W, F160W) was chosen over the (F775W, F110W) pair may indicate that measures in the V- and H-equivalent bands provide the intrinsically best combination of variables. This can be explained by the fact that the (F555W, F160W) filters pair provides a rich stellar sample across a dynamic range in colours, which is wider than those of any other combination of these four filters. The wide colour spread of the data allows a clearer distinction between PMS and LMS stars on the CMD. In any case these results provide strong indications that {\sl near-infrared measurements may be very important to the identification of PMS stars with {machine-learning} classification techniques}.

	\begin{table*}
		\centering
		\caption{Overview of our {machine-learning} experiments for the
			identification of PMS stars in the Tarantula Nebula. The table lists (1)
			the investigated combinations of variables (features) for training and
			prediction, {(2) the available training data for each individual feature}, (3) the applied algorithms, (4) the available data in both
			the training set and the HTTP survey for each features set, (5) the
			performance of each algorithm on a held-out test set {(the value in parenthesis gives the performance on the training set during cross-validation for comparison)}, and (6) the amount
			of resulting PMS candidates, i.e., stars identified with a predicted
			probability of being PMS of $\mathrm{p_{pms}} \geq 0.5$. Note that the
			number of available records is given only for the RF and SVM algorithms,
			because the decision tree algorithm can predict and train on data sets
			with incomplete attributes, i.e., on all records in the HTTP survey. {The performance on the test set is found to be comparable to that on the training set, which demonstrates that none of our models exhibit a case of overfitting.} }
		
		\begin{tabular}{l|r|c|c|c|c|c}
			\hline
			\hline
			{Feature} & {Available Training Data}  & \multicolumn{5}{c|}{{Features set No.}} \\
			& (out of 10443 stars)        & 1 		     & 2			 & 3			 & 4 			 & 5 			 \\
			
			\hline
			F275W (UV) &  2210  (21.2\%)   & \Checkmark    &               &               &               &               \\
			F336W (U)  &  4880  (46.7\%)   & \Checkmark    &               &               &               &               \\
			F555W (V)  &  10443 (100\%)    & \Checkmark    & \Checkmark    & \Checkmark    & \Checkmark    & \Checkmark    \\
			$\sigma_V$ &  10443 (100\%) &           &               &               & \Checkmark    &               \\
			F658N ($\mathrm{H}_\alpha$) & 4576  (43.8\%)   & \Checkmark    &               &               &               &               \\
			F775W (R)  & 10443 (100\%)   & \Checkmark    & \Checkmark    & \Checkmark    & \Checkmark    & \Checkmark    \\
			$\sigma_R$ & 10443 (100\%)  &             &               &               & \Checkmark    &               \\
			F110W (J)  & 9600  (91.9\%)  & \Checkmark    &               &               &               & \Checkmark    \\
			F160W (H)  & 9597  (91.9\%)  & \Checkmark    &               &               &               & \Checkmark    \\
			Extinction $\mathrm{A_v}$& 10443 (100\%)  &  &               & \Checkmark    &               &               \\
			\hline
			\multicolumn{2}{l|}{{Applied algorithms:}}      &               &               &               &               &               \\
			\multicolumn{2}{l|}{Decision Tree (DTRee)}           & \Checkmark    & \Checkmark    &               &               &               \\
			\multicolumn{2}{l|}{Random Forest (RF)}           &               & \Checkmark    & \Checkmark    & \Checkmark    & \Checkmark    \\
			\multicolumn{2}{l|}{Support Vector Machine (SVM)}   &               & \Checkmark    & \Checkmark    & \Checkmark    & \Checkmark    \\
			\hline
			\multicolumn{2}{l|}{{Available Data:}}		  	 &	             &               &               &               &               \\
			\multicolumn{2}{l|}{\hspace*{10pt} Training Total}           & 10443		 & 10443         & 10443         & 10443         & 9283          \\
			\multicolumn{2}{l|}{\hspace*{20pt} Reduced Training}        & 7310		 & 7310          & 7310          & 7310          & 6498          \\
			\multicolumn{2}{l|}{\hspace*{20pt} Held-out Test}           & 3133		 & 3133          & 3133          & 3133          & 2785          \\
			\multicolumn{2}{l|}{\hspace*{10pt }Prediction}               & 822204     & 403018		 & 403018  		 & 400229        & 287434        \\  
			\hline
			\hline
			\multicolumn{2}{l|}{{Performance on test (training) set ($\mathrm{p_{em}} \geq 0.85$):}} &               &               &               &               &               \\
			\hline
			\multicolumn{2}{l|}{Accuracy [in \%]:}        &               &               &               &               &               \\ 
			\multicolumn{2}{l|}{\hspace*{10pt} DTree}     & 86.11 (89.23) & 87.80 (89.67) &  -            & -             & -             \\
			\multicolumn{2}{l|}{\hspace*{10pt} RF}        &   -           & 94.16 (94.36) & 95.60 (95.85) & 94.54  (94.77) & 92.78 (93.00)\\
			\multicolumn{2}{l|}{\hspace*{10pt} SVM}       &   -           & 95.18 (95.36) & 97.29 (97.47) & 94.80  (94.26) & 94.82 (94.64)\\
			\multicolumn{2}{l|}{Balanced Accuracy [in \%]:}        &               &               &               &               &               \\ 
			\multicolumn{2}{l|}{\hspace*{10pt} DTree}     & 82.60 (87.15) & 85.83 (87.92) &  -            & -             & -             \\
			\multicolumn{2}{l|}{\hspace*{10pt} RF}        &   -           & 92.06 (93.09) & 94.24 (94.67) & 92.71  (93.15) & 90.88 (91.16)\\
			\multicolumn{2}{l|}{\hspace*{10pt} SVM}       &   -           & 93.39 (94.53) & 96.61 (95.89) & 93.06  (92.63) & 93.05 (93.90)\\
 			\multicolumn{2}{l|}{ROC AUC:}                 &               &               &               &               &               \\
 			\multicolumn{2}{l|}{\hspace*{10pt} DTree}     & 0.845 (0.899) & 0.852 (0.885) & -             & -             & -             \\
 			\multicolumn{2}{l|}{\hspace*{10pt} RF}        &   -           & 0.983 (0.984) & 0.990 (0.990) & 0.986 (0.986) & 0.977 (0.978) \\
 			\multicolumn{2}{l|}{\hspace*{10pt} SVM}       &   -           & 0.988 (0.989) & 0.994 (0.994) & 0.985 (0.983) & 0.989 (0.986) \\
 			\hline
 			\hline
 			\multicolumn{2}{l|}{{Number of PMS candidates}} &               &               &               &               &               \\
 			\hline
			 \multicolumn{2}{l|}{DTree}                   & 74375         & 73006         & -             & -             & -             \\
			 \multicolumn{2}{l|}{RF}                      & -             & 21306        & 20996         & 20923         & 15898        \\
			 \multicolumn{2}{l|}{SVM}                     & -             & 21550        & 19487         & 21554         & 16655        \\
			\hline
			\hline
		\end{tabular}

		\label{tab:train_parameters}
	\end{table*}

\begin{figure*}
	\centering
	\includegraphics[width = \columnwidth]{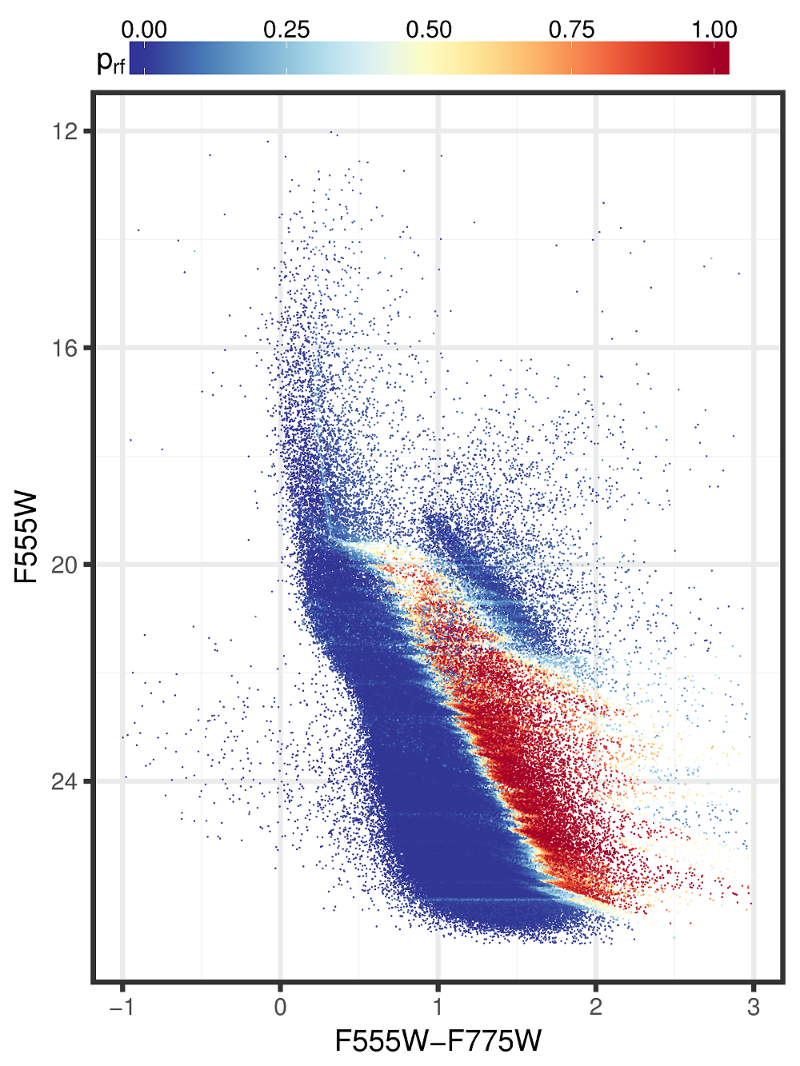}
	\includegraphics[width = \columnwidth]{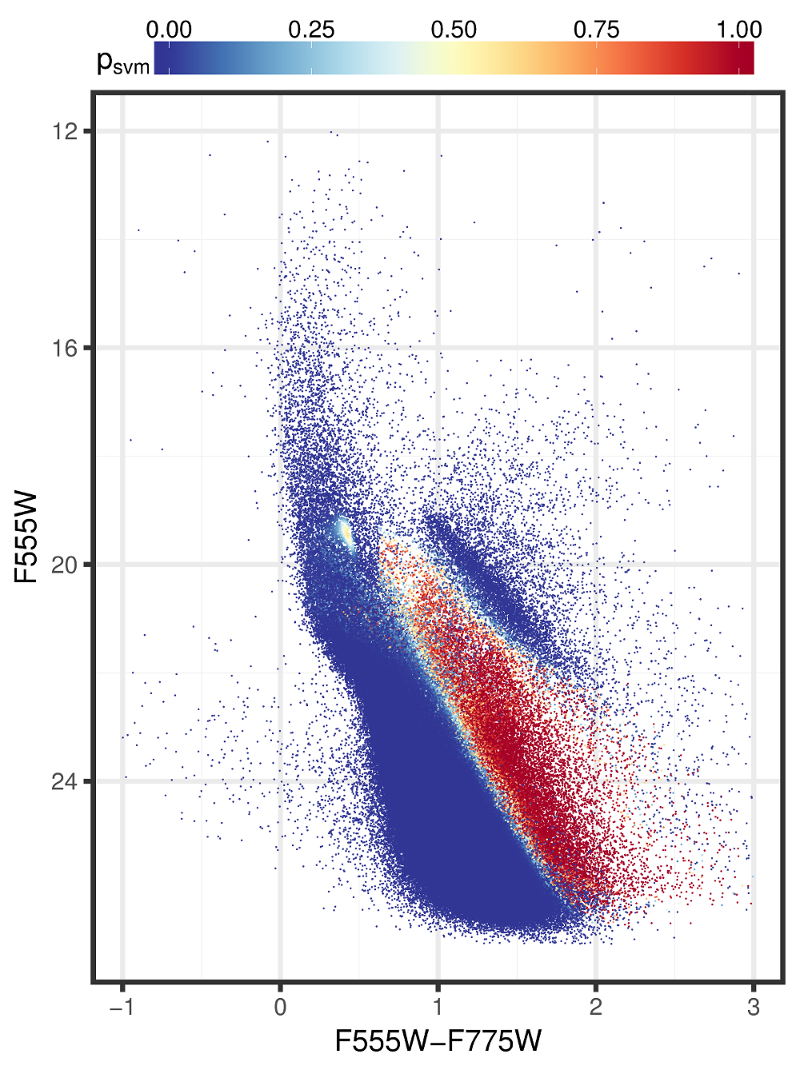}
	\caption{Optical CMDs of the HTTP stars, colour-coded according to their probability of being PMS stars, as predicted by the RF algorithm ({\em left}), and the SVM algorithm ({\em right}), trained on their measurements in the F555W and F775W bands and their extinction $A_v$.}
	\label{fig:rf_svm_VR_AV}
\end{figure*}
 
Considering that our primary dataset is that with the best spatial coverage across the observed FoV, provided by stars observed in at least one of the F555W and F775W filters, we trained a decision tree also on this stellar sample. This tree performed decently on the held-out test set, with an accuracy of 87.8\% and AUC of 0.852, comparable to the tree model based on the whole stellar sample, and still below our expectations for successful classification. Moreover, we identified few issues related to the decision tree classification. Specifically, there was a large number of misclassifications of LMS and {RC} stars after applying the algorithm on the entirety of the HTTP dataset, and the outcome of the classification itself appeared quite unrealistic, with the identified PMS candidates being aligned in prominent ``zigzag'' patterns across the CMD. We explain this phenomenon as the result of the binary splits the decision tree algorithm performs in order to make a decision. 

Another issue with the decision trees that were trained on the whole sample is that since the constructed trees use the observations in the F555W filter as one of the primary prediction parameters, they get heavily confused in the regions, where there is no coverage in this filter (see, e.g., Figure\,\ref{fig:http_spatial_coverage1}), by predicting an unrealistically large amount of PMS stars in these regions. This shows that the surrogate split method to compensate for incomplete feature vectors provides limited support to our classification goal. Based on these experiments, and due to the issues mentioned above, we assess that while the decision tree algorithm provides evidence for the importance of infrared measurements, in general it is not suited for the purpose of this study. As a consequence we did not proceed with any further tests of the decision tree, beyond these preliminary experiments. Our further tests were focused on the more sophisticated RF and SVM algorithms.

\subsection{Classification with Random Forests}
\label{sec:ClassRF}
A number of classification trees can be combined into a collection known as {\sl Decision Tree Forest}, or simply Random Forest, which is one of the most successful machine-learning classifiers. In contrast to a single decision tree that is grown in size and complexity as it is trained on the available data, the efficiency of the RF relies on the fact that the algorithm is a collection of smaller simpler trees that together reflect the data's complexity (see Appendix\,\ref{app:RF} for a detailed description). We applied our {machine-learning} method on the reduced training set in three steps: (1) We employed cross-validation to train 10 RF classification models, (2) we chose the best model based on its AUC during cross-validation, and (3) we evaluated the performance of the final model independently on the held-out test set. Two basic arguments in the implementation of the RF algorithm is the number of trees the ``forest'' consists of, and the number of variables to be sampled in each node. Due to the unavoidably low number of available variables in each features set (Table\,\ref{tab:train_parameters}), all of them were used in the training process for each considered set. The number of trees per forest, which should not be less than 200, was tested for values between 500 and 10,000. However, it appears that the algorithm's performance is not very sensitive to the number of trees, since all models provided AUC values with differences of the order of $0.0001$. Nevertheless, the best trade-off between performance and computational demand was achieved with the models for 500 trees, with that for features set no. 2 having an AUC of 0.9660. 

We evaluated the influence of our selection for the EM-derived PMS probability threshold ($\mathrm{p_{em}}$), introduced during the construction of our training dataset, on the classification performance of the RF algorithm by testing its performance in a range of limiting values, varying from $\mathrm{p_{em}} \geq 0.7$ to $\mathrm{p_{em}} \geq 0.9$ for the prominent PMS candidates in the training dataset. On our primary features set (set no. 3, i.e., F555W, F775W measurements and $A_V$), we found that apart from an overall excellent performance (AUC $\geq$ 0.984), there is a trend of increasing performance with higher limiting threshold. The best model was constructed for the highest considered threshold ($\mathrm{p_{em}} \geq 0.9$), achieving an AUC of 0.991, with a difference of only 0.007 larger than that for $\mathrm{p_{em}} \geq 0.7$. This indicates that the performance of the RF algorithm appears to be also not sensitive to the considered input sample of best PMS candidates. It should be noted, though, that choosing a threshold of $\mathrm{p_{em}} = 0.9$ might already be critical in terms of maintaining a good balance between positive and negative examples in the training set, since this threshold accounts for only $\sim$\,19\,percent of positive examples in the dataset. Based on this, and in order to achieve a trade-off between correct training of the algorithms and reasonable selection of the best positive examples in the training set, we applied a threshold of $\mathrm{p_{em}} \geq 0.85$ for the best PMS input sample. 

	\begin{figure*}
		\centering
		\includegraphics[width = \columnwidth]{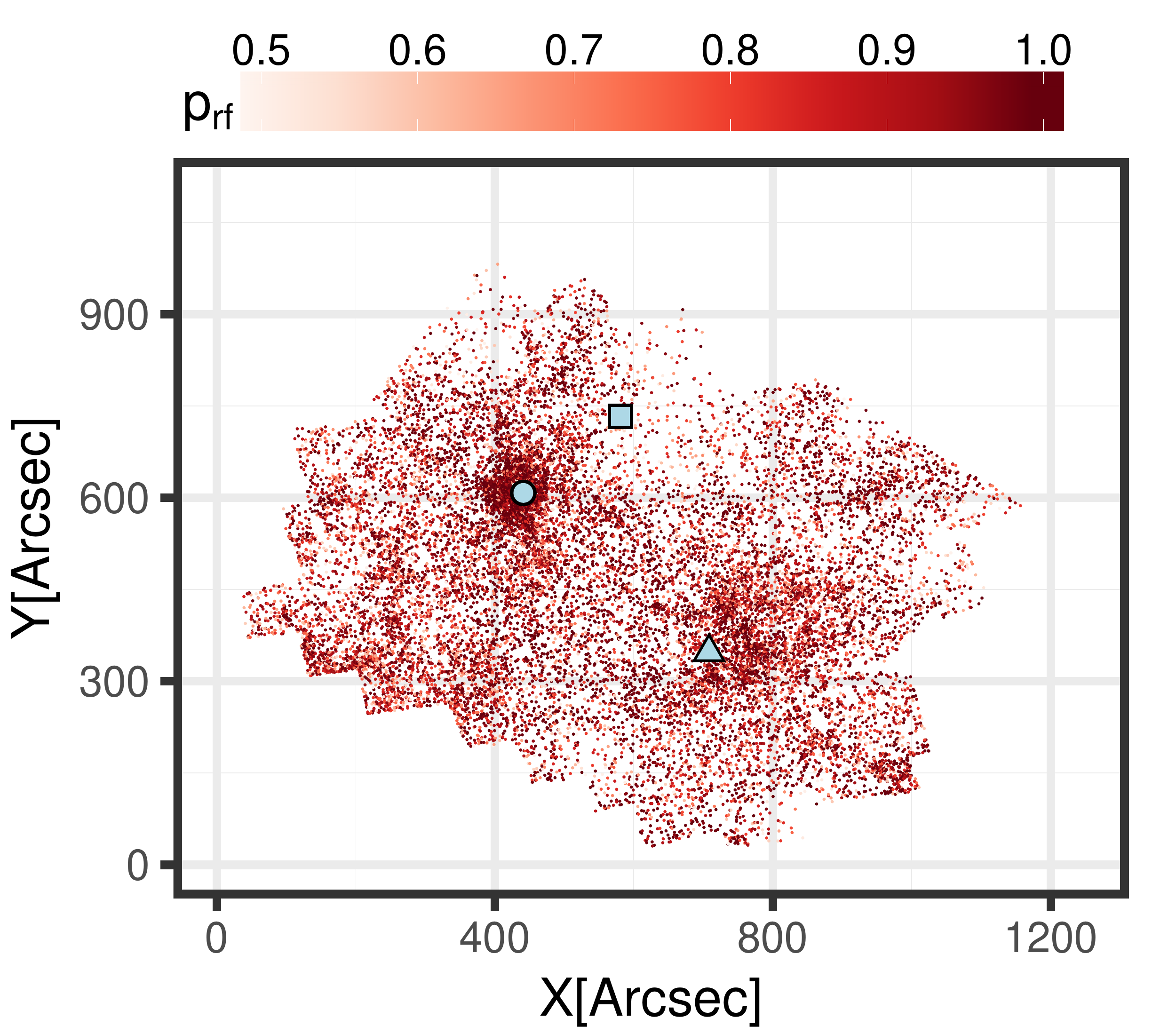}
		\includegraphics[width =\columnwidth]{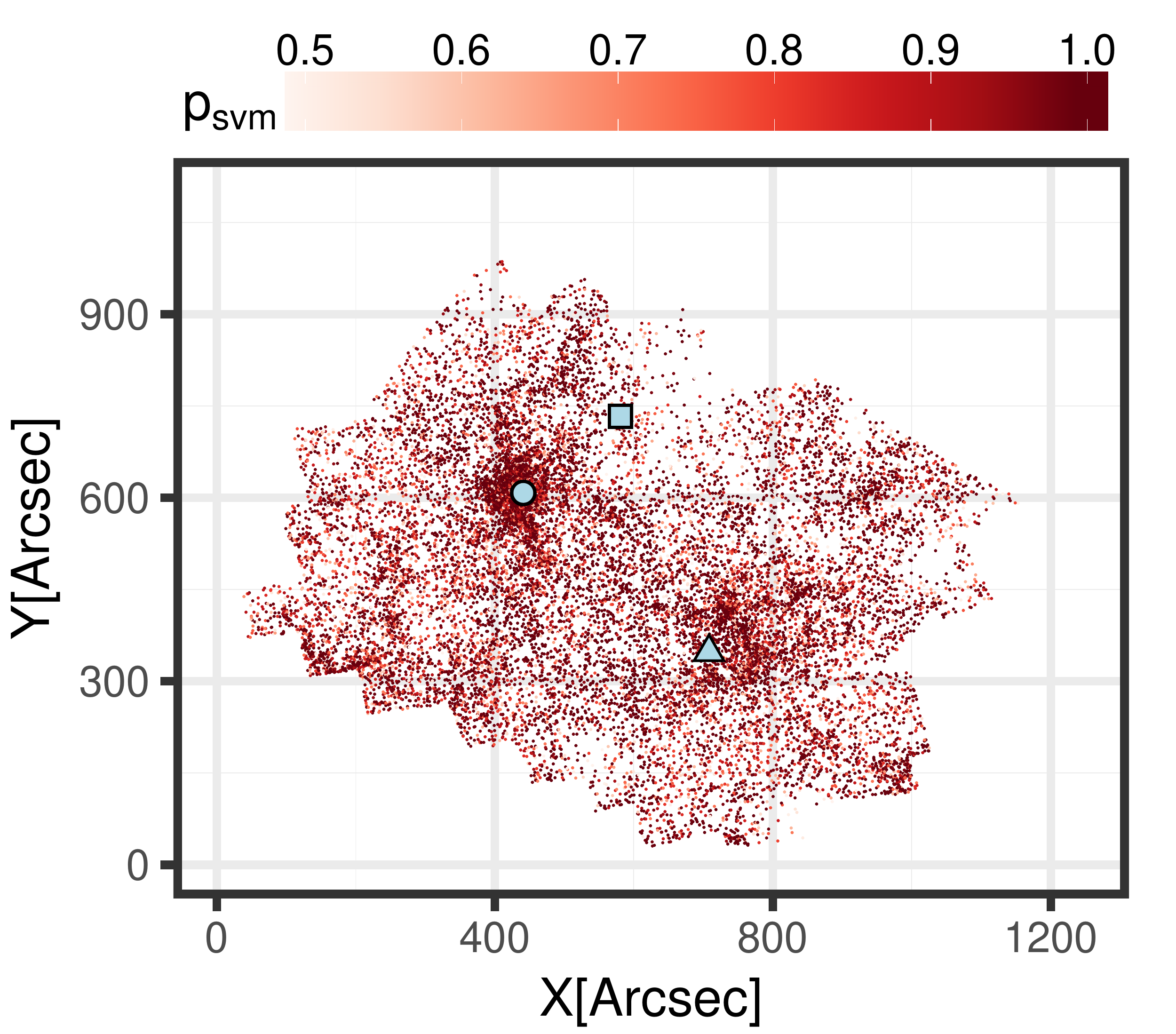}
		\caption{Spatial distribution of PMS candidates, i.e. all stars with $\mathrm{p_{rf/svm}}\geq 0.5$, coloured according to their probability of being PMS stars, as predicted by the RF algorithm ({\em left}), and the SVM algorithm ({\em right}), trained on their measurements in the F555W and F775W bands and their extinction $A_v$. For guidance, the positions of R136, Hodge 301 
		and NGC2060 
		are indicated by a large circle, square and triangle, respectively.}
	\label{fig:rf_svm_VR_AV2}
	\end{figure*}
		
With these settings the RF achieves an excellent accuracy and AUC of 95.6\% and 0.990 respectively for the primary features set (set no. 3), providing the best performance of the RF across all the feature combinations we have tested. In our implementation of RF, the predictions of the constructed models return, apart from the class of each star (PMS, non-PMS) a probability that this star is indeed a PMS star, $\mathrm{p_{rf}}$, determined from the proportion of votes of the trees in the ensemble. Classifying the whole available HTTP data on the features set no. 3 (403,018 stars, c.f. table \ref{tab:train_parameters}), the model predicts 20,996 stars, with probabilities $\mathrm{p_{rf}} \geq 0.5$. Figure \ref{fig:rf_svm_VR_AV} (left) shows the optical CMD of the PMS candidates, coloured according to their probabilities $\mathrm{p_{rf}}$. This CMD demonstrates that the RF drastically improves the zigzag pattern of the decision tree, albeit a smoothed such pattern can be still observed, apparently the outcome of the underlying tree nature of the method. The classification result further exhibits a mixture of PMS and Non-PMS classifications in the region where old, field MS turn-off and subgiant stars could potentially overlap with turn-on stars, clearly indicating that the algorithm distinguishes these two types of stars, in contrast to the decision tree, which in our tests tended to classify all stars in the region as PMS. The RF algorithm is also more successful than the decision trees in avoiding classification of {RC} as PMS stars, although there might still be a few misclassifications of the faintest {RC} stars. Figure\,\ref{fig:rf_svm_VR_AV2} (left) depicts the spatial distribution of the PMS candidates across the Tarantula Nebula, coloured according to their probabilities $\mathrm{p_{rf}}$. This map demonstrates, in agreement with our expectation, a large abundance of PMS stars in the regions of NGC\,2070 and NGC\,2060, as well as in less prominent compact stellar clusters and in features that appear almost filamentary. In this map we also mark for guidance the positions of R136 \citep{2016ApJS..222...11S}, Hodge 301 \citep{Glatt2010} and NGC2060 \citep{Cutri2003}. 

%
%
%

The classification of the RF models trained on the other three considered features sets (sets no. 2, 4 and 5 in Table\,\ref{tab:train_parameters}) appear to be overall similar to that of set no. 3, both in terms of performance, returning only 1.06\,percent to 2.82\,percent less accuracy, and in terms of spatial distribution of the identified PMS stars. The spatial distribution of the RF model based on features set no. 5 was somewhat different than the rest, due to the drastically reduced amount of data to be classified and the smaller available spatial coverage of stars found in both optical and infrared bands. The PMS stellar samples identified with the models of sets no. 2 to 4 are essentially identical, with $\sim$\,20,000 common identifications. Interestingly, the RF model trained on set no. 3 identifies fewer candidate PMS stars than those predicted by the model of set no. 2 in regions of lower extinction. This indicates that the algorithm intrinsically assumes a spatial correlation between PMS stars and larger extinction, possibly due to the region it was trained on. The marginal differences between the models for the features sets no. 2 and 4 indicate that the RF method was not sensitive to the enlargement of the feature space with the addition of the photometric errors, possibly because the photometric errors add small decision power to the models.

\subsection{Classification with Support Vector Machines}\label{sec:SVM_results}

	The third classification algorithm we experimented with is the SVM (see Appendix\,\ref{app:SVM} for a description). In our experiments the general purpose {\sl Gaussian Radial Basis Kernel} was chosen as the SVM kernel:
	\begin{equation}
		K(x,x') = \exp(-\sigma ||x-x'||^2),
	\end{equation}
The SVM model parameters, i.e. the cost $C$ and the kernel width $\sigma$, are determined again via a 10-fold cross-validation, choosing the best model according to its AUC. The influence of the chosen $\mathrm{p_{em}}$ threshold for the best PMS candidates in the training sample was evaluated in the same way as for the RF. As in the case of the RF, we found that the performance of the SVM increases with a higher threshold, but only slightly, indicating that the performance of the modeling is not sensitive to this threshold. For the features set no. 3, the best AUC of 0.995 was achieved when training on PMS stars with $\mathrm{p_{em}} \geq 0.9$, being only slightly larger though than the AUC values derived with other $\mathrm{p_{em}}$ thresholds, varying between 0.7 and 0.9. With the same reasoning as for the final RF classification we use the condition of $\mathrm{p_{em}} \geq 0.85$ for PMS candidacy here, also allowing a direct comparison between the results of the two methods. The corresponding accuracy of this SVM model is exceptionally good, equal to 97.29\%. The classification of the available HTTP data returns 19487 PMS candidates with outcome probabilities $\mathrm{p_{svm}} \geq 0.5$.

Figure\,\ref{fig:rf_svm_VR_AV} (right) shows the optical CMD of the PMS candidates, coloured according to their probabilities $\mathrm{p_{svm}}$. This CMD demonstrates that the SVM overall constructs a much smoother decision boundary than the RF and avoids successfully the misclassification of {RC} stars and objects with poor photometry as PMS stars. We also found that the SVM performs equally well as the RF in distinguishing possible turn-on and subgiant stars in their overlap CMD-region. There is, however, a small isolated patch of SVM-classified PMS candidates at $(\mathrm{F555W} \sim 19.5, \mathrm{F555W-F775W} \sim 0.5)$, which could possibly be young stars still on the turn-on, but they fit mostly to the UMS. This patch is likely the result of the SVM being prone to over-fit outliers, when there are very few records to train on, as is the case for this CMD region in our training dataset (e.g., Figure\,\ref{fig:training_set}). Figure\,\ref{fig:rf_svm_VR_AV2} (right) shows the spatial distribution of the PMS candidates, as classified by the SVM. This map displays the same prominent spatial features as the PMS stars found with the RF, i.e., high concentrations of stars in the regions of NGC\,2070 and NGC\,2060, and indications of substructure in the space between the clusters.

We found somewhat larger variations in the SVM than in the RF across the various investigated features sets. The SVM model constructed from the features set no. 2 results in a very smooth two-dimensional decision boundary. As a consequence for this features set the algorithm does not distinguish possible turn-on and subgiant stars. Instead it classifies the entire CMD region up to the brightest limit of our training dataset as PMS stars, albeit at a low overall probability. This certainly introduces a number of misclassifications of older subgiant stars and explains the larger number of identified PMS stars in comparison to the RF. Enlarging the features space with the inclusion of photometric errors in the F555W and F775W bands (i.e., features set no. 4) delivers a comparable performance and number of PMS candidates in comparison with features sets no. 2. Similar to the RF the overall classification outcome does not appear to be overly sensitive to the inclusion of photometric errors, given the $\sim 19,000$ commonly identified PMS candidates, although the decision boundaries in the CMD appear slightly broadened. In an earlier test with a training set that did not account for highly extincted 
giant and subgiant stars, we found that the SVM trained on this features set misclassified a large amount of very faint blue LMS stars with larger photometric errors as PMS stars. Even though this does not happen to the same degree (there are still a few such misclassifcations of faint LMS stars) with the more refined training set, described in this paper, we came to the conclusion, keeping those earlier tests in mind, that including the photometric errors is potentially compromising rather than helping the successful classification of PMS stars.

The classification result for features set no. 5, i.e., including the infrared measurements, is similar to the result for set no. 2, exhibiting a relatively smooth decision boundary in the CMD. There is a slight improvement, though, over features set no. 2 in the distinction between 
giant/subgiant and turn-on stars, indicating that infrared bands hold information that may be very useful in distinguishing possible MS turn-off and giant/subgiant stars 
from PMS turn-on stars, using {machine-learning} techniques. Nevertheless, we choose features set no. 3 as the primary set to base our final classification on due to the far more complete spatial coverage of the Tarantula Nebula over the other sets.

	\begin{figure}
		\centering
		\includegraphics[width = \linewidth]{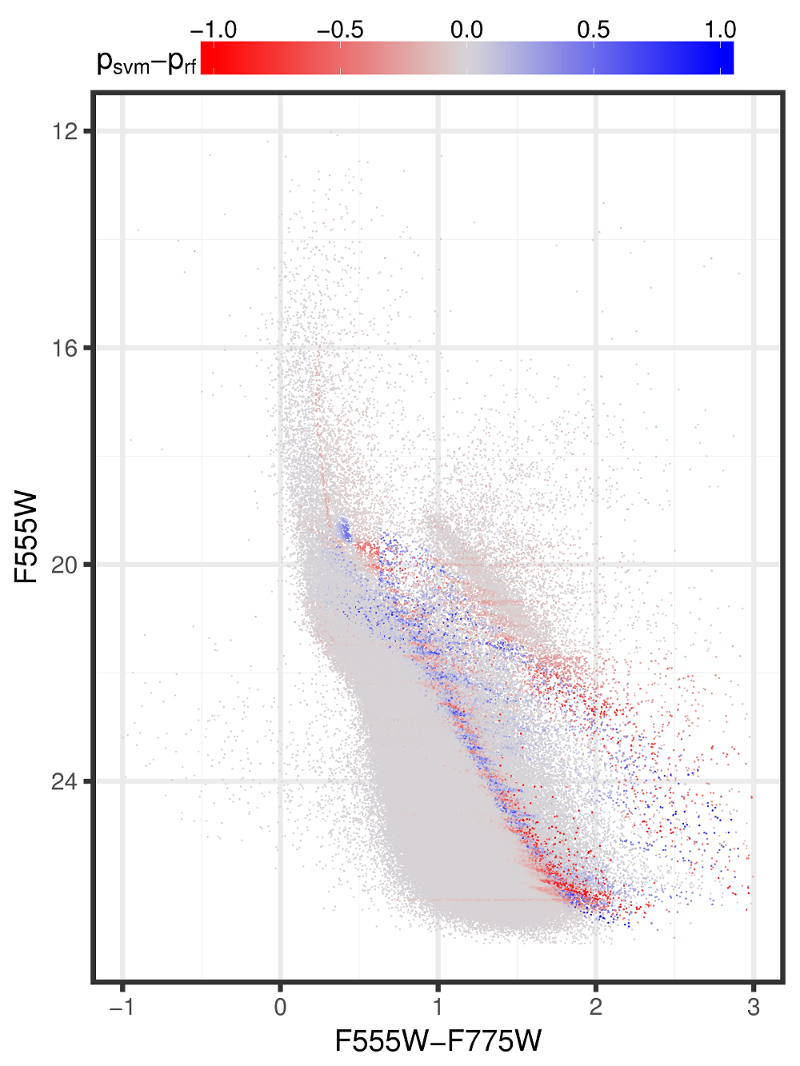}
		\caption{ Optical CMD of the HTTP data, where each star is coloured according to the difference $\mathrm{p_{svm}} - \mathrm{p_{rf}}$ of the predicted probability to belong to the PMS from SVM and RF, respectively.}
		\label{fig:COMB_VR_AV}
	\end{figure}
	
	\begin{figure}
		\centering
		\includegraphics[width = \linewidth]{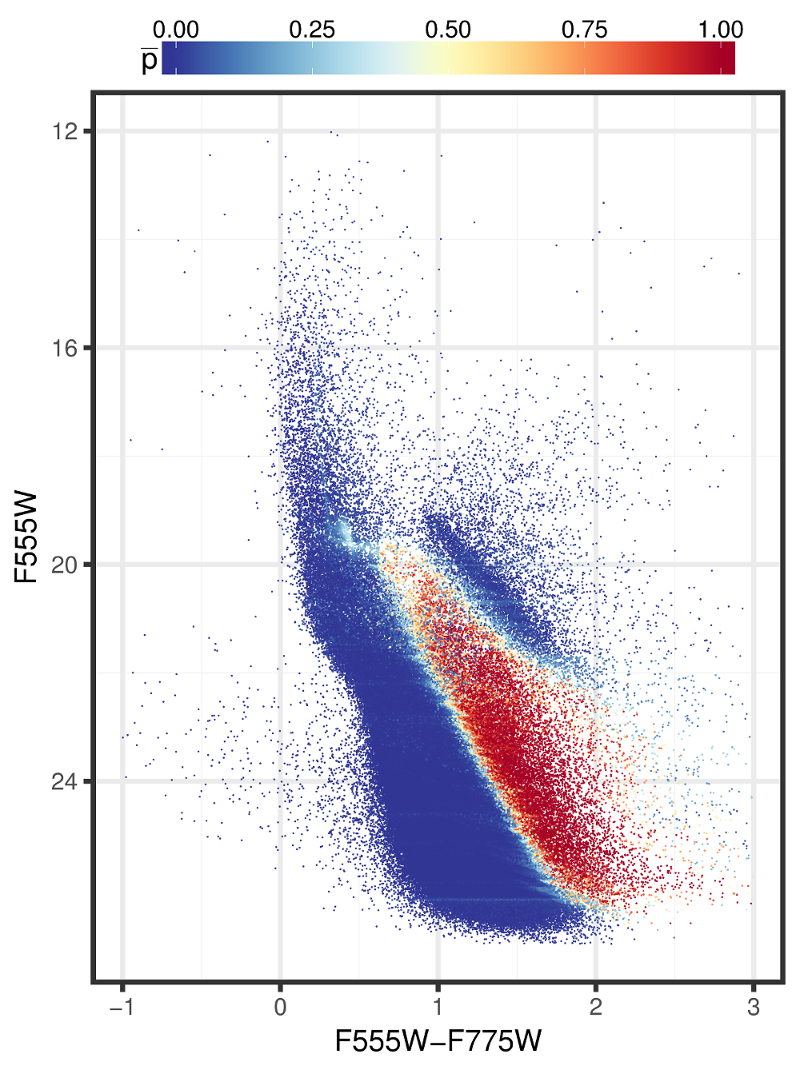}
		\caption{Optical CMD of the HTTP data, where each star is coloured according to the mean $\mathrm{\bar{p}}$ of the predicted PMS candidateship probability of the SVM and RF.}
		\label{fig:COMB_VR_AV2}
	\end{figure}

\subsection{Comparison and Combination}
	\label{sec:ComparisonCombination}

Both the RF and SVM methods scored an excellent performance on the held-out test dataset for our primary training features set (set no. 3, F555W, F775W, $A_V$), with the SVM providing the best modeling across all our experiments in terms of accuracy. Despite their performance, we identified individual shortcomings in the classification results of both methods, as described in the previous two sections. The RF method inherited from its decision trees the trend to produce a zig-zag pattern (faint nevertheless) at the LMS-PMS CMD border, while the SVM includes in its sample of best PMS candidates the few members of an isolated patch at the faint part of the UMS. In order to overcome these shortcomings and identify the most accurate PMS stellar census across the entire Tarantula nebula in terms of eliminated misclassifications, we combine the results of both methods. 

A comparison between the results of the RF and the SVM methods is shown in Figure\,\ref{fig:COMB_VR_AV}, where each star in the optical CMD is colour-coded according to the difference of the probabilities derived from each method that this is indeed a PMS star, $\delta \mathrm{p} = \mathrm{p_{svm}} - \mathrm{p_{rf}}$. An instance where the star is classified by the SVM as an excellent PMS candidate but not identified at all by the RF would have positive $\delta p$ with a value close to $1$, while in the opposite case this difference would be close to $-1$. On the other hand, instances in which both methods agree on the predicted probability of the stars being PMS stars would have differences $\delta p$ close to zero. These are the records with the best prediction about their nature as PMS stars. Both methods provided the same classification and probabilities for the vast majority of the identified PMS stars, 17728 stars in total. The optical CMD with the stars colour-coded according to their $\delta p$ value is shown in Figure\,\ref{fig:COMB_VR_AV}. The red points in this CMD indicate the stars identified by the RF as PMS candidates but not by the SVM, while the blue points indicate stars classified by the SVM but not by RF. Discrepancies of the two methods are located around their respective decision boundaries between the PMS and Non-PMS classes, but they are very few in comparison to the total classified PMS stars. 
This CMD indicates that the classification of the two methods agrees well for the majority of the classified objects. 

	\begin{figure*}
		\centering
		\includegraphics[width = \textwidth]{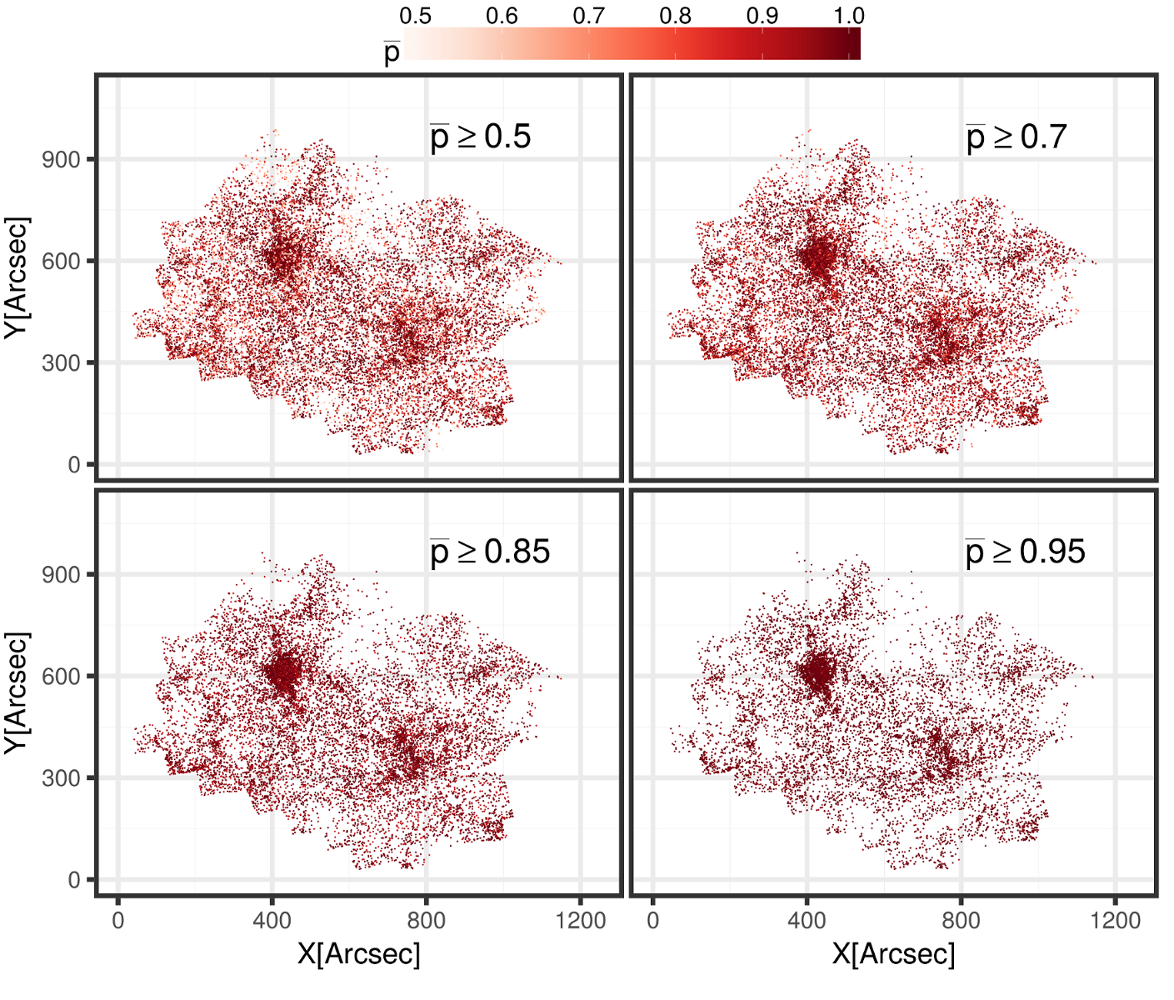}
		\caption{Series of spatial distribution plots of PMS candidates, determined by different thresholds of the mean predicted probability $\mathrm{\bar{p}}$ of SVM and RF. The respective threshold is indicated in each plot, where all stars are coloured according to $\mathrm{\bar{p}}$.} 
		\label{fig:COMB_VR_AV_spatial}
	\end{figure*}

In general the SVM classification appears to treat the faintest part of the RC, falsely identified by the RF as PMS stars, better than the RF, while the RF classifies the lower UMS patch of stars, classified as PMS by the SVM, as negative instances. Each method, thus, ``corrects'' for the shortcomings of the other. As a consequence, in order to achieve the most robust solution for the PMS stellar content of the Tarantula Nebula {(in terms of producing the purest possible sample)} we combine the individual classifications of each star by averaging the predicted PMS candidateship probabilities derived from both methods. This approach effectively compensates for the individual shortcomings of SVM and RF by assigning low mean PMS candidateship probabilities $\mathrm{\bar{p}}$ to the likely misclassified objects, such as the UMS stars for the SVM and the {RC} stars for the RF. The CMD of Figure\,\ref{fig:COMB_VR_AV2} shows the improved final classification provided by this averaging. We construct the final catalogue of PMS candidate stars across the entire Tarantula Nebula, by providing the original HTTP photometric data of the stars, their predicted PMS candidateship probabilities, derived from both the SVM and RF methods, the difference of the probabilities $\delta p$, and the mean PMS candidateship probability $\mathrm{\bar{p}}$. In this catalogue we include all stars, identified by at least one of the two methods as PMS candidates, delivering in total 22755 possible PMS stars for the entire Tarantula Nebula. Imposing thresholds on the mean probability to distinguish the most probable PMS stellar population of the complex, this catalogue entails 19831 candidates with $\bar{p}\geq 0.5$, 16696 with $\bar{p} \geq0.7$, 13526 for $\bar{p} \geq 0.85$ and 9636 with $\bar{p} \geq 0.95$. The corresponding spatial distributions are shown in Figure\,\ref{fig:COMB_VR_AV_spatial}. This series of thresholding the mean predicted PMS candidateship probability exhibits that the most probable PMS candidates, identified with our classification approach, also mark the most spatially confined structures across the entire Tarantula Nebula. 

\begin{figure}
	\centering
	\includegraphics[width = \columnwidth]{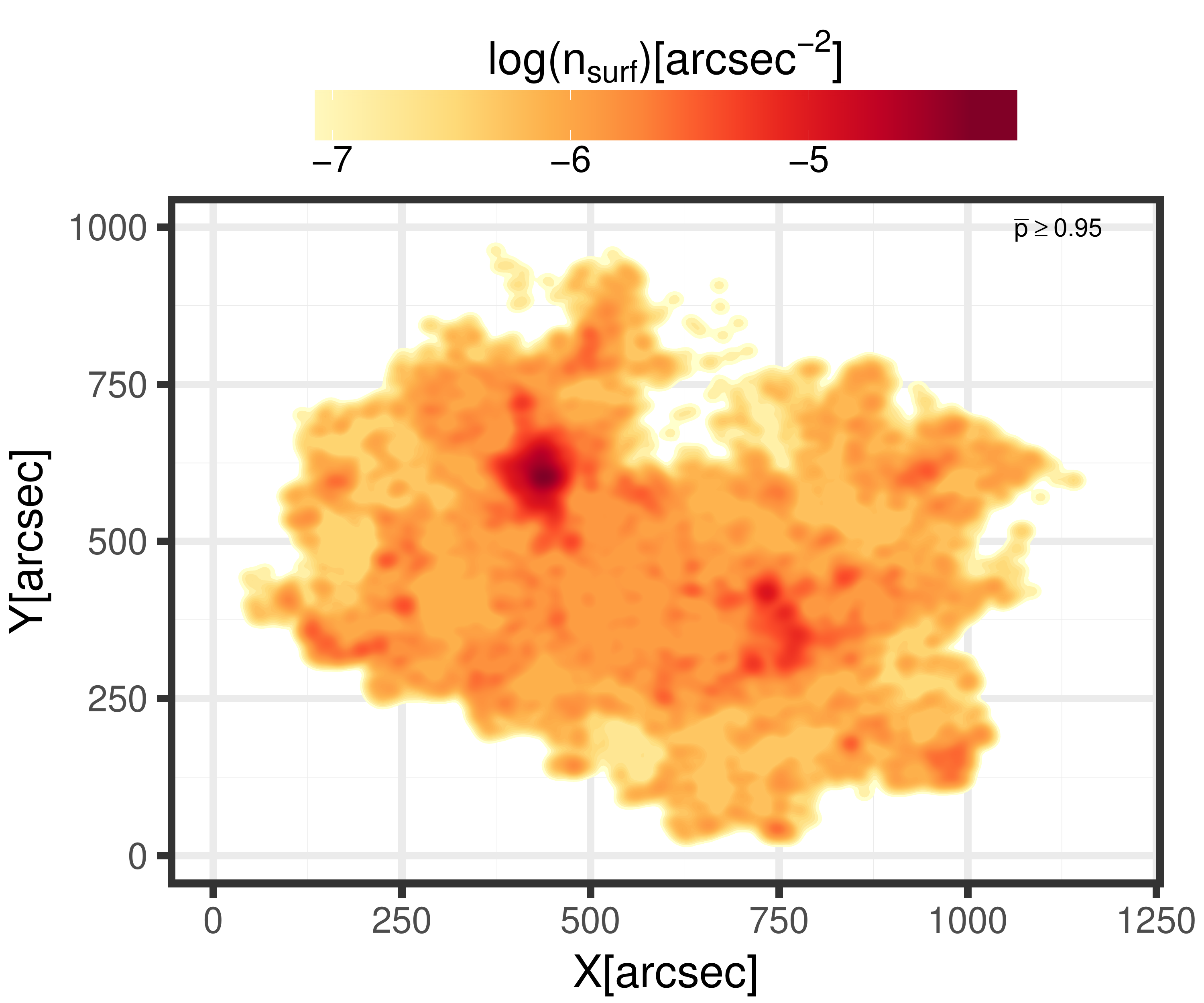}
	\caption{Surface density plot of the PMS candidate stars with $\bar{p} \geq 0.95$, i.e. the population in the bottom right panel in Figure\,\ref{fig:COMB_VR_AV_spatial}.} 
	\label{fig:PMS95_dens}
\end{figure}

{Figure\,\ref{fig:PMS95_dens} shows the surface density map of the most probable PMS candidates ($\bar{p} \geq 0.95$). This map is not qualitatively different from those constructed for stars with different probability limits, so that the general clustering of PMS stars does not appear 
to be very sensitive to the threshold on $\bar{p}$.}
The remarkable coincidence of the maps of Figure\,\ref{fig:COMB_VR_AV_spatial} (independently of the considered candidateship probability threshold) {and the density map of Figure\,\ref{fig:PMS95_dens}}
with the spatial distribution of the UMS stars (shown in Figure\,\ref{fig:http_extinction}) indicates that PMS stars are preferably clustered in regions of high concentrations of UMS stars. This is in agreement with results discussed in the literature concerning the clustering of PMS stars around massive young stars \citep{2015ApJ...811...76C, StephensGouliermis2017}, 
and provides an additional confirmation of the validity of our PMS identifications.

\section{Summary and future prospects}\label{s:summary}

In this paper we present our analysis with the employment of
machine-learning classification techniques for the identification of PMS
stars across the entire star-forming complex of the Tarantula Nebula in
the LMC. For this classification we extracted a robust training subset
from the observational data of the Hubble Tarantula Treasury Project,
which provides deep panchromatic Hubble imaging of the whole nebula, in
order to teach Naive Bayes classifier, decision tree, random forest (RF) and
support vector machine (SVM) classifiers to categorise the stars of the
entire HTTP catalogue into the classes ``PMS" and ``Non-PMS". To
construct this training dataset we selected a high-surface-density
region within the Tarantula Nebula, corresponding to the R136 starburst cluster
at the heart of NGC\,2070, based on the assumption that PMS
stars are more likely to be located in the most clustered regions of the
nebula. To account for differential extinction across the nebula, we
used UMS stars as extinction probes and derived extinction measures for
each individual star in the HTTP catalogue using a distance weighted
average of the extinction of the 20 nearest UMS neighbours.

After attributing extinction to the NGC\,2070 subset, improving
upon the approach of \cite{2012ApJ...748...64G}, we developed a robust
method to distinguish the cluster PMS stars from the field LMS stars in
the training dataset. This method is based on fitting bimodal Gaussian
mixture models to the distance of all stars from the apparent gap on
the CMD between these two populations via the maximum likelihood
expectation maximisation algorithm. From these mixture models we derived
a probability $\mathrm{p_{em}}$ for each star in the training set to
be PMS. We finalised the training set by adding further examples of
evolved populations, such as {RC} stars, and subgiant stars in low
and high extinction areas of the field of the Tarantula Nebula, as
``Non-PMS". We assigned the labels ``PMS" and ``Non-PMS" to the stars
depending on various selected thresholds of $\mathrm{p_{em}}$, and after
training the classification algorithms with this training set, we
evaluated their performance for different variables (features)
combinations. The findings of these experiments can be summarised as
follows:

\begin{enumerate}[leftmargin = *]
	
	\item During our preliminary tests neither the Naive Bayes nor the
	Decision Tree method were able to achieve adequate performance,
	providing accuracies not higher than $\sim$\,60 and $\sim$\,84\,percent,
	respectively. Consequently, both methods exhibited significant issues in
	classifying the entire HTTP catalogue, although the Decision Tree still
	provided valuable insights on the importance of specific features. It
	strongly suggests that near infrared measurements (e.g. in the F160W
	filter) are very useful to a classification approach for PMS stars.
	
	\item The best combination of features, in terms of stellar numbers,
	spatial coverage and algorithm performance, included the photometric
	measurements in the F555W and F775W filters in combination with the
	extinction values $A_V$, which we derived for each star using the UMS
	stars as extinction probes. Including the infrared wavebands resulted in
	a comparable performance of the classification algorithms, but,
	since the features set of the F555W, F775W, F110W and F160W filters
	suffered from the smaller spatial coverage of the infrared observations,
	it was not suited for finding the most complete PMS stellar census of
	the Tarantula Nebula. Extending the optical bands feature space by
	adding the photometric errors did not seem to provide any useful
	information for the RF, and it even compromised the classification
	ability of SVM.
	
	\item The best performance of both the RF and SVM methods was achieved
	when stars with $\mathrm{p_{em}} \geq 0.9$ were selected as the best PMS
	examples in the training set. However, the best trade-off between
	algorithm performance and balance between the numbers of positive and
	negative examples in the training set was achieved with the use of a
	threshold $\mathrm{p_{em}} = 0.85$ for labelling the training stars.
	
\end{enumerate}

Both the RF and SVM methods performed excellently on our primary
features set (F555W, F775W, $A_V$), achieving accuracies of $95.6$ and
$97.3$\,percent, and ROC AUCs of 0.990 and 0.994 respectively. The
classification outcomes of both methods on the entire HTTP data also met
the required expectations, except for minor shortcomings. Specifically,
the RF algorithm misclassified a few faint RC stars as PMS stars, and the
SVM did so for a small patch of likely UMS stars. A direct comparison
between the outcomes of the methods showed that they compensate each
other's shortcomings. As a consequence, the most robust classification
is achieved by combining the predicted PMS candidateship probability of
each star derived from both methods.

The combination of the results of both RF and SVM methods resulted in
22,755 stars, identified as PMS by at least one of the methods.
Among these sources, 19,831 stars have an average predicted PMS probability of
$\mathrm{\bar{p}} \geq 0.5$ and 9,636 have $\mathrm{\bar{p}} \geq 0.95$.
There is a number of studies that can be performed with the use of this catalog of the most probable low-mass PMS stars in the Tarantula nebula region. We identify three science cases, each deserving its own independent investigation.

1) PMS stars with emission lines due to accretion. The use of the H$\alpha$ filter permits the identification of several types of young stellar sources, such as massive main-sequence and supergiant emission-line stars (Oe, Be and B[e] stars), as well as PMS stars with strong emission lines, such as classical T\,Tauri stars \citep{AppenzellerMundt1989, Bertout1989}. The excess in these stars implies that the photospheric lines are not as deep as those of main sequence stars of the same spectral type \citep[e.g.,][]{Hartigan1995, Gullbring1998}.. With our classification we find that $\sim$\,60\,per\,cent of the PMS candidates are also detected in the H$\alpha$ band, while only $\sim$\,30\,per\,cent of the stars classified as non-PMS are found in this band. Our preliminary study of these stars indicates that most of the PMS stars do not show strong accretion,which is expected for Weak Line T\,Tauri stars that show very weak, if any emission lines \citep[e.g.,][]{Montmerle1993}. A more detailed analysis of the H$\alpha$ excess of all these stellar types in order to determine their mass accretion rates and investigate their variations across the nebula will be the topic of a separate study.

2) Physical characteristics of PMS stars. The use of the HTTP multi-band photometry will allow us to construct the Spectral Energy Distributions (SEDs) of our low-mass PMS candidates in order to establish their masses and ages through dedicated SED-fitting techniques. This study for the determination of physical parameters for the identified PMS stars is currently under development with the use of the {\sl Bayesian Extinction and Stellar Tool} \citep[BEAST,][]{Gordon2016BEAST}, appropriately tailored to our photometric data (Ksoll et al., in prep). These results will further allow the characterisation of the stellar Initial Mass Function (IMF) across its whole dynamic range and its variability across the HTTP FoV by combining mass estimates of our PMS stars with those of the UMS populations \citep[][]{Evans2011FLAMES}. We will be further able to investigate the propagation of
star formation in time with the investigation of spatial distributions of ages of the PMS stars.

3) The clustering pattern of star formation. The spatial distribution of the classified PMS stars shows well defined sub-structures within the regions of the clusters NGC\,2070 and NGC\,2060, as well as compact and loose -- occasionally filamentary -- clusterings
across the whole observed FoV. An elaborate investigation of the clustering behaviour of the PMS stars in the Tarantula nebula, based on the results of this study, and the quantification of the spatial cross-correlation between PMS and UMS stars is currently being performed in a separate forthcoming study (Gouliermis et al. in prep). 


\section*{Acknowledgments}
V.F.K acknowledges support by the Heidelberg Graduate School of
Mathematical and Computational Methods for the Sciences (HGS MathComp),
founded by DFG grant GSC\,220 in the German Universities Excellence
Initiative. D.A.G. acknowledges support by the German Research
Foundation (DFG) within program {\sl Clustered Star Formation on various
	scales} under grant GO\,1659/3-2, and by the German Aerospace Center
(DLR) within project 50OR1801 {\sl Mapping Young Stars in Space \&
	Time}. This research has made use of the NASA/IPAC Extragalactic
Database (NED), which is operated by the Jet Propulsion Laboratory,
California Institute of Technology, under contract with the National
Aeronautics and Space Administration.
This research has made use of the SIMBAD database, operated at CDS,
Strasbourg, France.
Based on observations made with the NASA/ESA {\sl Hubble Space
	Telescope}, obtained from the data archive at the
Space Telescope Science Institute (STScI). STScI is operated by the
Association of Universities for Research in Astronomy, Inc.\ under
NASA contract NAS 5-26555. These observations are associated with
program GO-13364. Support for Program 13364 was provided by NASA
through grants from STScI. This research made use of the
TOPCAT\footnote{TOPCAT is available at the permalink
	\href{http://www.starlink.ac.uk/topcat/}{http://www.starlink.ac.uk/topcat/
	}} application \citep{topcat2005}, the R environment for statistical
	computing
	and graphics\footnote{The R Project for Statistical Computing:
		\href{https://www.R-project.org/}{https://www.R-project.org/ }}
	\citep{Rlanguage},
	and NASA Astrophysics Data System (ADS) bibliographic
	services\footnote{Accessible at
		\href{http://adswww.harvard.edu/}{http://adswww.harvard.edu/}
		and \href{http://cdsads.u-strasbg.fr/}{http://cdsads.u-strasbg.fr/}}.



\begin{thebibliography}{}
	\makeatletter
	\relax
	\def\mn@urlcharsother{\let\do\@makeother \do\$\do\&\do\#\do\^\do\_\do\%\do\~}
	\def\mn@doi{\begingroup\mn@urlcharsother \@ifnextchar [ {\mn@doi@}
		{\mn@doi@[]}}
	\def\mn@doi@[#1]#2{\def\@tempa{#1}\ifx\@tempa\@empty \href
		{http://dx.doi.org/#2} {doi:#2}\else \href {http://dx.doi.org/#2} {#1}\fi
		\endgroup}
	\def\mn@eprint#1#2{\mn@eprint@#1:#2::\@nil}
	\def\mn@eprint@arXiv#1{\href {http://arxiv.org/abs/#1} {{\tt arXiv:#1}}}
	\def\mn@eprint@dblp#1{\href {http://dblp.uni-trier.de/rec/bibtex/#1.xml}
		{dblp:#1}}
	\def\mn@eprint@#1:#2:#3:#4\@nil{\def\@tempa {#1}\def\@tempb {#2}\def\@tempc
		{#3}\ifx \@tempc \@empty \let \@tempc \@tempb \let \@tempb \@tempa \fi \ifx
		\@tempb \@empty \def\@tempb {arXiv}\fi \@ifundefined
		{mn@eprint@\@tempb}{\@tempb:\@tempc}{\expandafter \expandafter \csname
			mn@eprint@\@tempb\endcsname \expandafter{\@tempc}}}
	
	\bibitem[\protect\citeauthoryear{{Appenzeller} \& {Mundt}}{{Appenzeller} \&
		{Mundt}}{1989}]{AppenzellerMundt1989}
	{Appenzeller} I.,  {Mundt} R.,  1989, \mn@doi [Astronomy and Astrophysics
	Review] {10.1007/BF00873081}, \href
	{http://cdsads.u-strasbg.fr/abs/1989A%26ARv...1..291A} {1, 291}
		
		\bibitem[\protect\citeauthoryear{{Beaumont}, {Goodman}, {Kendrew}, {Williams}
			\& {Simpson}}{{Beaumont} et~al.}{2014}]{Beaumont2014}
		{Beaumont} C.~N.,  {Goodman} A.~A.,  {Kendrew} S.,  {Williams} J.~P.,
		{Simpson} R.,  2014, \mn@doi [\apjs] {10.1088/0067-0049/214/1/3}, \href
		{http://cdsads.u-strasbg.fr/abs/2014ApJS..214....3B} {214, 3}
		
		\bibitem[\protect\citeauthoryear{Benaglia, Chauveau, Hunter  \& Young}{Benaglia
			et~al.}{2009}]{UBHD-}
		Benaglia T.,  Chauveau D.,  Hunter D.,   Young D.,  2009, J. Stat. Softw., 32,
		1
		
		\bibitem[\protect\citeauthoryear{{Bertout}}{{Bertout}}{1989}]{Bertout1989}
		{Bertout} C.,  1989, \mn@doi [Annual review of astronomy and astrophysics]
		{10.1146/annurev.aa.27.090189.002031}, \href
		{http://cdsads.u-strasbg.fr/abs/1989ARA%26A..27..351B} {27, 351}
			
			\bibitem[\protect\citeauthoryear{{Bodenheimer}}{{Bodenheimer}}{2011}]{2011psf..book.....B}
			{Bodenheimer} P.~H.,  2011, {Principles of Star Formation},
			\mn@doi{10.1007/978-3-642-15063-0.
			}
			
			\bibitem[\protect\citeauthoryear{{Brandner}, {Grebel}, {Barb{\'a}}, {Walborn}
				\& {Moneti}}{{Brandner} et~al.}{2001}]{Brandner2001}
			{Brandner} W.,  {Grebel} E.~K.,  {Barb{\'a}} R.~H.,  {Walborn} N.~R.,
			{Moneti} A.,  2001, \mn@doi [The Astronomical Journal] {10.1086/322065},
			\href {http://adsabs.harvard.edu/abs/2001AJ....122..858B} {122, 858}
			
			\bibitem[\protect\citeauthoryear{Breiman}{Breiman}{2001}]{Breiman01RandomForests}
			Breiman L.,  2001, Machine Learning, 45, 5
			
			\bibitem[\protect\citeauthoryear{{Breiman}, {Friedman}, {Olshen}  \&
				{Stone}}{{Breiman} et~al.}{1984}]{Breiman84ClassificationTrees}
			{Breiman} L.,  {Friedman} J.~H.,  {Olshen} R.~A.,   {Stone} C.~J.,  1984,
			Classification and Regression Trees.
			Statistics/Probability Series, Wadsworth Publishing Company, Belmont,
			California, U.S.A.
			
			\bibitem[\protect\citeauthoryear{{Bressan}, {Marigo}, {Girardi}, {Salasnich},
				{Dal Cero}, {Rubele}  \& {Nanni}}{{Bressan} et~al.}{2012}]{Bressan2012}
			{Bressan} A.,  {Marigo} P.,  {Girardi} L.,  {Salasnich} B.,  {Dal Cero} C.,
			{Rubele} S.,   {Nanni} A.,  2012, \mn@doi [\mnras]
			{10.1111/j.1365-2966.2012.21948.x}, \href
			{http://adsabs.harvard.edu/abs/2012MNRAS.427..127B} {427, 127}
			
			\bibitem[\protect\citeauthoryear{{Cignoni}, {Tosi}, {Sabbi}, {Nota},
				{Degl'Innocenti}, {Prada Moroni}  \& {Gallagher}}{{Cignoni}
				et~al.}{2010}]{Cignoni2010}
			{Cignoni} M.,  {Tosi} M.,  {Sabbi} E.,  {Nota} A.,  {Degl'Innocenti} S.,
			{Prada Moroni} P.~G.,   {Gallagher} J.~S.,  2010, \mn@doi [\apjl]
			{10.1088/2041-8205/712/1/L63}, \href
			{http://cdsads.u-strasbg.fr/abs/2010ApJ...712L..63C} {712, L63}
			
			\bibitem[\protect\citeauthoryear{{Cignoni} et~al.,}{{Cignoni}
				et~al.}{2015}]{2015ApJ...811...76C}
			{Cignoni} M.,  et~al., 2015, \mn@doi [\apj] {10.1088/0004-637X/811/2/76}, \href
			{http://adsabs.harvard.edu/abs/2015ApJ...811...76C} {811, 76}
			
			\bibitem[\protect\citeauthoryear{{Cignoni} et~al.,}{{Cignoni}
				et~al.}{2016}]{2016ApJ...833..154C}
			{Cignoni} M.,  et~al., 2016, \mn@doi [\apj] {10.3847/1538-4357/833/2/154},
			\href {http://adsabs.harvard.edu/abs/2016ApJ...833..154C} {833, 154}
			
			\bibitem[\protect\citeauthoryear{Cortes \& Vapnik}{Cortes \&
				Vapnik}{1995}]{Cortes95SVM}
			Cortes C.,  Vapnik V.,  1995, Machine Learning, 20, 273
			
			\bibitem[\protect\citeauthoryear{{Cutri} et~al.,}{{Cutri}
				et~al.}{2003}]{Cutri2003}
			{Cutri} R.~M.,  et~al., 2003, VizieR Online Data Catalog, \href
			{http://adsabs.harvard.edu/abs/2003yCat.2246....0C} {2246}
			
			\bibitem[\protect\citeauthoryear{{Da Rio}, {Gouliermis}  \& {Henning}}{{Da Rio}
				et~al.}{2009}]{DaRio2009LH95}
			{Da Rio} N.,  {Gouliermis} D.~A.,   {Henning} T.,  2009, \mn@doi [\apj]
			{10.1088/0004-637X/696/1/528}, \href
			{http://cdsads.u-strasbg.fr/abs/2009ApJ...696..528D} {696, 528}
			
			\bibitem[\protect\citeauthoryear{{Da Rio}, {Gouliermis}  \& {Gennaro}}{{Da Rio}
				et~al.}{2010}]{DaRio2010LH95}
			{Da Rio} N.,  {Gouliermis} D.~A.,   {Gennaro} M.,  2010, \mn@doi [\apj]
			{10.1088/0004-637X/723/1/166}, \href
			{http://cdsads.u-strasbg.fr/abs/2010ApJ...723..166D} {723, 166}
			
			\bibitem[\protect\citeauthoryear{{De Marchi} et~al.,}{{De Marchi}
				et~al.}{2016}]{2016MNRAS.455.4373D}
			{De Marchi} G.,  et~al., 2016, \mn@doi [\mnras] {10.1093/mnras/stv2528}, \href
			{http://adsabs.harvard.edu/abs/2016MNRAS.455.4373D} {455, 4373}
			
			\bibitem[\protect\citeauthoryear{{De Marchi}, {Panagia}  \& {Beccari}}{{De
					Marchi} et~al.}{2017}]{DeMarchi2017}
			{De Marchi} G.,  {Panagia} N.,   {Beccari} G.,  2017, \mn@doi [The
			Astrophysical Journal] {10.3847/1538-4357/aa85e9}, \href
			{http://adsabs.harvard.edu/abs/2017ApJ...846..110D} {846, 110}
			
			\bibitem[\protect\citeauthoryear{Dempster, Laird  \& Rubin}{Dempster
				et~al.}{1977}]{Dempster77maximumlikelihood}
			Dempster A.~P.,  Laird N.~M.,   Rubin D.~B.,  1977, J. Royal Stat. Soc., Series
			B, 39, 1
			
			\bibitem[\protect\citeauthoryear{{Dieleman}, {Willett}  \& {Dambre}}{{Dieleman}
				et~al.}{2015}]{Dieleman2015}
			{Dieleman} S.,  {Willett} K.~W.,   {Dambre} J.,  2015, \mn@doi [\mnras]
			{10.1093/mnras/stv632}, \href
			{http://cdsads.u-strasbg.fr/abs/2015MNRAS.450.1441D} {450, 1441}
			
			\bibitem[\protect\citeauthoryear{{Elorrieta} et~al.,}{{Elorrieta}
				et~al.}{2016}]{Elorrieta2016}
			{Elorrieta} F.,  et~al., 2016, \mn@doi [\aap] {10.1051/0004-6361/201628700},
			\href {http://cdsads.u-strasbg.fr/abs/2016A%26A...595A..82E} {595, A82}
				
				\bibitem[\protect\citeauthoryear{{Evans} et~al.,}{{Evans}
					et~al.}{2011}]{Evans2011FLAMES}
				{Evans} C.~J.,  et~al., 2011, \mn@doi [Astronomy & Astrophysics]
				{10.1051/0004-6361/201116782}, \href
				{http://cdsads.u-strasbg.fr/abs/2011A%26A...530A.108E} {530, A108}
					
					\bibitem[\protect\citeauthoryear{{Glatt}, {Grebel}  \& {Koch}}{{Glatt}
						et~al.}{2010}]{Glatt2010}
					{Glatt} K.,  {Grebel} E.~K.,   {Koch} A.,  2010, \mn@doi [Astronomy and
					Astrophysics] {10.1051/0004-6361/201014187}, \href
					{http://adsabs.harvard.edu/abs/2010A%26A...517A..50G} {517, A50}
						
						\bibitem[\protect\citeauthoryear{{Gordon} et~al.,}{{Gordon}
							et~al.}{2016}]{Gordon2016BEAST}
						{Gordon} K.~D.,  et~al., 2016, \mn@doi [The Astrophysical Journal]
						{10.3847/0004-637X/826/2/104}, \href
						{http://cdsads.u-strasbg.fr/abs/2016ApJ...826..104G} {826, 104}
						
						\bibitem[\protect\citeauthoryear{{Gouliermis}}{{Gouliermis}}{2012}]{2012SSRv..169....1G}
						{Gouliermis} D.~A.,  2012, \mn@doi [Space Sci. Rev.]
						{10.1007/s11214-012-9868-2}, \href
						{http://adsabs.harvard.edu/abs/2012SSRv..169....1G} {169, 1}
						
						\bibitem[\protect\citeauthoryear{{Gouliermis}, {Brandner}  \&
							{Henning}}{{Gouliermis} et~al.}{2006}]{Gouliermis2006LH52}
						{Gouliermis} D.,  {Brandner} W.,   {Henning} T.,  2006, \mn@doi [\apjl]
						{10.1086/500209}, \href {http://cdsads.u-strasbg.fr/abs/2006ApJ...636L.133G}
						{636, L133}
						
						\bibitem[\protect\citeauthoryear{{Gouliermis}, {Henning}, {Brandner},
							{Dolphin}, {Rosa}  \& {Brandl}}{{Gouliermis}
							et~al.}{2007}]{Gouliermis2007LH95}
						{Gouliermis} D.~A.,  {Henning} T.,  {Brandner} W.,  {Dolphin} A.~E.,  {Rosa}
						M.,   {Brandl} B.,  2007, \mn@doi [\apjl] {10.1086/521224}, \href
						{http://cdsads.u-strasbg.fr/abs/2007ApJ...665L..27G} {665, L27}
						
						\bibitem[\protect\citeauthoryear{{Gouliermis} et~al.,}{{Gouliermis}
							et~al.}{2010}]{Gouliermis2010ESA}
						{Gouliermis} D.~A.,  et~al., 2010, \mn@doi [Astrophysics and Space Science
						Proceedings] {10.1007/978-90-481-3400-7_14}, \href
						{http://cdsads.u-strasbg.fr/abs/2010ASSP...15...71G} {15, 71}
						
						\bibitem[\protect\citeauthoryear{{Gouliermis}, {Schmeja}, {Dolphin}, {Gennaro},
							{Tognelli}  \& {Prada Moroni}}{{Gouliermis}
							et~al.}{2012}]{2012ApJ...748...64G}
						{Gouliermis} D.~A.,  {Schmeja} S.,  {Dolphin} A.~E.,  {Gennaro} M.,  {Tognelli}
						E.,   {Prada Moroni} P.~G.,  2012, \mn@doi [\apj]
						{10.1088/0004-637X/748/1/64}, \href
						{http://adsabs.harvard.edu/abs/2012ApJ...748...64G} {748, 64}
						
						\bibitem[\protect\citeauthoryear{{Gouliermis}, {Hony}  \&
							{Klessen}}{{Gouliermis} et~al.}{2014}]{Gouliermis2014}
						{Gouliermis} D.~A.,  {Hony} S.,   {Klessen} R.~S.,  2014, \mn@doi [\mnras]
						{10.1093/mnras/stu228}, \href
						{http://cdsads.u-strasbg.fr/abs/2014MNRAS.439.3775G} {439, 3775}
						
						\bibitem[\protect\citeauthoryear{{Grebel} \& {Chu}}{{Grebel} \&
							{Chu}}{2000}]{Grebel2000}
						{Grebel} E.~K.,  {Chu} Y.-H.,  2000, \mn@doi [The Astronomical Journal]
						{10.1086/301218}, \href {http://adsabs.harvard.edu/abs/2000AJ....119..787G}
						{119, 787}
						
						\bibitem[\protect\citeauthoryear{{Gullbring}, {Hartmann}, {Brice{\~n}o}  \&
							{Calvet}}{{Gullbring} et~al.}{1998}]{Gullbring1998}
						{Gullbring} E.,  {Hartmann} L.,  {Brice{\~n}o} C.,   {Calvet} N.,  1998,
						\mn@doi [The Astrophysical Journal] {10.1086/305032}, \href
						{http://cdsads.u-strasbg.fr/abs/1998ApJ...492..323G} {492, 323}
						
						\bibitem[\protect\citeauthoryear{{Hartigan}, {Edwards}  \&
							{Ghandour}}{{Hartigan} et~al.}{1995}]{Hartigan1995}
						{Hartigan} P.,  {Edwards} S.,   {Ghandour} L.,  1995, \mn@doi [Astrophysical
						Journal] {10.1086/176344}, \href
						{http://cdsads.u-strasbg.fr/abs/1995ApJ...452..736H} {452, 736}
						
						\bibitem[\protect\citeauthoryear{{Haschke}, {Grebel}  \& {Duffau}}{{Haschke}
							et~al.}{2011}]{Haschke2011}
						{Haschke} R.,  {Grebel} E.~K.,   {Duffau} S.,  2011, \mn@doi [The Astronomical
						Journal] {10.1088/0004-6256/141/5/158}, \href
						{http://adsabs.harvard.edu/abs/2011AJ....141..158H} {141, 158}
						
						\bibitem[\protect\citeauthoryear{Hastie, Tibshirani  \& Friedman}{Hastie
							et~al.}{2009}]{ElOfStatLearn}
						Hastie T.,  Tibshirani R.,   Friedman J.~H.,  2009, The elements of statistical
						learning, 2nd edn.
						Springer series in statistics, Springer, New York, NY
						
						\bibitem[\protect\citeauthoryear{{Hony} et~al.,}{{Hony}
							et~al.}{2015}]{Hony2015}
						{Hony} S.,  et~al., 2015, \mn@doi [\mnras] {10.1093/mnras/stv107}, \href
						{http://cdsads.u-strasbg.fr/abs/2015MNRAS.448.1847H} {448, 1847}
						
						\bibitem[\protect\citeauthoryear{{Hunter}, {Shaya}, {Holtzman}, {Light},
							{O'Neil}  \& {Lynds}}{{Hunter} et~al.}{1995}]{Hunter1995}
						{Hunter} D.~A.,  {Shaya} E.~J.,  {Holtzman} J.~A.,  {Light} R.~M.,  {O'Neil}
						Jr. E.~J.,   {Lynds} R.,  1995, \mn@doi [Astrophysical Journal]
						{10.1086/175950}, \href {http://adsabs.harvard.edu/abs/1995ApJ...448..179H}
						{448, 179}
						
						\bibitem[\protect\citeauthoryear{James, Witten, Hastie  \& Tibshirani}{James
							et~al.}{2014}]{James2014}
						James G.,  Witten D.,  Hastie T.,   Tibshirani R.,  2014, An Introduction to
						Statistical Learning: With Applications in R.
						Springer Publishing Company, Incorporated
						
						\bibitem[\protect\citeauthoryear{Karatzoglou, Smola, Hornik  \&
							Zeileis}{Karatzoglou et~al.}{2004}]{kernlab}
						Karatzoglou A.,  Smola A.,  Hornik K.,   Zeileis A.,  2004, J. Stat. Softw.,
						11, 1
						
						\bibitem[\protect\citeauthoryear{Kuhn}{Kuhn}{2017}]{caret}
						Kuhn M.,  2017, caret: Classification and Regression Training.
						\url {https://CRAN.R-project.org/package=caret}
						
						\bibitem[\protect\citeauthoryear{Liaw \& Wiener}{Liaw \&
							Wiener}{2002}]{randomForest}
						Liaw A.,  Wiener M.,  2002, R News, 2, 18
						
						\bibitem[\protect\citeauthoryear{{Massey} \& {Hunter}}{{Massey} \&
							{Hunter}}{1998}]{MasseyHunter1998}
						{Massey} P.,  {Hunter} D.~A.,  1998, \mn@doi [The Astrophysical Journal]
						{10.1086/305126}, \href {http://adsabs.harvard.edu/abs/1998ApJ...493..180M}
						{493, 180}
						
						\bibitem[\protect\citeauthoryear{McLachlan \& Peel}{McLachlan \&
							Peel}{2000}]{FiniteMixtureModels}
						McLachlan G.~J.,  Peel D.~A.,  2000, Finite mixture models.
						Wiley series in probability and statistics : Applied probability and statistics
						section, Wiley, New York ; Weinheim [u.a.]
						
						\bibitem[\protect\citeauthoryear{Meyer, Dimitriadou, Hornik, Weingessel  \&
							Leisch}{Meyer et~al.}{2015}]{e1071}
						Meyer D.,  Dimitriadou E.,  Hornik K.,  Weingessel A.,   Leisch F.,  2015,
						e1071: Misc Functions of the Department of Statistics, Probability Theory
						Group (Formerly: E1071), TU Wien.
						\url {https://CRAN.R-project.org/package=e1071}
						
						\bibitem[\protect\citeauthoryear{{Montmerle}, {Feigelson}, {Bouvier}  \&
							{Andre}}{{Montmerle} et~al.}{1993}]{Montmerle1993}
						{Montmerle} T.,  {Feigelson} E.~D.,  {Bouvier} J.,   {Andre} P.,  1993, in
						{Levy} E.~H.,  {Lunine} J.~I.,  eds, Protostars and Planets III. pp 689--717
						
						\bibitem[\protect\citeauthoryear{{Nota} et~al.,}{{Nota}
							et~al.}{2006}]{Nota2006}
						{Nota} A.,  et~al., 2006, \mn@doi [\apjl] {10.1086/503301}, \href
						{http://cdsads.u-strasbg.fr/abs/2006ApJ...640L..29N} {640, L29}
						
						\bibitem[\protect\citeauthoryear{{Panagia}, {Gilmozzi}, {Macchetto}, {Adorf}
							\& {Kirshner}}{{Panagia} et~al.}{1991}]{Panagia1991}
						{Panagia} N.,  {Gilmozzi} R.,  {Macchetto} F.,  {Adorf} H.-M.,   {Kirshner}
						R.~P.,  1991, \mn@doi [The Astrophysical Journal] {10.1086/186164}, \href
						{http://adsabs.harvard.edu/abs/1991ApJ...380L..23P} {380, L23}
						
						\bibitem[\protect\citeauthoryear{{Panagia}, {Romaniello}, {Scuderi}  \&
							{Kirshner}}{{Panagia} et~al.}{2000}]{Panagia2000}
						{Panagia} N.,  {Romaniello} M.,  {Scuderi} S.,   {Kirshner} R.~P.,  2000,
						\mn@doi [The Astrophysical Journal] {10.1086/309212}, \href
						{http://adsabs.harvard.edu/abs/2000ApJ...539..197P} {539, 197}
						
						\bibitem[\protect\citeauthoryear{Platt}{Platt}{1999}]{Platt99probabilisticoutputs}
						Platt J.~C.,  1999, in ADVANCES IN LARGE MARGIN CLASSIFIERS. MIT Press, pp
						61--74
						
						\bibitem[\protect\citeauthoryear{{R Core Team}}{{R Core
								Team}}{2013}]{Rlanguage}
						{R Core Team} 2013, R: A Language and Environment for Statistical Computing.
						R Foundation for Statistical Computing, Vienna, Austria, \url
						{http://www.R-project.org/}
						
						\bibitem[\protect\citeauthoryear{{Romaniello}, {Panagia}, {Scuderi}  \&
							{Kirshner}}{{Romaniello} et~al.}{2002}]{Romaniello2002}
						{Romaniello} M.,  {Panagia} N.,  {Scuderi} S.,   {Kirshner} R.~P.,  2002,
						\mn@doi [The Astronomical Journal] {10.1086/338430}, \href
						{http://adsabs.harvard.edu/abs/2002AJ....123..915R} {123, 915}
						
						\bibitem[\protect\citeauthoryear{Russell, Norvig  \& Davis}{Russell
							et~al.}{2016}]{RussellNorvigAI}
						Russell S.~J.,  Norvig P.,   Davis E.,  2016, Artificial intelligence, third
						edition, global edition edn.
						Always learning, Pearson, Boston ; Munich
						
						\bibitem[\protect\citeauthoryear{{Sabbi} et~al.,}{{Sabbi}
							et~al.}{2007}]{Sabbi2007}
						{Sabbi} E.,  et~al., 2007, \mn@doi [\aj] {10.1086/509257}, \href
						{http://cdsads.u-strasbg.fr/abs/2007AJ....133...44S} {133, 44}
						
						\bibitem[\protect\citeauthoryear{{Sabbi} et~al.,}{{Sabbi}
							et~al.}{2008}]{Sabbi2008}
						{Sabbi} E.,  et~al., 2008, \mn@doi [\aj] {10.1088/0004-6256/135/1/173}, \href
						{http://cdsads.u-strasbg.fr/abs/2008AJ....135..173S} {135, 173}
						
						\bibitem[\protect\citeauthoryear{{Sabbi} et~al.,}{{Sabbi}
							et~al.}{2013}]{2013AJ....146...53S}
						{Sabbi} E.,  et~al., 2013, \mn@doi [\aj] {10.1088/0004-6256/146/3/53}, \href
						{http://adsabs.harvard.edu/abs/2013AJ....146...53S} {146, 53}
						
						\bibitem[\protect\citeauthoryear{{Sabbi} et~al.,}{{Sabbi}
							et~al.}{2016}]{2016ApJS..222...11S}
						{Sabbi} E.,  et~al., 2016, \mn@doi [\apjs] {10.3847/0067-0049/222/1/11}, \href
						{http://adsabs.harvard.edu/abs/2016ApJS..222...11S} {222, 11}
						
						\bibitem[\protect\citeauthoryear{{Schmalzl}, {Gouliermis}, {Dolphin}  \&
							{Henning}}{{Schmalzl} et~al.}{2008}]{Schmalzl2008}
						{Schmalzl} M.,  {Gouliermis} D.~A.,  {Dolphin} A.~E.,   {Henning} T.,  2008,
						\mn@doi [\apj] {10.1086/588722}, \href
						{http://cdsads.u-strasbg.fr/abs/2008ApJ...681..290S} {681, 290}
						
						\bibitem[\protect\citeauthoryear{{Schmeja}, {Gouliermis}  \&
							{Klessen}}{{Schmeja} et~al.}{2009}]{Schmeja2009}
						{Schmeja} S.,  {Gouliermis} D.~A.,   {Klessen} R.~S.,  2009, \mn@doi [\apj]
						{10.1088/0004-637X/694/1/367}, \href
						{http://cdsads.u-strasbg.fr/abs/2009ApJ...694..367S} {694, 367}
						
						\bibitem[\protect\citeauthoryear{{Schulz}}{{Schulz}}{2012}]{2012fees.book.....S}
						{Schulz} N.~S.,  2012, {The Formation and Early Evolution of Stars},
						\mn@doi{10.1007/978-3-642-23926-7.
						}
						
						\bibitem[\protect\citeauthoryear{{Stahler} \& {Palla}}{{Stahler} \&
							{Palla}}{2005}]{2005fost.book.....S}
						{Stahler} S.~W.,  {Palla} F.,  2005, {The Formation of Stars}
						
						\bibitem[\protect\citeauthoryear{{Stephens} et~al.,}{{Stephens}
							et~al.}{2017}]{StephensGouliermis2017}
						{Stephens} I.~W.,  et~al., 2017, \mn@doi [The Astrophysical Journal]
						{10.3847/1538-4357/834/1/94}, \href
						{http://adsabs.harvard.edu/abs/2017ApJ...834...94S} {834, 94}
						
						\bibitem[\protect\citeauthoryear{Szeliski}{Szeliski}{2011}]{ComputerVision}
						Szeliski R.,  2011, Computer Vision: Algorithms and Applications.
						Springer
						
						\bibitem[\protect\citeauthoryear{{Taylor}}{{Taylor}}{2005}]{topcat2005}
						{Taylor} M.~B.,  2005, in {Shopbell} P.,  {Britton} M.,   {Ebert} R.,  eds,
						Astronomical Society of the Pacific Conference Series Vol. 347, Astronomical
						Data Analysis Software and Systems XIV. p.~29
						
						\bibitem[\protect\citeauthoryear{Therneau, Atkinson  \& Ripley}{Therneau
							et~al.}{2017}]{rpart}
						Therneau T.,  Atkinson B.,   Ripley B.,  2017, rpart: Recursive Partitioning
						and Regression Trees.
						\url {https://CRAN.R-project.org/package=rpart}
						
						\bibitem[\protect\citeauthoryear{Venables \& Ripley}{Venables \&
							Ripley}{2002}]{MoApStatwS}
						Venables W.~N.,  Ripley B.~D.,  2002, Modern applied statistics with S, 4th
						edn.
						Statistics and computing, Springer, New York ; Berlin ; Heidelberg [u.a.]
						
						\bibitem[\protect\citeauthoryear{{Walborn} \& {Blades}}{{Walborn} \&
							{Blades}}{1997}]{WalbornBlades1997}
						{Walborn} N.~R.,  {Blades} J.~C.,  1997, \mn@doi [The Astrophysical Journal
						Supplement Series] {10.1086/313043}, \href
						{http://adsabs.harvard.edu/abs/1997ApJS..112..457W} {112, 457}
						
						\bibitem[\protect\citeauthoryear{{Zaritsky}, {Harris}, {Thompson}, {Grebel}  \&
							{Massey}}{{Zaritsky} et~al.}{2002}]{Zaritsky2002}
						{Zaritsky} D.,  {Harris} J.,  {Thompson} I.~B.,  {Grebel} E.~K.,   {Massey} P.,
						2002, \mn@doi [The Astronomical Journal] {10.1086/338437}, \href
						{http://adsabs.harvard.edu/abs/2002AJ....123..855Z} {123, 855}
\end{thebibliography}





\appendix

\section{Classification Algorithms} \label{app:MLCA}
		  
			  The following sections will give an overview over the three classification algorithms; decision tree, random forest classifier and support vector machine; which are used in Section \ref{chap:Classification} for the identification of PMS stars in the Tarantula Nebula, as well as two performance measures - the confusion matrix and the receiver operating characteristic curve - for these algorithms. All of the described methods are so-called {\em supervised learning techniques}, because they require a labelled data set to be trained on in order to perform a classification of new data. Here ``labelled" means that this training set consists of example observations, for which the class is known. In a two class scenario the examples of the class of primary interest, in our case the PMS stars, are often called ``{\em positives}", and examples that do not belong to that class are called ``{\em negative}" (i.e. lower main-sequence, LMS, upper main-sequence, UMS, red clump stars, etc., for us). In this context the space spanned by all possible values of the attributes of an object is usually referred to as the {\em feature space}, where an individual instance is represented by a feature vector, a vector that contains all its respective attributes.

The algorithms are applied with implementations in {\sl R}, a popular language and environment for statistical computing and graphics \citep{Rlanguage}. Specifically, packages {\tt e1071} \citep{e1071} and {\tt rpart} \citep{rpart} are used for the application of {\sl Naive Bayes}, and {\sl Decision Trees} respectively. For the application of {\sl Random Forests} and {\em Support Vector Machines} we used the {\sl R} package {\tt caret} \citep{caret}, which invokes package {\tt randomForest} \citep{randomForest} for the former and package {\tt kernlab} \citep{kernlab} for the latter method.

\begin{figure*}
\centering
\includegraphics[width = 0.8\linewidth]{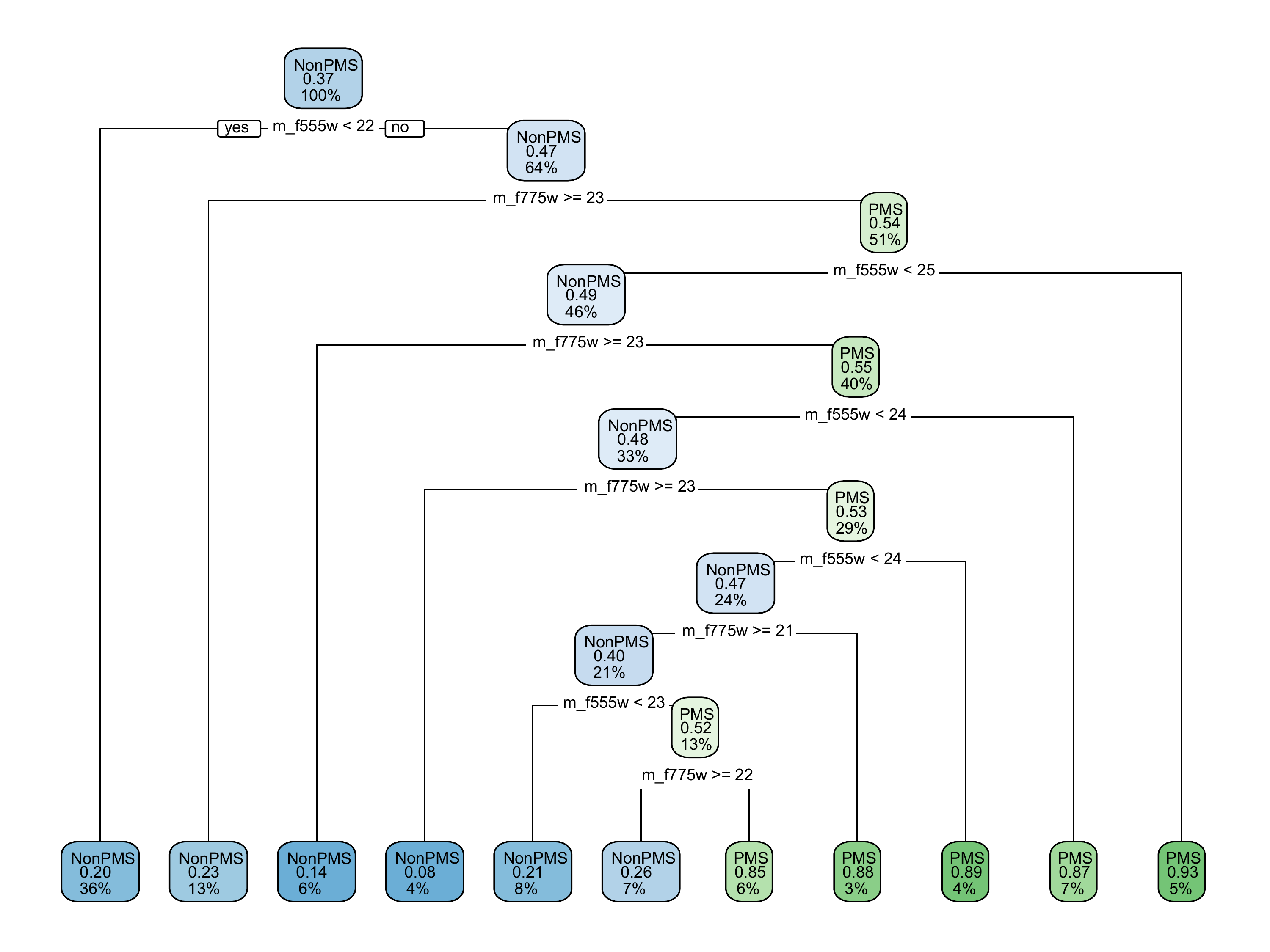}
\caption{Schematic representation of a pruned decision tree model, trained on the V- and R-equivalent magnitudes of the stars. Each node is labelled according to the majority class of observations in the node and shows furthermore the prediction probability for PMS, as well as the percentage of total observations assigned to the node.}
\label{fig:DTree_VR}
\end{figure*}

\subsection{Decision Trees} \label{app:dtree}

The general idea of a {\em decision tree classifier} or {\em classification tree} \citep{Breiman84ClassificationTrees} is that, ideally, it is possible to partition the feature space such that all object instances will be correctly classified. Thus, the result corresponds to a hierarchical partition of the feature space. This partition is represented by the end point (called {\em leaves} or {\em terminal nodes}) of a tree, where each node of the tree splits the feature space according to the value of a certain attribute. Interpreting the tree as a probability model, each node $i$ of the tree possesses a probability distribution $p_{ik}$ over the classes $k$. After building the tree, each case in the training set is assigned to one leaf, so that each leaf has a random sample $N_{ik}$ from the distribution $p_{ik}$ \citep{MoApStatwS}. 

The decision tree is constructed by recursively splitting the feature space until a stopping criterion is reached. At a certain node a split (usually a binary split, separating a continuous variable $x_j$ into $x_j <t$ and $x_j \geq t$) is chosen according to a measure of its value. Most commonly a measure of {\em impurity} is defined for each node. A node is considered to be pure if it only contains instances of a single class. A widely used impurity measure is the \textit{Gini index}:
\begin{equation}
G = \sum_{j \neq k}p_{ij}p_{ik}= 1 - \sum_{k} p_{ik}^2,
\label{eq:Gini_index}
\end{equation}
which measures the training error rate of classifying object instances in a node to class $k$ instead of the majority class of the node. For a pure node the Gini index is zero. Based on this impurity measure, at each node the split that reduces most the average impurity is performed. Stopping criteria can be when all nodes become pure, the tree reaches a maximum predefined depth or further splitting cannot reduce the average impurity more than a given minimal threshold. If the tree construction is stopped before all nodes become pure, the terminal nodes assume the majority class of their assigned training instances \citep{MoApStatwS}. Figure \ref{fig:DTree_VR} shows an example decision tree model, constructed on our training set using only V and R magnitudes (see Section \ref{sec:ClassDtree}).

\subsubsection*{Prediction}

To classify a new object instance, the decision tree propagates it according to its attributes along the tree, starting from the root, until a corresponding terminal node is reached, assigning the respective nodes class to the new object instance \citep{ElOfStatLearn}.
To some degree the decision tree can compensate for missing attributes of an object, by either assigning the majority class of the deepest non-terminal node reached with the available attributes or using {\em surrogate splits}. In the latter method each non-terminal node keeps a list of surrogate splits during the tree construction. During classification, if the primary split attribute is missing, one of these surrogate attributes is used to propagate the object instance further along the tree. Surrogate splits are constructed as follows:
During the construction of the tree when considering a certain attribute for a split only those training instances are considered, which are not missing that attribute. Afterwards a list of surrogate attributes and split points is generated, sorted according to how well this surrogate split approximates the split by the primary attribute. During prediction, surrogate splits are considered in that same order \citep{ElOfStatLearn}.

\subsubsection*{Tree Pruning}
If the training data are noisy, i.e. the class distributions overlap in feature space, a decision tree might 
{overfit} and thus perform badly on a set of new object instances. In order to avoid this one employs a method called \textit{cost-complexity pruning} \citep{James2014}. In this approach rooted subtrees of the decision tree are constructed by removing terminal subtrees. Then each of these subtrees is assigned a value $R$, which is the sum of some measure $R_i$ of the leaves of the tree. The size of these trees is equal to their number of leaves. One can now show that a set of rooted subtrees of tree T, which minimises the cost-complexity measure:
\begin{equation}
R_\alpha = R + \alpha \times \mathrm{size}
\end{equation}
is a nested tree. With increasing $\alpha$ one can find the optimal trees by a series of snip (i.e. cutting terminal subtrees) operations on the current tree, producing a sequence of trees with sizes of T down to just the root node. To choose the desired degree of pruning one computes an impurity measure versus $\alpha$ for the pruned tree and finds the smallest tree close to the minimum of the impurity measure when predicting on a separate validation set or using cross-validation \citep{MoApStatwS}.

\subsection{Random Forest}
\label{app:RF}

The basic concept of the {\em Random forest classifier} \citep{Breiman01RandomForests} is the so-called {\em bagging}, a general purpose procedure for variance reduction of statistical models through averaging many models of high variance and low bias \citep{James2014}. Decision tree models suffer from high variance, e.g., trees that fit to randomly determined halves of the same training data could vary significantly from one another, but provide low bias, if grown deep enough. Deep un-pruned decision trees are the underlying model of the {\em random forest classifier}. Following the principles of bagging, a random forest is constructed by building $B$ individual trees, which are grown by {\em bootstrapping} from the training data, i.e. taking repeated samples from the single training set, generating $B$ different bootstrapped training subsets \citep{James2014, ElOfStatLearn}. 

Improving upon a simple bagging of decision trees, the random forest further increases the variance reduction by a modification of the tree construction procedure. Instead of choosing the split attribute that reduces the impurity measure the most, $m$ random attributes out of the available $p$ are selected and the best variable and split point are determined out of those. 
{A small value for $m$ is typically helpful if a large number of the attributes are correlated. For classification purposes a general choice is $m = \sqrt{p}$ \citep[see, e.g.,][]{ElOfStatLearn}.}
The tree growth then proceeds until a minimum node size is reached. This procedure {\em decorrelates} the trees by preventing strong predictor attributes to dominate the split selection in all trees grown\footnote{{Suppose there is a single strong predictor attribute along with a number of moderately strong ones. If we grow $B$ decision trees with the standard procedure this strong predictor attribute would always be considered for the splits. With the random sampling procedure, however, on average $(p-m)/p$ splits will not even consider the dominant predictor for the split.}}, thus increasing the overall variance reduction of the bagging approach by averaging many {\em uncorrelated} models \citep{James2014}. In order to classify a new object instance the random forest classifier casts a {\em majority vote} over all the trees it has grown, i.e. each individual tree classifies the object, counting the results and assigning the class most voted for \citep{ElOfStatLearn}.

\begin{figure*}
	\centering
	\includegraphics[width = 0.33 \linewidth]{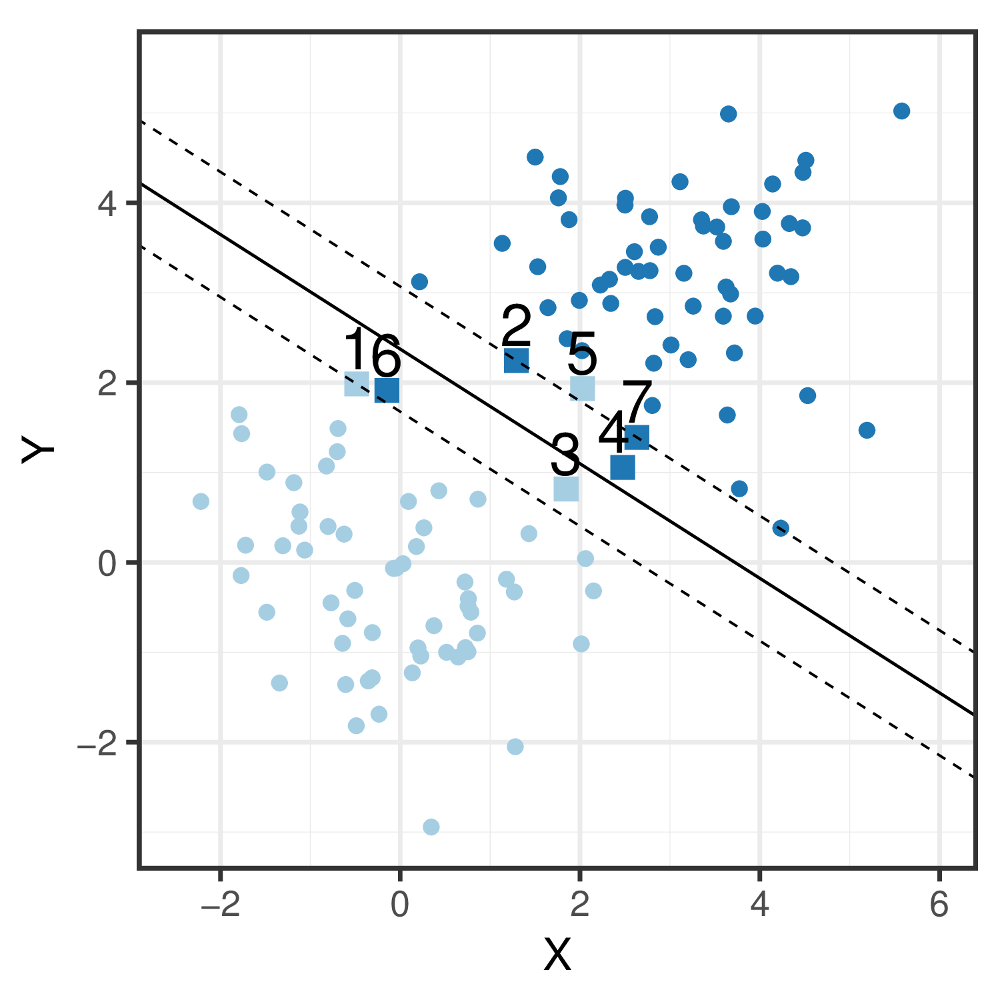}
	\includegraphics[width = 0.33 \linewidth]{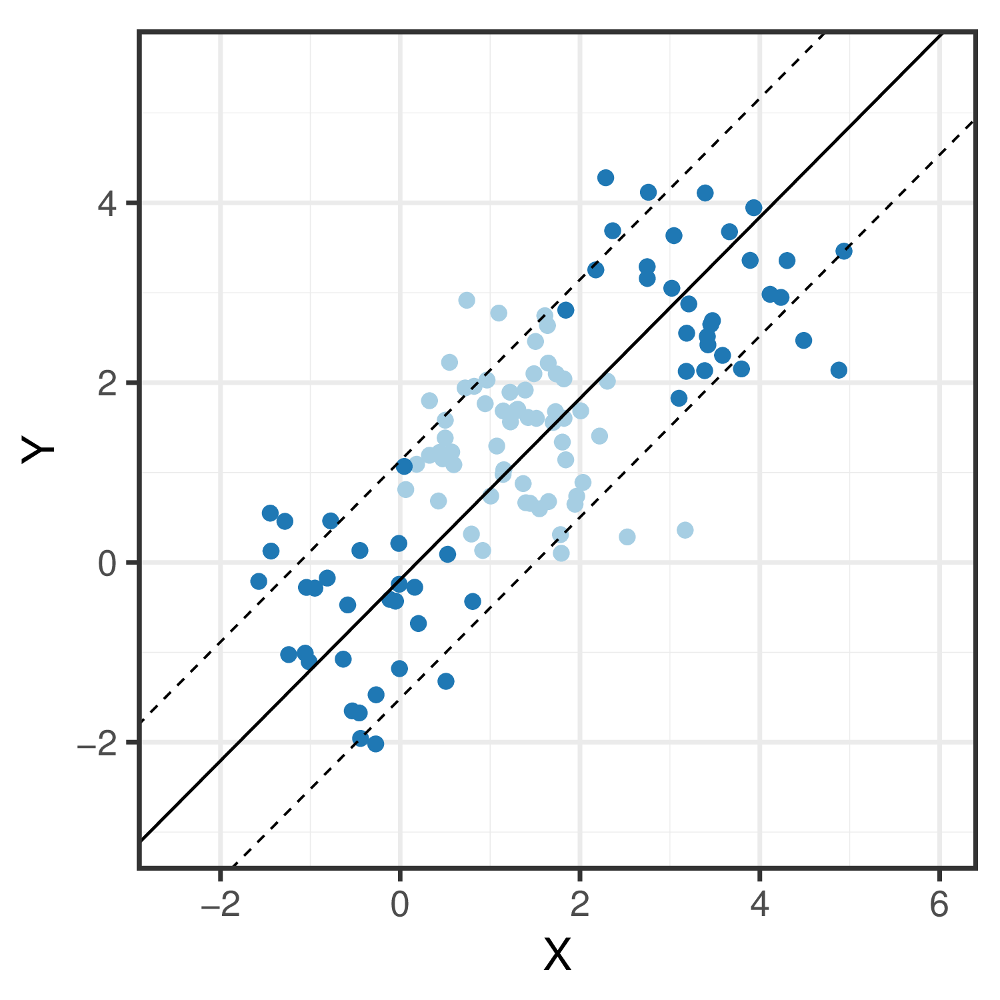}
	\includegraphics[width = 0.33 \linewidth]{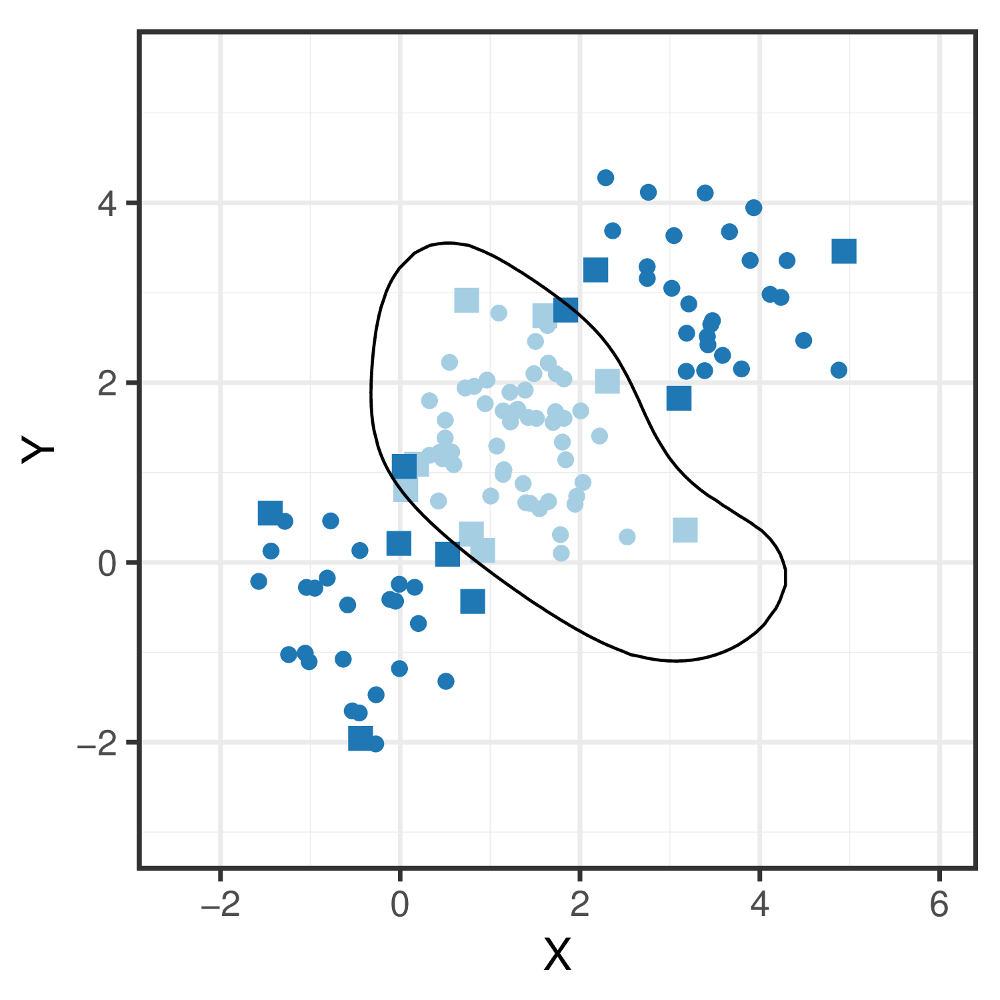}
	\caption{{\em Left:} Example of a support vector classifier fit to a small data set, distinguishing two classes of points in a 2d space. The black solid and dashed lines mark the constructed separating hyperplane and boundaries of the margin respectively. The coloured squares indicate the support vectors of the classifier. The instances 1 and 2 are support vectors lying {\em on} the margin, while 3 and 4 are examples for instances being on the {wrong side} of the margin of their respective class. 5 and 6 are instances that are on the wrong side of the margin {\em and} the separating hyperplane. {\em Middle:} Support vector classifier fit to a small data set, where the class boundaries are non-linear. Here the linear support vector classifier performs poorly. {\em Right:} Support Vector Machine using a radial basis kernel fit to the same data set. The solid black line indicates the non-linear decision boundary of the constructed SVM, while the coloured squares mark the support vectors.}
	\label{fig:SVC_SVM}
\end{figure*}

\subsection{Support Vector Machine}
\label{app:SVM}
	The support vector machine \citep{Cortes95SVM} is a classifier that produces a non-linear decision boundary in feature space by constructing a linear boundary in a transformed version of the feature space. It is a generalisation of the {\em support vector classifier}, which itself is based on the {\em maximal margin classifier} \citep{James2014,ElOfStatLearn}.
	\subsubsection*{Maximal Margin and Support Vector Classifier}
		The underlying concept of the maximal margin classifier is the {\em optimal separating hyperplane}, a hyperplane being a flat affine $p-1$ dimensional subspace of a p-dimensional space (e.g. a line in 2d or a plane in 3d space), 
		{describing the solution space to a set of linear equations}
		\begin{equation}
			x^T \beta + \beta_0 = 0,
		\end{equation}
		{where $\beta$ and $\beta_0$ denote a vector of coefficients and a constant vector respectively.}
		Given a set of $n$ $p$-dimensional training instances $x_i$, which fall into two classes with labels $y_i \in \{-1,1\}$, in a classification context a {\em seperating hyperplane} describes a hyperplane constructed such that it perfectly separates all training instances according to their class labels, i.e. having the property
		\begin{equation}
			y_i (\beta_0 + x_i^T\beta) > 0
		\end{equation}
		for all $i = 1, ..., n$ \citep{James2014}. Such a hyperplane induces a natural classification rule for a new test instance $x_*$ by assigning a class depending on which side of the hyperplane it is located, i.e. classifying $x_*$ based on the sign
		\begin{equation}
			G(x_*) = \mathrm{sign}[x_*^T \beta + \beta_0].
			\label{eq:MMclassifier}
		\end{equation} 
		The {\em margin} is defined as the minimum of the (perpendicular) distances of all training instances to a given hyperplane (e.g. the distance from the plane of points 1,2, and 7, ignoring points 3-6, as shown in Figure \ref{fig:SVC_SVM}, left panel). The {\em optimal separating} or {\em maximal margin hyperplane} is the separating hyperplane for which the margin is the largest, i.e. the hyperplane that has the farthest minimum distance to the training instances. A classifier Eq. \eqref{eq:MMclassifier} based on this hyperplane is a {\em maximal margin classifier}. Training instances that are equidistant from the maximal margin hyperplane and lie on the margin are called {\em support vectors}, as they ``support" the hyperplane in the sense that a variation of their position would change the hyperplane as well \citep{James2014}. To build this classifier one has to find the maximal margin hyperplane as the solution to the optimisation problem
		\begin{equation}
			\max_{\beta,\beta_0,||\beta||=1} M,
			\label{eq:MMC_OP1}
		\end{equation}
			subject to 
		\begin{equation}
			y_i (\beta_0 + x_i^T * \beta) \geq M \,\,\,\forall i = 1,...,n,
			\label{eq:MMC_OP2}
		\end{equation}
		where $M$, $M>0$, represents the width of the margin. 
		
		The {\em support vector classifier} is a generalisation of the maximal margin classifier for the case in which the training data is not linearly separable, i.e. when there is no solution to the optimisation problem with $M > 0$ (Figure \ref{fig:SVC_SVM}, left panel). The basic concept behind this method is a {\em soft margin}, which means that instead of constructing a hyperplane that {\em perfectly} separates the training instances, a hyperplane is built that allows some instances to be on the incorrect side of the margin or even of the hyperplane, i.e. a hyperplane that {\em almost} separates the classes. Such a hyperplane is the solution to the optimisation problem 
		\begin{equation}
			\max_{\beta,\beta_0, \epsilon_1, ..., \epsilon_n} M,  \,\,\,\,\,\,\,\,\,\, ||\beta||=1
			\label{eq:SVC1}
		\end{equation}
			subject to 
		\begin{equation}
		\begin{split}
			y_i (\beta_0 + x_i^T * \beta) &\geq M (1-\epsilon_i) \,\,\,\,\,\,\,\,\,\, \epsilon_i \geq 0,\,\\ \sum_{i=1}^{n}\epsilon_i &\leq C,
			\label{eq:SVC2}
		\end{split}
		\end{equation}
		where $C$, $C>0$, is a tuning parameter, $M$ is again the width of the margin, which is to be made as large as possible, and $\epsilon_1,...,\epsilon_n$ are {\em slack variables}, allowing individual instances to fall on the wrong side of the margin or hyperplane. A value $\epsilon_i = 0$ signifies that the $i$th training instance is on the correct side of the margin, while $\epsilon_i > 0$ indicates that it is on the wrong side of the margin ({\em violating} the margin, e.g. points 3 and 4 in Figure \ref{fig:SVC_SVM}, left panel). A value of $\epsilon_i >1$ indicates that the instance is on the wrong side of the hyperplane (e.g., points 5 and 6 in the figure). The tuning parameter $C$ ({often called \textit{cost}}), bounding the sum $\sum \epsilon_i$, signifies a {\em budget} that determines the number and severity of tolerated margin violations. Consequently a large $C$, tolerating many margin violations, results in a wider margin, while a smaller $C$ narrows it. Both this hyperplane, built by the support vector classifier, and the classifier itself are only dependent on training instances that lie directly on the margin (points 1,2,7 in \ref{fig:SVC_SVM}, left panel) or are violating it, i.e. its {\em support vectors} \citep{James2014}. 
	\subsubsection*{Support Vector Machine}
		The support vector classifier is an effective tool for a two-class setting if the two classes can be divided by a linear boundary. In the scenario of non-linear class boundaries, however, it will perform poorly without modification, as indicated by the example shown in Figure \ref{fig:SVC_SVM} (middle panel). To create such non-linear class boundaries with a support vector classifier one has to enlarge the feature space, by e.g. adding quadratic functions of the features. While the classifier is still linear within the enlarged feature space it was built in, it corresponds to a non-linear class boundary in the original feature space. This is the basic concept behind the {\em support vector machine} (SVM). The support vector classifier, i.e. the solution to the optimisation problem of Eqs. \eqref{eq:SVC1}, \eqref{eq:SVC2}, can be written as
		\begin{equation}
			f(x) = \beta_0 + \sum_{i=1}^{n} \alpha_i \langle x, x_i \rangle
		\end{equation}
		where $\langle x_i, x_{i'} \rangle = \sum_{j=1}^{p} x_{ij} x_{i'j}$ is the inner product and $\alpha_i$ ($i = 1,...,n$) are $n$ parameters, one per training instance, which in the solution are only nonzero for support vectors. One can now {\em generalise} the inner product with a {\em kernel function} $K(x,x_i)$. By choosing a {\em linear kernel} $K(x_i, x_{i'}) = \sum_{j=1}^{p} x_{ij} x_{i'j})$ we retrieve the normal support vector classifier, but if instead a {\em polynomial} or {\em radial} kernel function is chosen, we essentially fit a support vector classifier in a higher-dimensional space, constructing a non-linear class boundary in the original feature space \citep[see e.g][for a full presentation of the calculation]{ElOfStatLearn}. 
		
		This combination of a support vector classifier with a non-linear kernel function is a {\em support vector machine}. The kernel ``trick" has the advantage of not only being computationally efficient, but also avoiding the necessity for an explicit transformation to the enlarged feature space and even allowing the latter to become infinite-dimensional, as e.g. is the case for the radial kernel \citep{James2014, ElOfStatLearn}. Figure \ref{fig:SVC_SVM} (right panel) shows an example of the non-linear class boundary constructed by a SVM with a radial kernel on the data set, where the linear support vector classifier failed to construct a meaningful class boundary. To provide class probabilities instead of class labels, when using a SVM, one can use Platt's posterior probabilities, which fit a sigmoid function to the decision value $f$ of the support vector machine
		\begin{equation}
		P(y = 1| f) = \frac{1}{1+\exp(Af + B)},
		\end{equation}
		where A and B are estimated by minimizing the negative log-likelihood function \citep{kernlab, Platt99probabilisticoutputs}.

\subsection{Training and Performance Measures}
	\label{app:Performance}
			  \subsubsection{Training with Cross-Validation}
				  \label{app:CV}
				  Cross-validation is the most commonly used method for training classification models and estimating their prediction
				  error. Typically, a $k$-fold cross-validation is applied by 1) partitioning the dataset into $k$ equal-sized subsets, 2) training the algorithm on the total data of the $k-1$ subsets, while holding out the remaining subset to test its performance on, and 3) repeating step 2) $k$ times, while holding out each of the subsamples for testing in each iteration. As a result, none of the $k$ produced models has made predictions on its own training data. The model that predicts best among them is considered as the final classification model \citep[see, e.g.,][]{ElOfStatLearn}. While cross-validation is usually applied to test the modeling process, the evaluation of the performance of the {\em final model} is done via a Train/Test split of the training dataset.
			  	
			  \subsubsection{Confusion Matrix}
 				  \begin{table}
  				  	\centering
  				  	\caption{Example of a confusion matrix.}
  				  	\begin{tabular}{c|cc}
  				  		\hline
  				  		& Actual Positive & Actual Negative \\
  				  		\hline
  				  		Predicted Positive  & TP			   &  FP             \\
  				  		Predicted Negative  & FN 			   &  TN             \\
  				  		\hline
  				  	\end{tabular}
  				  	\label{tab:confmatrix}
 				  \end{table} 
				  The {\em confusion matrix} is a way to summarise the performance of a classification algorithm when predicting on a test set with known labels. It contains the following quantities:
				  \begin{enumerate}[leftmargin = *]
				  	\item {\em True Positives (TP)}: number of instances that are correctly predicted to be positives.
				  	\item {\em False Positives (FP)}: number of instances that are incorrectly predicted to be positives.
				  	\item {\em True Negatives (TN)}: number of instances that are correctly predicted to be negatives.
				  	\item {\em False Negatives (FN)}: number of instances that are incorrectly predicted to be negatives.
				  \end{enumerate}
				  An example of the confusion matrix is given in Table \ref{tab:confmatrix}. From the confusion matrix one can derive the {\em accuracy} performance measure by dividing the trace by the sum of all entries or calculating:
				  \begin{equation}
					  \mathrm{ACC} = \frac{\mathrm{TP + TN}}{\mathrm{P + N}},
					  \label{eq:accuracy}
				  \end{equation}
				  where P and N denote the number of positive and negative instances in the training set, respectively. 
				  {Further diagnostics that can be derived from the confusion matrix are the {\em true positive rate} (TPR), also called {\em Sensitivity}, and the {\em false positive rate} (FPR), {\em Specificity}, which are defined by
				  \begin{equation}
				    \begin{split}
				  		  \mathrm{TPR} &= \frac{\mathrm{TP}}{\mathrm{TP}+\mathrm{FN}} = \frac{\mathrm{TP}}{\mathrm{P}} \\
				  		  \mathrm{FPR} &= \frac{\mathrm{FP}}{\mathrm{FP}+\mathrm{TN}} = \frac{\mathrm{FP}}{\mathrm{N}}. \\
				  	  \end{split}
				  \end{equation}
				  The {\em Balanced Accuracy} is an accuracy measurement that accounts for an imbalance in the number of positive and negative instances in the training set. Therefore it allows for an assessment of the class-specific accuracy and is defined as the mean of Sensitivity and Specificity \citep[see, e.g.,][]{ComputerVision}
				  \begin{equation}
					  \mathrm{BACC} = \frac{1}{2} \left(\frac{\mathrm{TP}}{\mathrm{P}} +  \frac{\mathrm{FP}}{\mathrm{N}}\right).
				  \end{equation}}
			  \subsubsection{Receiver Operating Characteristic Curve}

				  \begin{figure}
					  \centering
					  \includegraphics[width = 0.9\columnwidth]{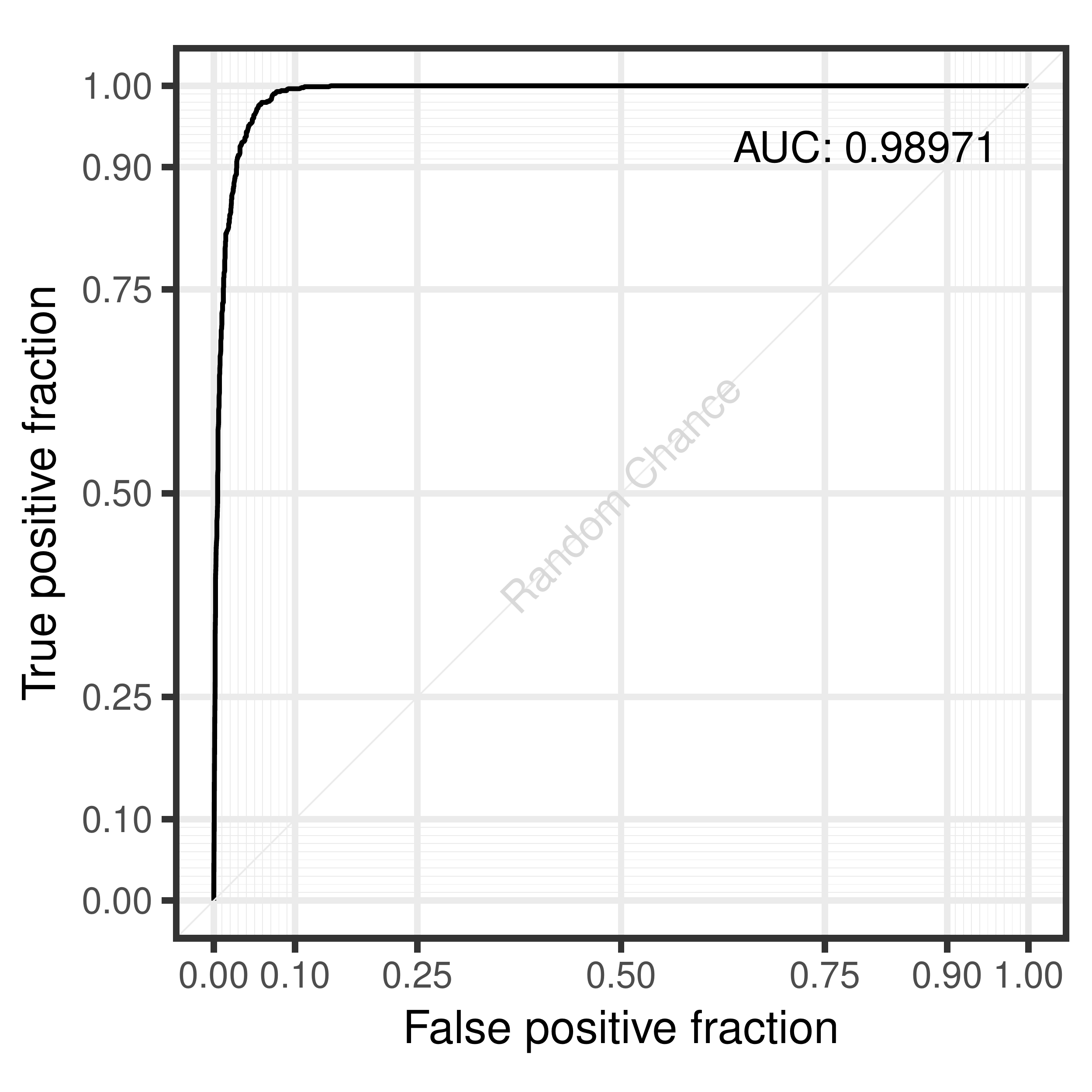}
					  \caption{Example ROC curve of the random forest on our features set no. 3 (see. Section \ref{chap:Classification}). }
					  \label{fig:ROC_example}
				  \end{figure}

				  The {\em Receiver Operating Characteristic (ROC) Curve} relates the {\em true positive rate} 
				  to the {\em false positive rate} 
				  for different parameters of the classification rule (such as the decision threshold). The closer the curve is to the top left corner, i.e. the larger the {\em area under the curve (AUC)}, the better the algorithm performs. The ROC curve of a randomly guessing algorithm corresponds to a straight line with unit slope. Consequently the AUC is a commonly used quantitative summary of performance of an algorithms \citep{ElOfStatLearn, ComputerVision}. Figure \ref{fig:ROC_example} shows an example ROC curve from our analysis (see Section \ref{sec:ClassRF}).


\section{Mixture models and the EM algorithm}\label{sec:em}

\label{sec:EM_algorithm}
Mixture Models are a useful method for density estimation, with the most popular being the Gaussian mixture model of the form
\begin{equation}
\label{eq:mixture}
f(x) = \sum_{m=1}^{M} \alpha_m \Phi(x; \mu_m, \mathbf{\Sigma}_m),
\end{equation}
where $\alpha_m$ denotes the mixing proportions, subject to $\sum_{m}^{}\alpha_m = 1$, $M$ marks the total number of components and the individual Gaussian densities have a mean $\mu_m$ and covariance matrix $\mathbf{\Sigma}_m$. These parameters are usually fit by maximum likelihood with, for instance, the {\em Expectation Maximisation (EM) algorithm} \citep{Dempster77maximumlikelihood}. A mixture model can then be used to provide an estimate of the posterior probability that a certain observation $i$ belongs to a component $m$, given by
\begin{equation}
\label{eq:posterior}
r_{im} = \frac{\alpha_m \Phi(x_i; \mu_m, \Sigma_m)}{\sum_{k=1}^{M} \alpha_k \Phi(x_i; \mu_k, \Sigma_k)}.
\end{equation}

When fitting a finite mixture model like Eq. \eqref{eq:mixture} to an observed random sample $\mathbf{x} = (x_1, ..., x_n)$ the log-likelihood from the data, which is to be maximised to retrieve the parameters of the model, takes the form
\begin{equation}
\label{eq:log_lik}
l(\theta; \mathbf{x}) = \sum_{j=1}^{n} \log(\sum_{i=1}^{M} \alpha_i \Phi_\theta(x_j; \theta_i))
\end{equation}
where $\theta = (\alpha_1, ..., \alpha_M, \theta_1, ..., \theta_M)$ denotes all parameters of the model and $\theta_i = (\mu_i, \Sigma_i)$ the parameters of mixture component $i$. In practice, maximising equation \eqref{eq:log_lik} can be complicated numerically due to the sum in the logarithm.
To alleviate this problem the {\em EM algorithm} treats it as an {\em incomplete data} problem. The observed data vector $\mathbf{x}$ is assumed to be incomplete, missing a set of associated component-label vectors $\mathbf{z} = (\mathbf{z}_1, ..., \mathbf{z}_n)$, where each $\mathbf{z}_j$ is a M-dimensional vector with $z_{ij} = 1$ or $0$, according to whether $x_j$ belongs to component $i$. Thus, the complete data vector is $ \mathbf{x}_c = (\mathbf{x}, \mathbf{z})$ with log-likelihood
\begin{equation}
\label{eq:log_lik_complete}
l(\theta;\mathbf{x_c}) = \sum_{i=1}^{M}\sum_{j=1}^{n} z_{ij}[\log\alpha_i + \log\Phi_\theta(x_j; \theta_i)].
\end{equation}
Based on this incomplete data assumption the EM algorithm proceeds iteratively, alternating between the {\em expectation (E)} and the {\em maximisation (M)} step \citep[see, e.g.,][]{FiniteMixtureModels, ElOfStatLearn, UBHD-}.

In the {\em E-step} the conditional expectation of the complete-data log-likelihood, based on the observed data $\mathbf{x}$ and the current fit $\theta^{(k)}$, expressed as the operator
\begin{equation}
\label{eq:E_operator}
Q(\theta'; \theta^{(k)}) = E(l(\theta', \mathbf{x}_c)|\mathbf{x},\theta^{(k)})
\end{equation}
is computed.
For the finite mixture model equation \eqref{eq:log_lik_complete} shows that the complete-data log-likelihood is linear in the latent data $z_{ij}$, so that the E-step in iteration k+1 only requires to calculate the current conditional expectation of $Z_{ij}$ (the random variable corresponding to $z_{ij}$) given the observations $\mathbf{x}$
\begin{equation}
E(Z_{ij}|\mathbf{x}, \theta^{(k)}) = p(Z_{ij}=1 | \mathbf{x};\theta^{(k)}) = r_i(x_j;\theta^{(k)}),
\end{equation}
where following \eqref{eq:posterior}
\begin{equation}
\label{eq:post2}
r_{i}(x_j; \theta^{(k)}) = \frac{\alpha_i^{(k)} \Phi_\theta(x_j;\theta_i^{(k)} )}{\sum_{h=1}^{M} \alpha_h^{(k)} \Phi_\theta(x_j;\theta_h^{(k)})}.
\end{equation}
With \eqref{eq:post2} the operator \eqref{eq:E_operator} becomes:
\begin{equation}
\label{eq:E_operator2}
Q(\theta'; \theta^{(k)}) =\sum_{i=1}^{M}\sum_{j=1}^{n} r_{i}(x_j; \theta^{(k)})[\log\alpha_i + \log\Phi_\theta(x_j; \theta_i)].
\end{equation}

In the {\em M-step} in iteration k+1, $Q(\theta'; \theta^{(k)})$ is globally maximised with respect to $\theta'$ to update the estimate of the parameters:
\begin{equation}
\label{eq:M_step}
\theta^{(k+1)} = \arg\max_{\theta'}	Q(\theta'; \theta^{(k)}).
\end{equation}
For {\em finite} mixture models the updated estimates of the mixing proportions $\alpha_i^{(k+1)}$ and the component parameters $\theta_j^{k+1}$ can be determined independently. The maximum likelihood estimate of the mixing proportions takes the form
\begin{equation}
\label{eq:M_step_props}
\alpha_i^{(k+1)} = \sum_{j=1}^{n} \frac{r_{i}(x_j; \theta^{(k)})}{n},
\end{equation}
while the update for the component parameters can be deduced by solving
\begin{equation}
\label{eq:mle_mstep}
\sum_{i=1}^{M}\sum_{j=1}^{n} r_{i}(x_j; \theta^{(k)}) \partial \log \Phi_\theta(x_j; \theta_i)/\partial \mathbf{\Theta} = 0,
\end{equation}
where $\Theta = (\theta_i, ..., \theta_M)$.

The algorithm stops once the difference $l(\theta^{(k+1)};\mathbf{x}) - l(\theta^{(k)};\mathbf{x})$ is smaller than some threshold provided that the sequence of likelihood values of the incomplete data $\{l(\theta^{(k)};\mathbf{x})\}$ converges. The EM-algorithm works because the EM-iteration does not decrease the log-likelihood of the incomplete data, i.e., 
\begin{equation}
l(\theta^{(k+1)};\mathbf{x}) \geq l(\theta^{(k)};\mathbf{x}).
\end{equation}
In Section \ref{sec:EM_fit} the EM algorithm was employed to fit a {\em mixture of two Gaussian normal distributions}. In this case the EM algorithm operates as follows \citep[see, e.g.,][]{FiniteMixtureModels, ElOfStatLearn, UBHD-}:
\begin{enumerate}[leftmargin = *]
	\item{Initially guess the parameters $\theta^{(0)}$.}
	\item{{\em E-step}: Compute equation \eqref{eq:E_operator2} via \eqref{eq:post2}.}
	\item{{\em M-step}: Update parameters according to \eqref{eq:M_step}, i.e calculate \eqref{eq:M_step_props} and \eqref{eq:M_step_bimodal}. In this case the solutions to equation \eqref{eq:mle_mstep} have the closed forms:
		\begin{equation}
		\label{eq:M_step_bimodal}
		\begin{split}
		\mu_i^{(k+1)} &= \frac{\sum_{j=1}^{n}r_{ij}^{(k)} x_j}{\sum_{j=1}^{n}r_{ij}^{(k)}} \\
		\sigma_i^{(k+1)} & = \frac{\sum_{j=1}^{n}r_{ij}^{(k)} (x_j-\mu_i^{(k+1)})^2}{\sum_{j=1}^{n}r_{ij}^{(k)}},\\
		\end{split}
		\end{equation}
		where $ r_{ij}^{(k)} = r_{i}(x_j; \theta^{(k)})$.}
	\item{Repeat steps 2 and 3 until convergence is reached.}
\end{enumerate}
\label{lastpage}

\end{document}